\documentclass[aps,prb,twocolumn,superscriptaddress,eqsecnum]{revtex4-1}

\usepackage[dvips]{graphicx}
\usepackage[pdftex]{color} 
\usepackage{multirow}
\usepackage{simplewick}
\usepackage{graphicx}
\usepackage{amssymb,bbm,amsmath,braket,indentfirst,bm}
\usepackage{hyperref}
\usepackage{color}
\usepackage{ulem} 

\def\infinity{\infty}
\def\t#1{\textrm{#1}}
\def\n{\nonumber \\ }
\def\tensor{\otimes}
\def\braket#1{{\langle #1 \rangle}}
\def\psibar{\bar{\psi}}

\begin{document}

\title{Anderson localization and the topology of classifying spaces}

\author{Takahiro Morimoto}
\affiliation{Condensed Matter Theory Laboratory, 
RIKEN, Wako, Saitama, 351-0198, Japan}

\author{Akira Furusaki}
\affiliation{Condensed Matter Theory Laboratory, 
RIKEN, Wako, Saitama, 351-0198, Japan}
\affiliation{RIKEN Center for Emergent Matter Science (CEMS), 
Wako, Saitama, 351-0198, Japan}

\author{Christopher Mudry}
\affiliation{Condensed Matter Theory Group,
  Paul Scherrer Institute, CH-5232 Villigen PSI, Switzerland}

\date{\today}

\begin{abstract}
We construct the generic phase diagrams encoding the topologically distinct
localized and delocalized phases of noninteracting fermionic quasiparticles
for any symmetry class from the tenfold way in one, two, and three dimensions.
To this end, we start from a massive Dirac Hamiltonian perturbed by
a generic disorder for any dimension of space and for any one of the ten
symmetry classes from the tenfold way. The physics of Anderson localization
is then encoded by a two-dimensional phase diagram that we deduce from
the topology of the space of normalized Dirac masses.
This approach agrees with previously known results and gives 
an alternative explanation for the even-odd effect
in the one-dimensional chiral symmetry classes. We also give a qualitative
explanation for the Gade singularity and Griffiths effects 
in the density of states using the first homotopy group of 
the normalized Dirac masses in two dimensions. 
Finally, this approach is used to analyze the stability 
of massless Dirac fermions on the surface of three-dimensional
topological crystalline insulators.
\end{abstract}


\maketitle


\section{Introduction} 
\label{sec: Introduction}

Anderson localization is the nonperturbative phenomenon by which the plane-wave
solutions to a linear differential equation become exponentially localized 
upon the breaking of translation symmetry by a local random potential.%
~\cite{Anderson58,1Footnote-on-local-disorder}
Anderson localization always prevails for metals in one-dimensional (1D)
space and for metals without spin-orbit coupling in two-dimensional (2D)
space, however small the disorder.%
~\cite{Lee85,Evers-Mirlin-rmp} 
It requires the disorder to be large enough, of the order 
of the Fermi energy, otherwise.%
~\cite{Lee85,Evers-Mirlin-rmp} 
Until the experimental discovery of the integer quantum Hall effect (IQHE)
in 1980,~\cite{Klitzing80}
the most important challenge brought about
by Anderson localization had been to understand the metal-insulator transition.
      
The IQHE is defined by the quantized value of the Hall conductivity
at fixed filling fraction and by a sharp (quantum) transition between
two consecutive quantized values of the Hall conductivity,
the plateau transition in short,
when the filling fraction is tuned. It is
explained by the topological character of the Hall conductivity
when the chemical potential is in between two consecutive Landau levels
of a two-dimensional electron gas subjected to a uniform magnetic field
and by the fact that this topological character is retained
in the regime of Anderson localization.%
~\cite{Laughlin81,Halperin82,TKNN82,Thouless84,Niu85}
The IQHE teaches two important lessons.
First, there can be topologically distinct insulating
phases of electronic matter. 
Second, direct continuous (quantum) transitions between 
these phases are possible. Several approaches have been used to study the
plateau transition from an analytical and a computational point of view.
Effective models such as non-linear sigma models (NLSMs),%
~\cite{Khmelnitskii83,Levine83,Pruisken84}
quantum network models,%
~\cite{Chalker88}
and Dirac fermions%
~\cite{ludwig94}
were proposed. In parallel, the two-parameter scaling theory of
Khmelnitskii and Levine \textit{et al}.\ was verified through numerous 
large-scale numerical simulations.%
~\cite{Huckestein95}
In this paper, we will generalize the approach pioneered by Ludwig
\textit{et al}.\ in Ref.~\onlinecite{ludwig94} 
by which they showed that the minimal continuum model that captures
the IQHE in both the clean and disordered limits is a 
Dirac Hamiltonian with random mass and gauge fields.

Integer quantum Hall states are examples of
topological insulators and superconductors defined as follows.
Topological insulators and superconductors are states of matter
such that fermionic quasiparticles are gapped in the bulk
but gapless on the boundary.%
~\cite{hasan-kane10,qi-zhang-rmp11} 
There are many materials that realize topological insulators, say
Bi$_{2}$Se$_{3}$.%
~\cite{hasan-kane10,qi-zhang-rmp11} 
Topological crystalline insulators differ from topological insulators
in that it is a spatial symmetry as opposed to a ``generic'' symmetry 
such as time-reversal symmetry that endows the boundary states with 
a topological protection. SnTe materials are examples of topological
crystalline insulators supporting an even number of massless Dirac cones 
on their mirror-symmetric boundaries.%
~\cite{hsieh12,dziawa2012topological,xu2012observation,tanaka2012experimental}
Weyl and Dirac semimetals are yet another family of noninteracting
fermionic Hamiltonians displaying gap-closing points, although 
in the Brillouin zone now, that are topologically stable.%
~\cite{murakami-semimetal07,Wan-semimetal11,burkov-balents11,Young-Dirac-semimetal12}
        
A defining property of (strong) topological insulators and superconductors
in $d$-dimensional space
is that their boundary states are immune to Anderson localization provided 
the disorder strength is smaller than the bulk gap.%
~\cite{2Footnote-on-bulk-gap}
This defining property was used by Schnyder \textit{et al}.\ to establish 
a classification of topological insulators and superconductors as follows.%
~\cite{schnyder-ryu08,schnyder-aip-proc09,ryu-njp10}
For a local disorder that is not too strong, its effect
on extended boundary states dispersing through the gap of the bulk
states can be treated as a problem of Anderson localization
in $(d-1)$ dimensions. Following the classification 
by Altland and Zirnbauer of metallic or superconducting quantum dots
(zero-dimensional space)
in terms of ten universality classes,%
~\cite{Zirnbauer96,Altland96,AZ-classes}
Schnyder \textit{et al}.\ examined the possibility of adding 
either a Wess-Zumino-Witten (WZW) term or
a $\mathbb{Z}^{\,}_{2}$ topological term to the NLSM describing 
$(d-1)$-dimensional transport along a boundary under the assumption
that the presence of such topological terms prevents Anderson localization
of the boundary states.
There follows a table with the ten Altland-Zirnbauer (AZ) symmetry classes
as rows and the dimensionality $d$ of space as columns 
(see Table~\ref{table: AZ classes}). 
The entries of this table, the tenfold way of topological insulators
and superconductors, are identified as either 
a trivial insulator/superconductor when no WZW and no 
$\mathbb{Z}^{\,}_{2}$ topological term are allowed,
a $\mathbb{Z}$ topological insulator/superconductor when a WZW term is allowed,
or a $\mathbb{Z}^{\,}_{2}$ topological insulator/superconductor 
when a $\mathbb{Z}^{\,}_{2}$ topological term is allowed.%
~\cite{schnyder-ryu08,schnyder-aip-proc09,ryu-njp10}
        
An alternative derivation of this table was proposed independently by
Kitaev using K theory.~\cite{kitaev09}
K theory enables a systematic study of a homotopical structure 
of gapped Hamiltonians in the bulk (upon the imposition of 
periodic boundary conditions, say) in the clean limit.
Appropriately normalized gapped Hamiltonians 
define ``a classifying space'' 
for each AZ symmetry class.
The tenfold way of topological insulators 
follows from the homotopy groups of the classifying spaces,
with the additional insight that it obeys
a periodicity that originates from the Bott periodicity in K theory. 
Since the periodic table is also obtained by using a representation theory of 
massive Dirac Hamiltonians,%
~\cite{ryu-njp10}
the approach relying on K theory also delivers 
a systematic classification of Dirac masses entering Dirac Hamiltonians
satisfying the AZ symmetry constraints.%
~\cite{morimoto-clifford13}

The goal of this paper is to reverse the logic of Schnyder \textit{et al}.
Instead of classifying topological insulators as an exercise in evading
Anderson localization on the boundary, 
we seek to understand how topology determines
Anderson localization in the bulk of a disordered insulator.
To this end, we marry the approach pioneered by Ludwig
\textit{et al}.\ for the IQHE with that of Kitaev.
In any dimension $d$ and for
any AZ symmetry class, we study all random Dirac Hamiltonians
that support a Dirac mass. In particular, we show that the homotopy groups
of the classifying spaces, regarded as the space of normalized Dirac masses
in each AZ symmetry class, are essential to establish the phase diagram 
encoding Anderson localization for these random Dirac Hamiltonians.

Our approach gives a qualitative understanding of a generic global
phase diagram of Anderson localization for any AZ symmetry class in
one, two, three, etc.\ dimensions.
Our phase diagram is consistent with previously known results
and gives a simple and alternative explanation for the even-odd
effect in the one-dimensional chiral classes.%
~\cite{Brouwer98,MBF-AIII-99,BMF-PRL00,BrouwerMudryFurusaki00}
We can also give a qualitative explanation for the Gade singularity%
~\cite{gade-wegner,gade1993anderson,guruswamy2000gl,mudry-gade-singularity03}
and Griffiths effects%
~\cite{Motrunich-2D-02}
in two dimensions for five out of the ten AZ symmetry classes
in terms of the first homotopy group of the classifying spaces.

Our approach can also be adapted to analyze the stability of
massless Dirac fermions on the surface of
3D topological crystalline insulators.
We establish the conditions for the stability of the boundary states of
statistical topological crystalline insulators (STCIs)
that are protected by a crystalline (here a reflection) symmetry on average 
for the ten AZ symmetry classes.
In this way, we show that
the alloy Sn${}^{\,}_{1-x}$Pb${}^{\,}_{x\vphantom{1}}$Te 
has surface states that are protected by the local symmetry class AII
and the average reflection symmetry.

This paper is organized as follows.
In Sec.~\ref{sec: Classifying spaces of normalized Dirac masses},
we review the classification of topological insulators and superconductors 
using the K theory and present classifying spaces of normalized Dirac masses 
in Dirac Hamiltonians.
In Sec.~\ref{sec: Anderson localization and the zeroth homotopy group of ...},
we relate the topology of classifying spaces to Anderson localization.
In Secs.~\ref{sec: Application to 1D space},
\ref{sec: Application to 2D space},
and \ref{sec: Application to 3D space}, 
we apply our method to 1D, 2D, and 3D systems, respectively,
to infer global phase diagrams.
In Sec.~\ref{sec: Boundaries of topological insulators}, 
we briefly explain how one may use the classifying spaces 
to study the stability of boundary states of topological insulators.
In Sec.~\ref{sec: Surface stability of reflection topological insulators},
we study the stability of the boundary states in a STCI.
We present a summary in Sec.~\ref{sec: Summary}.
In Appendix A we present explicit constructions of 
1D massive Dirac Hamiltonians and discuss some topological
properties of the space of normalized Dirac masses.
Finally, we discuss one-loop renormalization-group flows
of random masses and potentials for 2D Dirac Hamiltonians in Appendix B.

\section{Classifying spaces of normalized Dirac masses} 
\label{sec: Classifying spaces of normalized Dirac masses}

\begin{table*}[tb]
\begin{center}
\caption{\label{table: AZ classes}
The ten Altland-Zirnbauer (AZ) symmetry classes and their
topological classification in terms of the zeroth homotopy groups
of their classifying spaces.
Two complex and eight real symmetry classes are characterized
by the presence or the absence of 
time-reversal symmetry ($T$), 
particle-hole symmetry ($C$), 
and chiral symmetry ($\Gamma$).
Their presence is complemented
by the sign multiplying the identity 
in $T^{2}=\pm1$ or $C^{2}=\pm1$, 
and by $1$ for $\Gamma$.
Their absence is indicated by 0.
For each symmetry class and for the dimension $d$ of space,
the relevant extension problem with its solution in terms of the 
classifying space 
$V^{\,}_{d}\equiv\lim_{N\to\infty}\,V^{\,}_{d,r^{\,}_{\mathrm{min}}\,N}$
are given in the fifth and sixth columns, respectively. 
The zeroth homotopy groups of the classifying spaces
for $d=0,\ldots,7$ are given in the last eight columns.
Each entry with a nontrivial zeroth homotopy group of the classifying space
defines a noninteracting topological (superconductor) insulator.
\\
}
\begin{tabular}[t]{ c c c c c c c c c c c c c c}
\hline \hline
Class 
& 
$T$ 
& 
$C$ 
& 
$\Gamma$ 
& 
Extension 
& 
$V^{\,}_{d}$ 
& 
$\pi^{\,}_{0}(V^{\,}_{d=0})$
& 
$\pi^{\,}_{0}(V^{\,}_{d=1})$ 
&
$\pi^{\,}_{0}(V^{\,}_{d=2})$ 
&
$\pi^{\,}_{0}(V^{\,}_{d=3})$ 
&
$\pi^{\,}_{0}(V^{\,}_{d=4})$
& 
$\pi^{\,}_{0}(V^{\,}_{d=5})$ 
&
$\pi^{\,}_{0}(V^{\,}_{d=6})$ 
&
$\pi^{\,}_{0}(V^{\,}_{d=7})$ 
\\
\hline
A   
& 
0 
& 
0    
&  
0     
& 
$Cl^{\,}_{d}\to Cl^{\,}_{d+1}$ 
& 
$C^{\,}_{0+d}$ 
& 
$\mathbb{Z}$   
& 
0
&
$\mathbb{Z}$   
&
0
& 
$\mathbb{Z}$   
& 
0
&
$\mathbb{Z}$   
&
0
\\
AIII  
& 
0 
& 
0    
&  
1     
& 
$Cl^{\,}_{d+1}\to Cl^{\,}_{d+2}$ 
& 
$C^{\,}_{1+d}$ 
& 
0 
&
$\mathbb{Z}$   
&
0
&
$\mathbb{Z}$ 
& 
0 
&
$\mathbb{Z}$   
&
0
&
$\mathbb{Z}$ 
\\
\hline
AI   
& 
$+1$ 
& 
0    
&  
0     
& 
$Cl^{\,}_{0,d+2}\to Cl^{\,}_{1,d+2}$ 
& 
$R^{\,}_{0-d}$ 
& 
$\mathbb{Z}$
&
0
&
0
&
0
&
$\mathbb{Z}$
&
0
&
$\mathbb{Z}^{\,}_{2}$
&
$\mathbb{Z}^{\,}_{2}$
\\
BDI  
& 
$+1$ 
& 
$+1$ 
&  
1     
& 
$Cl^{\,}_{d+1,2}\to Cl^{\,}_{d+1,3}$ 
& 
$R^{\,}_{1-d}$ 
& 
$\mathbb{Z}^{\,}_{2}$ 
&
$\mathbb{Z}$
&
0
&
0
&
0
&
$\mathbb{Z}$
&
0
&
$\mathbb{Z}^{\,}_{2}$
\\
D    
& 
0    
& 
$+1$ 
&  
0     
& 
$Cl^{\,}_{d,2}\to Cl^{\,}_{d,3}$   
& 
$R^{\,}_{2-d}$ 
& 
$\mathbb{Z}^{\,}_{2}$ 
&
$\mathbb{Z}^{\,}_{2}$ 
&
$\mathbb{Z}$
&
0
&
0
&
0
&
$\mathbb{Z}$
&
0
\\
DIII 
& 
$-1$ 
& 
$+1$ 
&  
1     
& 
$Cl^{\,}_{d,3}\to Cl^{\,}_{d,4}$   
& 
$R^{\,}_{3-d}$ 
& 
0              
&
$\mathbb{Z}^{\,}_{2}$ 
&
$\mathbb{Z}^{\,}_{2}$ 
&
$\mathbb{Z}$
&
0
&
0
&
0
&
$\mathbb{Z}$
\\
AII  
& 
$-1$ 
& 
0    
&  
0     
& 
$Cl^{\,}_{2,d}\to 
Cl^{\,}_{3,d}$   
& 
$R^{\,}_{4-d}$ 
& 
$\mathbb{Z}$   
&
0
&
$\mathbb{Z}^{\,}_{2}$ 
&
$\mathbb{Z}^{\,}_{2}$ 
&
$\mathbb{Z}$
&
0
&
0
&
0
\\
CII  
& 
$-1$ 
& 
$-1$ 
&  
1     
& 
$Cl^{\,}_{d+3,0}\to Cl^{\,}_{d+3,1}$ 
& 
$R^{\,}_{5-d}$ 
& 
0              
&
$\mathbb{Z}$
&
0
&
$\mathbb{Z}^{\,}_{2}$
&
$\mathbb{Z}^{\,}_{2}$ 
&
$\mathbb{Z}$
&
0
&
0
\\
C    
& 
0    
& 
$-1$ 
&  
0     
& 
$Cl^{\,}_{d+2,0}\to Cl^{\,}_{d+2,1}$ 
& 
$R^{\,}_{6-d}$ 
& 
0              
&
0
&
$\mathbb{Z}$
&
0
&
$\mathbb{Z}^{\,}_{2}$
&
$\mathbb{Z}^{\,}_{2}$
&
$\mathbb{Z}$
&
0 
\\
CI   
& 
$+1$
& 
$-1$ 
&  
1     
& 
$Cl^{\,}_{d+2,1}\to Cl^{\,}_{d+2,2}$ 
& 
$R^{\,}_{7-d}$ 
& 
0              
&
0
&
0
&
$\mathbb{Z}$
&
0
&
$\mathbb{Z}^{\,}_{2}$ 
&
$\mathbb{Z}^{\,}_{2}$ 
&
$\mathbb{Z}$ 
\\
\hline \hline
\end{tabular}
\end{center}
\end{table*}

The key object of this paper
is the set of allowed normalized Dirac masses for each dimension $d$ 
of space, rank of the Dirac matrices, and AZ symmetry classes.
This set can be
attached a topology, i.e., it becomes a topological space.
As a topological space, it
is given by a classifying space associated with the extension problem 
of a Clifford algebra, which we review in this section.
Classifying spaces are characterized by homotopy groups 
that obey Bott periodicity, a property that carries over to
the allowed normalized Dirac masses.
The homotopy groups characterizing the allowed normalized Dirac masses
are used to characterize the physics of Anderson localization in 
Secs.~\ref{sec: Anderson localization and the zeroth homotopy group of ...},
\ref{sec: Application to 1D space},
and
\ref{sec: Application to 2D space}.

This section is devoted to the systematic construction 
of the classifying spaces $V^{\,}_{d}$ that are given in the sixth column
of Table~\ref{table: AZ classes}. 
To this end, we shall proceed in the following steps.
First, we review the algebraic definition of Clifford algebras.
Second, we review the relation between Clifford algebras and Dirac Hamiltonians.
Third, we review the tenfold way for Clifford algebras.
Fourth, we review the tenfold way for classifying spaces.

\subsection{Definitions of Clifford algebras and their classifying spaces}
\label{subsec: Definitions of Clifford algebras}

The complex Clifford algebra
\begin{subequations}
\label{eq: def Clq}
\begin{equation}
Cl^{\,}_{q}\equiv
\left\{e^{\,}_{1},\ldots,e^{\,}_{q}\right\}
\label{eq: def Clq a}
\end{equation}
is a complex vector space $\mathbb{C}^{2^{q}}$ 
of dimension $2^{q}$ that is spanned by the basis with the basis elements
\begin{equation}
e^{\,}_{n^{\,}_{1}\ldots n^{\,}_{q}}\equiv
\prod_{\iota=1}^{q}(e^{\,}_{\iota})^{n^{\,}_{\iota}},
\qquad
n^{\,}_{1},\ldots,n^{\,}_{q}=0,1,
\label{eq: def Clq b}
\end{equation}
whereby the multiplication rule
\begin{equation}
\left\{e^{\,}_{\iota},e^{\,}_{\iota'}\right\}=
2\,\delta^{\,}_{\iota,\iota'}
\label{eq: def Clq c}
\end{equation} 
\end{subequations}
applies $\iota,\iota'=1,\ldots,q$. The vector space $Cl^{\,}_{q}$
is closed under multiplication of any two of its elements
owing to Eq.~(\ref{eq: def Clq c}). Hence 
$Cl^{\,}_{q}$ is also an associative algebra.

The real Clifford algebra
\begin{subequations}
\label{eq: def Clp,q}
\begin{equation}
Cl^{\,}_{p,q}\equiv
\left\{e^{\,}_{1},\ldots,e^{\,}_{p};e^{\,}_{p+1},\ldots,e^{\,}_{p+q}\right\}
\label{eq: def Clp,q a}
\end{equation}
is a real vector space $\mathbb{R}^{2^{p+q}}$ 
of dimension $2^{p+q}$ that is spanned by the basis with the basis elements
\begin{equation}
e^{\,}_{n^{\,}_{1}\ldots n^{\,}_{p+q}}\equiv
\prod_{\iota=1}^{p+q}(e^{\,}_{\iota})^{n^{\,}_{\iota}},
\qquad
n^{\,}_{1},\ldots,n^{\,}_{p+q}=0,1,
\label{eq: def Clp,q b}
\end{equation}
whereby the multiplication rule
\begin{equation}
\begin{split}
&
\left\{e^{\,}_{\iota},e^{\,}_{\iota'}\right\}=
2\,\eta^{\,}_{\iota,\iota'},
\\
&
\eta^{\,}_{\iota,\iota'}=
\mathrm{diag}\,
(
\overbrace{-1,\ldots,-1}^{\mbox{$p$ times}},
\overbrace{+1,\ldots,+1}^{\mbox{$q$ times}}
),
\end{split}
\label{eq: def Clp,q c}
\end{equation} 
\end{subequations}
applies for $\iota,\iota'=1,\ldots,p+q$.
The vector space $Cl^{\,}_{p,q}$
is closed under multiplication of any two of its elements
owing to Eq.~(\ref{eq: def Clp,q c}). Hence 
$Cl^{\,}_{p,q}$ is also an associative algebra.

Given a representation of the complex Clifford algebra $Cl^{\,}_{q}$,
``the extension problem'' denoted
\begin{equation}
Cl^{\,}_{q}\to Cl^{\,}_{q+1}
\end{equation}
consists in identifying ``the classifying space'' $C^{\,}_{q}$
that parametrizes the representation of the generator
$e^{\,}_{q+1}$ present in $Cl^{\,}_{q+1}$ but absent in
$Cl^{\,}_{q}$. Similarly, given a representation of
the real Clifford algebra $Cl^{\,}_{p,q}$, 
there are two possible extension problems.
There is the  extension problem
\begin{equation}
Cl^{\,}_{p,q}\to Cl^{\,}_{p,q+1}
\end{equation}
that consists in identifying the classifying space $R^{\,}_{q-p}$
that parametrizes the representation of the generator
$e^{\,}_{p+q+1}$ present in $Cl^{\,}_{p,q+1}$ 
and thus satisfying $e^{2}_{p+q+1}=+1$, but absent in
$Cl^{\,}_{p,q}$.
There is the  extension problem
\begin{equation}
Cl^{\,}_{p,q}\to Cl^{\,}_{p+1,q}
\label{eq: extension problem with solution R p-q+2}
\end{equation}
that consists in identifying the classifying space $R^{\,}_{p-q+2}$
that parametrizes the representation of the generator
$e^{\,}_{p+1}$ present in $Cl^{\,}_{p+1,q}$ 
and thus satisfying $e^{2}_{p+1}=-1$, but absent in
$Cl^{\,}_{p,q}$.
The latter extension problem
can be related to the former one through the homeomorphism
\begin{equation}
Cl^{\,}_{p,q}\otimes Cl^{\,}_{0,2}\simeq Cl^{\,}_{q,p+2},
\end{equation}
where $Cl^{\,}_{0,2}$ is a linear algebra of real two-dimensional matrices
and does not affect the extension problem.
Classifying spaces depend on the difference $q-p$ only because
of the homeomorphism 
\begin{equation}
Cl^{\,}_{p+1,q+1}\simeq 
Cl^{\,}_{p,q}\tensor Cl^{\,}_{1,1}\simeq Cl^{\,}_{p,q}\tensor Cl^{\,}_{0,2},
\end{equation}
where dropping $Cl^{\,}_{0,2}$ does not affect the extension problem. 

Now, K-theory makes the remarkable
prediction that there are two families of complex classifying spaces 
$C^{\,}_{0}$ and $C^{\,}_{1}$,
while there are eight families of real classifying spaces 
$R^{\,}_{0},\ldots,R^{\,}_{7}$.
In other words, the dependence on $q$ 
enters modulo two for the complex classifying spaces,
\begin{subequations}
\label{eq: consequences Bott periodicity}
\begin{equation}
C^{\,}_{q+2}\simeq C^{\,}_{q},
\end{equation}
while it enters modulo eight for the real classifying spaces,
\begin{equation}
R^{\,}_{q+8}\simeq R^{\,}_{q},
\end{equation}
\end{subequations}
according to Bott's periodicity.%
~\cite{karoubi}
All families of classifying spaces $V$
in Table~\ref{table: all homotopies classifying spaces}
are labeled by the integer number $N$ entering the rank 
\begin{equation}
r=
r^{\,}_{\mathrm{min}} N
\end{equation}
assumed for the representation of the Clifford algebras.

\begin{table*}[tb]
\begin{center}
\caption{\label{table: all homotopies classifying spaces}
Complex and real classifying spaces are a list of ten topological
spaces that are built out of
the compact Lie groups $U(N)$, $O(N)$, and $Sp(N)$
-- the unitary, orthogonal, and symplectic matrix groups, respectively --
and their quotients as is given in the column ``Classifying space''.
The symbols denoting them are given in the column ``Label''. 
The number $N$ is related to the rank $r=r^{\,}_{\mathrm{min}}\,N$
of the Dirac matrices that form the
Clifford algebras, i.e.,
$N=1,2,\ldots$ is the number of copies of the minimal massive
Dirac Hamiltonian of rank $r^{\,}_{\mathrm{min}}$.
Let $p=0,1,\ldots$ index the $p$th homotopy group. 
The complex classifying spaces obey the periodicity condition
$\pi^{\,}_{p}(C^{\,}_{q})=\pi^{\,}_{p+2}(C^{\,}_{q})$
for $q=0,1$.
The real classifying spaces obey the periodicity condition
$\pi^{\,}_{p}(R^{\,}_{q})=\pi^{\,}_{p+8}(R^{\,}_{q})$
for $q=0,\dots,7$. Hence an exhaustive list of
the homotopy groups of the classifying spaces is shown in the columns 
``$\pi^{\,}_{0}(V)$'', ``$\pi^{\,}_{1}(V)$'',...,``$\pi^{\,}_{7}(V)$''. 
In each homotopy column, the three entries $\mathbb{Z}$ 
hold for $N$ larger than an integer (infinity included)
that depends on the order
of the homotopy group and the classifying space, 
the two entries $\mathbb{Z}^{\,}_{2}$ 
hold for $N$ larger than an integer that also depends on the order
of the homotopy group and the classifying space.
The entry $0$ is a short hand for the group $\{0\}$
made of the single element $0$. 
        }
\begin{tabular}[t]{cccccccccc}
\hline \hline
Label 
& 
Classifying space $V$
& 
$\pi^{\,}_{0}(V)$ 
&
$\pi^{\,}_{1}(V)$ 
&
$\pi^{\,}_{2}(V)$
&
$\pi^{\,}_{3}(V)$ 
 &
$\pi^{\,}_{4}(V)$ 
&
$\pi^{\,}_{5}(V)$ 
&
$\pi^{\,}_{6}(V)$ 
&
$\pi^{\,}_{7}(V)$
\\
\hline
$C^{\,}_{0}$
&
$\cup_{n=0}^{N}\big\{U(N)/\big[U(n)\times U(N-n)\big]\big\}$
&
$\mathbb{Z}$
&
$0$
&
$\mathbb{Z}$
&
$0$
&
$\mathbb{Z}$
&
$0$
&
$\mathbb{Z}$
&
$0$
\\
~$C^{\,}_{1}$
&
~$U(N)$
&
$0$
&
$\mathbb{Z}$
&
$0$
&
$\mathbb{Z}$
&
$0$
&
$\mathbb{Z}$
&
$0$ 
&
$\mathbb{Z}$
\\
\hline 
~$R^{\,}_{0}$
&
~$\cup_{n=0}^{N}\big\{O(N)/\big[O(n)\times O(N-n)\big]\big\}$
& 
$\mathbb{Z}$
&
$\mathbb{Z}^{\,}_{2}$
&
$\mathbb{Z}^{\,}_{2}$
&
$0$
&
$\mathbb{Z}$
&
$0$
& 
$0$
&
$0$
\\
~$R^{\,}_{1}$
& 
~$O(N)$
& 
$\mathbb{Z}^{\,}_{2}$ 
&
$\mathbb{Z}^{\,}_{2}$
&
$0$
&
$\mathbb{Z}$
&
$0$
&
$0$
&
$0$
&
$\mathbb{Z}$
\\
~$R^{\,}_{2}$
& 
$O(2N)/U(N)$ 
& 
$\mathbb{Z}^{\,}_{2}$
&
$0$
&
$\mathbb{Z}$
&
$0$
&
$0$
&
$0$
&
$\mathbb{Z}$
&
$\mathbb{Z}^{\,}_{2}$
\\
~$R^{\,}_{3}$ 
& 
~$U(2N)/Sp(N)$
&
$0$
&
$\mathbb{Z}$
&
$0$
&
$0$
&
$0$   
&
$\mathbb{Z}$
&
$\mathbb{Z}^{\,}_{2}$
&
$\mathbb{Z}^{\,}_{2}$
\\
~$R^{\,}_{4}$
& 
~$\cup_{n=0}^{N}\big\{Sp(N)/\big[Sp(n)\times Sp(N-n)\big]\big\}$
& 
$\mathbb{Z}$
&
$0$
&
$0$
&
$0$ 
&
$\mathbb{Z}$
&
$\mathbb{Z}^{\,}_{2}$
&
$\mathbb{Z}^{\,}_{2}$
&
$0$
\\
~$R^{\,}_{5}$
&
~$Sp(N)$
& 
$0$
&
$0$         
&
$0$
&
$\mathbb{Z}$ 
&
$\mathbb{Z}^{\,}_{2}$
&
$\mathbb{Z}^{\,}_{2}$
&
$0$
&
$\mathbb{Z}$
\\
~$R^{\,}_{6}$
& 
~$Sp(N)/U(N)$
& 
$0$
&
$0$   
&
$\mathbb{Z}$         
&
$\mathbb{Z}^{\,}_{2}$
&
$\mathbb{Z}^{\,}_{2}$
&
$0$
&
$\mathbb{Z}$         
&
$0$
\\
~$R^{\,}_{7}$ 
& 
~$U(N)/O(N)$
& 
$0$
&
$\mathbb{Z}$  
&
$\mathbb{Z}^{\,}_{2}$
&
$\mathbb{Z}^{\,}_{2}$
&
$0$
&
$\mathbb{Z}$
&
$0$
&
$0$       
\\
\hline \hline
\end{tabular}
\end{center}
\end{table*}

\subsection{Definition of minimal massive Dirac Hamiltonians}
\label{subsec: Definition of minimal massive Dirac Hamiltonians}

We assume that space is $d$-dimensional. The kinetic part
of a translation-invariant Dirac Hamiltonian is
\begin{subequations}
\label{eq: def Dirac Hamiltonian}
\begin{equation}
\mathcal{H}^{\,}_{\mathrm{kin}}(\bm{k})=
\sum_{i=1}^{d}
k^{\,}_{i}\,
\alpha^{\,}_{i},
\label{eq: def Dirac Hamiltonian a}
\end{equation}
where $\bm{k}\equiv(k^{\,}_{1},\ldots,k^{\,}_{d})\in\mathbb{R}^{d}$ 
is the momentum and
$\bm{\alpha}\equiv(\alpha^{\,}_{1},\ldots,\alpha^{\,}_{d})$
are the (Hermitian) Dirac matrices that obey the algebra
\begin{equation}
\left\{
\alpha^{\,}_{i},
\alpha^{\,}_{j}
\right\}=
2\delta^{\,}_{i,j},
\qquad
i,j=1,\ldots,d.
\label{eq: def Dirac Hamiltonian b}
\end{equation}
On the one hand, we assume that the dimensionality of the representation of
the matrices $\bm{\alpha}$ is sufficiently large so that
there exists at least one Hermitian matrix
$\beta$ such that it anticommutes with all the components of
$\bm{\alpha}$ and it squares to the identity matrix,
\begin{equation}
\left\{
\beta,
\mathcal{H}^{\,}_{\mathrm{kin}}(\bm{k})
\right\}=0,
\qquad
\beta^{2}=1.
\label{eq: def Dirac Hamiltonian c}
\end{equation}
It is then possible to write the
translation-invariant massive Dirac Hamiltonian
\begin{equation}
\mathcal{H}(\bm{k})=
\sum_{i=1}^{d}
k^{\,}_{i}\,
\alpha^{\,}_{i}
+
m\,\beta,
\label{eq: def Dirac Hamiltonian d}
\end{equation}
\end{subequations}
where $m\in\mathbb{R}$ is a mass. However,
the matrix $\beta$ with its mass $m$ (i.e., mass matrix in short)
may not be unique. For example, 
if the Dirac matrices are chosen to be of rank two, then
there are two linearly-independent mass matrices 
anticommuting with each other in $d=1$,
one possible mass matrix in $d=2$,
and none in $d=3$. On the other hand,
the translation-invariant massive Dirac Hamiltonian
(\ref{eq: def Dirac Hamiltonian d})
becomes reducible for sufficiently large rank of the Dirac matrices.
In any of the AZ symmetry classes, 
we may start from a sufficiently large matrix representation of
the translation-invariant massive Dirac Hamiltonian
(\ref{eq: def Dirac Hamiltonian d}),
which we then reduce until we reach the rank of the
Dirac matrices below which we would lose all mass matrices.
In this way, one obtains for each AZ symmetry class
and dimension $d$ 
\textit{an irreducible translation-invariant massive Dirac Hamiltonian 
of the form (\ref{eq: def Dirac Hamiltonian d}) 
which is of minimum rank $r^{\,}_{\mathrm{min}}$} 
(not necessarily unique in that more than one distinct 
mass matrix may be possible).

To which AZ symmetry class
the Dirac Hamiltonian
\begin{equation}
\mathcal{H}=
\sum_{i=1}^{d}
\alpha^{\,}_{i}\,
\frac{\partial}{\mathrm{i}\partial x^{\,}_{i}}
+
\beta\,m(\bm{x})
\label{eq: Dirac Hamiltonian with m(x)}
\end{equation}
belongs depends on whether it is possible to 
construct a combination from the triplet of operations
\begin{subequations}
\begin{equation}
T\equiv \mathcal{T}\,\mathsf{K},
\qquad
C\equiv \mathcal{C}\,\mathsf{K},
\qquad
\Gamma,
\end{equation}
for time-reversal symmetry (TRS), particle-hole symmetry (PHS),
and chiral symmetry (CHS), respectively,
($\mathsf{K}$ denotes the operation of complex conjugation and
$\mathcal{T}$,
$\mathcal{C}$,
and $\Gamma$ are matrices sharing the same rank as the Dirac matrices
$\bm{\alpha}$)
such that
\begin{equation}
T^{2}=\pm1,
\quad
C^{2}=\pm1,
\quad
[T,C]=0,
\quad
\Gamma^{2}=1,
\end{equation}
\end{subequations}
and
\begin{subequations}
\label{eq: TRS, PHS, CHS on Dirac}
\begin{align}
&
\hbox{TRS:}\qquad
\left[T,\mathcal{H}\right]=0,
\\
&
\hbox{PHS:}\qquad
\left\{C,\mathcal{H}\right\}=0,
\\
&
\hbox{CHS:}\qquad
\left\{\Gamma,\mathcal{H}\right\}=0.
\end{align}
\end{subequations}
(We have performed a global choice of gauge 
for which $[T,C]=0$ holds.)
Equations (\ref{eq: TRS, PHS, CHS on Dirac})
are equivalent to
\begin{subequations}
\label{eq: TRS, PHS, CHS on Dirac bis}
\begin{align}
&
\hbox{TRS:}\qquad
\left[T,\beta\right]=\left\{T,\bm{\alpha}\right\}=0,
\\
&
\hbox{PHS:}\qquad
\left\{C,\beta\right\}=\left[C,\bm{\alpha}\right]=0,
\\
&
\hbox{CHS:}\qquad
\left\{\Gamma,\beta\right\}=\left\{\Gamma,\bm{\alpha}\right\}=0.
\end{align}
\end{subequations}
Observe here that the antiunitarity of $T$ and $C$ 
interchanges the action of commutators and anticommutators 
when acting on the mass relative to the kinetic part of the
Dirac Hamiltonian (\ref{eq: Dirac Hamiltonian with m(x)}).

\subsection{The tenfold way for the Clifford algebras}
\label{subsec: The tenfold way for the Clifford algebras}

We are ready to combine the Clifford algebras and their
classifying spaces from Sec.~\ref{subsec: Definitions of Clifford algebras}
with the AZ classification of massive Dirac Hamiltonians
from Sec.~\ref{subsec: Definition of minimal massive Dirac Hamiltonians}.

We associate with each AZ symmetry class,
with each dimension $d$ of space,
and with any rank of the Dirac matrices 
$\bm{\alpha}\equiv(\alpha^{\,}_{1},\ldots,\alpha^{\,}_{d})$
equal to or larger than the minimal rank as defined
below Eq.~(\ref{eq: def Dirac Hamiltonian d}),
a Clifford algebra according to the following rules.

The symmetry classes A and AIII are associated with
the complex Clifford algebra
\begin{subequations}
\label{eq: complex Clifford algebra}
\begin{align}
&
\hbox{A:}\hphantom{IIAAA}
Cl^{\,}_{d+1}=
\left\{\beta,\bm{\alpha}\right\},
\\
&
\hbox{AIII:}\hphantom{AAA}
Cl^{\,}_{d+2}=
\left\{\beta,\Gamma,\bm{\alpha}\right\},
\end{align}
\end{subequations}
respectively.

For the remaining eight symmetry classes, it is always possible to define
the operation $J$ that satisfies the relations
\begin{equation}
\{T,J\}=\{C,J\}=[\Gamma,J]=[\bm{\alpha},J]=[\beta,J]=0,
\label{eq: def imagimary unit in Clifford algebra}
\end{equation}
and  plays the role of an imaginary unit 
for the real Clifford algebras as $J^{2}=-1$.
The symmetry classes 
AI,  BDI, D, DIII, AII, CII, C, and CI are associated
with the real Clifford
algebras%
~\cite{morimoto-clifford13}
\begin{subequations}
\label{eq: real Clifford algebra}
\begin{align}
&\hbox{AI:}\hphantom{IIAAi}
Cl^{\,}_{1,d+2}=
\{J\beta;T,TJ,\bm{\alpha}\},
\\
&\hbox{BDI:}\hphantom{AAa}
Cl^{\,}_{d+1,3}=
\{J\bm{\alpha},TCJ;C,CJ,\beta\},
\\
&\hbox{D:}\hphantom{IIIAA}
Cl^{\,}_{d,3}=
\{J\bm{\alpha};C,CJ,\beta\},
\\
&\hbox{DIII:}\hphantom{AAa}
Cl^{\,}_{d,4}=
\{J\bm{\alpha};C,CJ,TCJ,\beta\},
\\
&\hbox{AII:}\hphantom{IAA\ }
Cl^{\,}_{3,d}=
\{J\beta,T,TJ;\bm{\alpha}\},
\\
&\hbox{CII:}\hphantom{IIAA}
Cl^{\,}_{d+3,1}=
\{J\bm{\alpha},C,CJ,TCJ;\beta\},
\\
&\hbox{C:}\hphantom{IIIAA}
Cl^{\,}_{d+2,1}=
\{J\bm{\alpha},C,CJ;\beta\}, 
\\
&\hbox{CI:}\hphantom{IIAA\ }
Cl^{\,}_{d+2,2}=
\{J\bm{\alpha},C,CJ;TCJ,\beta\},
\end{align}
\end{subequations}
respectively. 

\subsection{The tenfold way for the classifying spaces $V$}
\label{subsec: The tenfold way for the classifying spaces V}

The definition of the classifying space $V$
associated with the Dirac Hamiltonian 
(\ref{eq: Dirac Hamiltonian with m(x)})
depends on the AZ symmetry class to which it belongs,
the dimensionality $d$ of space,
and the rank $r$ of the Dirac matrices 
$\bm{\alpha}=(\alpha^{\,}_{1},\ldots,\alpha^{\,}_{d})$,
which is assumed larger than the minimal rank $r^{\,}_{\mathrm{min}}$
defined in 
Sec.~\ref{subsec: Definition of minimal massive Dirac Hamiltonians}.
The classifying space $V$ encodes the fact that
the mass matrix $\beta$ 
in Eq.~(\ref{eq: Dirac Hamiltonian with m(x)})
might not be unique for given $d$ and the rank of the Dirac matrices
$\bm{\alpha}\equiv(\alpha^{\,}_{1},\ldots,\alpha^{\,}_{d})$.
The construction of the classifying space $V$ proceeds in
the following steps. 

\textbf{Step 1:}
To each AZ symmetry class, 
we assign the following pair of Clifford algebras
differing by one generator, holding the dimensionality 
$d$ of space fixed and the rank of the Dirac Hamiltonian
fixed. The Clifford algebra to the left of the arrow
in the fifth column ``extension'' from Table~\ref{table: AZ classes}
is obtained after removing from the tenfold list of Clifford algebras
defined by Eqs.~(\ref{eq: complex Clifford algebra}) 
and (\ref{eq: real Clifford algebra})
one generator, namely the mass matrix $J\,\beta$ for the symmetry
classes AI and AII and the mass matrix $\beta$ otherwise. This gives
the tenfold list
\begin{subequations}
\label{eq: complex Clifford algebra bis}
\begin{align}
&
\hbox{A:}\hphantom{IIAAA}
Cl^{\,}_{d}=
\left\{\bm{\alpha}\right\},
\\
&
\hbox{AIII:}\hphantom{AAA}
Cl^{\,}_{d+1}=
\left\{\Gamma,\bm{\alpha}\right\},
\end{align}
\end{subequations}
and
\begin{subequations}
\label{eq: real Clifford algebra bis}
\begin{align}
&\hbox{AI:}\hphantom{IIAAi}
Cl^{\,}_{0,d+2}=
\{;T,TJ,\bm{\alpha}\},
\\
&\hbox{BDI:}\hphantom{AAa}
Cl^{\,}_{d+1,2}=
\{J\bm{\alpha},TCJ;C,CJ\},
\\
&\hbox{D:}\hphantom{IIIAA}
Cl^{\,}_{d,2}=
\{J\bm{\alpha};C,CJ\},
\\
&\hbox{DIII:}\hphantom{AAa}
Cl^{\,}_{d,3}=
\{J\bm{\alpha};C,CJ,TCJ\},
\\
&\hbox{AII:}\hphantom{IAA\ }
Cl^{\,}_{2,d}=
\{T,TJ;\bm{\alpha}\},
\\
&\hbox{CII:}\hphantom{IIAA}
Cl^{\,}_{d+3,0}=
\{J\bm{\alpha},C,CJ,TCJ;\},
\\
&\hbox{C:}\hphantom{IIIAA}
Cl^{\,}_{d+2,0}=
\{J\bm{\alpha},C,CJ;\}, 
\\
&\hbox{CI:}\hphantom{IIAA\ }
Cl^{\,}_{d+2,1}=
\{J\bm{\alpha},C,CJ;TCJ\}.
\end{align}
\end{subequations}

\textbf{Step 2:}
For each AZ symmetry class,
we seek \textit{all} distinct Hermitian matrices 
sharing the same rank as the Dirac matrices
$\bm{\alpha}\equiv(\alpha^{\,}_{1},\ldots,\alpha^{\,}_{d})$
such that they can be added to the list of generators
entering the corresponding Clifford algebra from the tenfold list
defined by Eqs.~(\ref{eq: complex Clifford algebra bis}) 
and (\ref{eq: real Clifford algebra bis})
so as to deliver the corresponding Clifford algebra from the tenfold list
defined by Eqs.~(\ref{eq: complex Clifford algebra}) 
and (\ref{eq: real Clifford algebra}).
These mass matrices form a set $V$, the classifying space.
Determining $V$ is an example of the extension problem in K-theory.
A characterization of the classifying space $V$ can be deduced from
K-theory.~\cite{kitaev09,karoubi} The outcome of this exercise is listed in
the second column of
Table~\ref{table: all homotopies classifying spaces} 
(owing to the Bott periodicity)
for the case when the rank $r$ of the Dirac matrices
$\bm{\alpha}\equiv(\alpha^{\,}_{1},\ldots,\alpha^{\,}_{d})$
is
\begin{equation}
r= 
r^{\,}_{\mathrm{min}}\, 
N,
\end{equation}
where $r^{\,}_{\mathrm{min}}$ is the rank of the minimal representation for
the Dirac Hamiltonian~(\ref{eq: Dirac Hamiltonian with m(x)})
and $N$ is an integer.
Explicit examples for the construction of $V$ can be found
in Ref.~\onlinecite{morimoto-clifford13}.

The zeroth homotopy group of the classifying spaces
is given in the third column of
Table~\ref{table: all homotopies classifying spaces}.
This homotopy group has the following physical consequences,
as shown by Kitaev in Ref.~\onlinecite{kitaev09}.
Imagine that $d$-dimensional space is divided into two 
halves. Both halves share a $(d-1)$-dimensional boundary.
Whenever the zeroth homotopy group of the classifying space
in the tenfold way is non-vanishing, consider the Dirac
Hamiltonian~(\ref{eq: Dirac Hamiltonian with m(x)})
with the mass term interpolating smoothly across
the $(d-1)$-dimensional boundary between two fixed elements
in the classifying space characterized by distinct values
of the zeroth homotopy group. By the very definition of
a homotopy group, this is only possible if the mass term
vanishes along the  $(d-1)$-dimensional boundary separating
the two halves of $d$-dimensional space. As was shown by
Jackiw and Rebbi for the symmetry class BDI when $d=1$,%
~\cite{Jackiw76}
the Dirac Hamiltonian~(\ref{eq: Dirac Hamiltonian with m(x)})
must then support a zero mode that is extended along the 
$(d-1)$-dimensional boundary
but exponentially localized away from it.
This is the defining property of a topological insulator (superconductor).
Hence, by combining the zeroth homotopy group of the classifying spaces
given in the third column of
Table~\ref{table: all homotopies classifying spaces}
with the Bott periodicity~(\ref{eq: consequences Bott periodicity}), 
one infers which of the AZ symmetry classes
allows a topological (superconductor) insulator for any
given dimensionality $d$ of space. The periodic table for 
topological (superconductors) insulators follows.%
~\cite{kitaev09,ryu-njp10}

Another application of the zeroth homotopy groups associated 
with the classifying spaces is relegated to the 
Appendix~\ref{appsec: Lessons from ten ...},
where we show in which sense one may declare that a mass matrix is 
``unique''.

All higher homotopy groups of the classifying spaces
are given in column four to ten
of Table~\ref{table: all homotopies classifying spaces},
for they obey the periodicities
\begin{subequations}
\label{eq: periodicities obeyed by the homotopies of classspaces}
\begin{equation}
\pi^{\,}_{p}(C^{\,}_{j})=
\pi^{\,}_{p+2}(C^{\,}_{j}),
\qquad
p=0,1,2,\dots,
\end{equation}
for the complex classes $j=0,1$ and 
\begin{equation}
\pi^{\,}_{p}(R^{\,}_{j})=
\pi^{\,}_{p+8}(R^{\,}_{j}),
\qquad
p=0,1,2,\dots,
\end{equation}
\end{subequations}
for the real classes $j=0,1,\ldots,7$.
They also obey the translation rules
\begin{subequations}
\label{eq: translation rules obeyed by the homotopies of classspaces}
\begin{equation}
\pi^{\,}_{p}(C^{\,}_{j})=
\pi^{\,}_{p+1}(C^{\,}_{j+1}),
\qquad
p=0,1,2,\dots,
\end{equation}
for the complex classes 
with the integers $j$ and $j+1$ defined modulo 2 and 
\begin{equation}
\pi^{\,}_{p}(R^{\,}_{j})=
\pi^{\,}_{p+1}(R^{\,}_{j-1}),
\qquad
p=0,1,2,\ldots,
\end{equation}
\end{subequations}
for the real classes 
with the integers $j$ and $j-1$ defined modulo 8.
If we combine these translations rules with the definition
of $V^{\,}_{d}$ given in Table \ref{table: AZ classes},
we find the relation
\begin{equation}
\pi^{\,}_{D}(V^{\,}_{d}) 
= 
\pi^{\,}_{0}(V^{\,}_{d-D})
\label{eq: piD is related to pi0}
\end{equation}
for any $D=0,1,2,\ldots,d$.

\begin{figure}[tb]
\begin{center}
\includegraphics[width=0.8\linewidth]{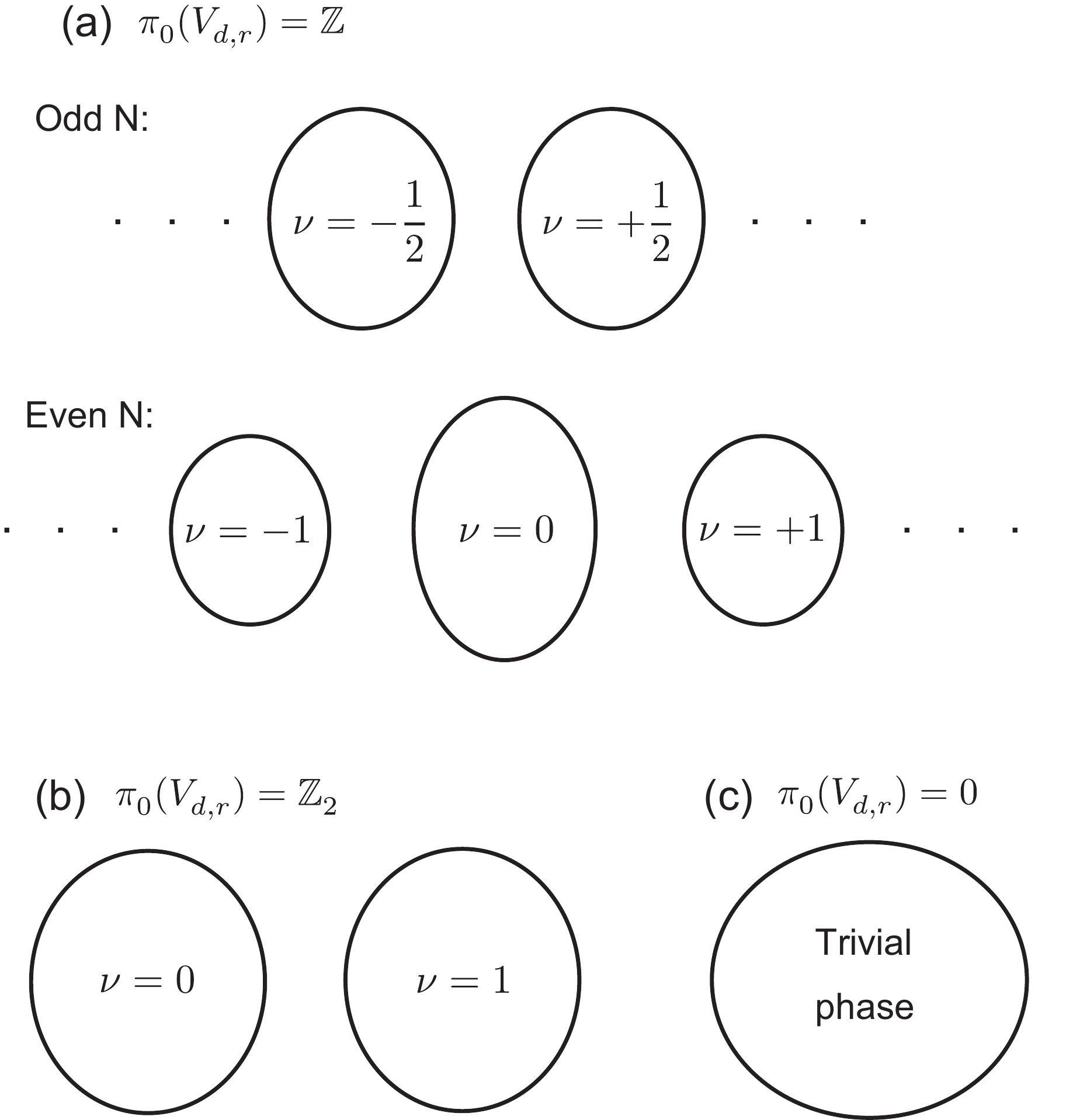}
\end{center}
\caption{\label{Fig: space of mass}
Connectedness of the compact topological space $V^{\,}_{d,r}$
parametrized by the normalized Dirac masses. The zeroth homotopy groups
$\pi^{\,}_{0}(V^{\,}_{d,r})$
index the disconnected parts of the compact topological space $V^{\,}_{d,r}$. 
The zeroth homotopy group
$\pi^{\,}_{0}(V^{\,}_{d,r^{\,}_{\mathrm{min}}\,N})$
is either (a) $\mathbb{Z}$ (in the limit $N\to\infty$), 
(b) $\mathbb{Z}^{\,}_{2}$, 
or (c) $\{0\}\equiv 0$
according to Table~\ref{table: all homotopies classifying spaces}.
(a) When  
$\pi^{\,}_{0}(\lim_{N\to\infty} V^{\,}_{d,r^{\,}_{\mathrm{min}}\,N})
=\mathbb{Z}$,
$V^{\,}_{d,r^{\,}_{\mathrm{min}}\,N}$ is disconnected and given by the union
of path-connected and compact topological subspaces 
indexed by the half-integers or integers defined in Eq.~(\ref{eq: Z index}).
For odd $N$, $V^{\,}_{d,r^{\,}_{\mathrm{min}}\,N}$ is the union of
path-connected and compact topological subspaces labeled
by negative and positive half integers. 
The total ``volume'' of the union of all subspaces
with negative half-integer labels equals that of
the union of all subspaces with positive half-integer labels
when the mean values for the Dirac masses vanish.
For even $N$, $V^{\,}_{d,r^{\,}_{\mathrm{min}}\,N}$ is the union of 
path-connected and compact topological subspaces labeled by the integers%
~(\ref{eq: Z index}).
The total ``volume'' of the union of all subspaces with strictly negative 
integer labels equals that of the union of all subspaces
with strictly positive integer labels
when the mean values for the Dirac masses vanish.
The subspace labeled by the integer 0
has a distinct (larger) ``volume''
when the mean values for the Dirac masses vanish.
(b) When  
$\pi^{\,}_{0}(V^{\,}_{d,r})=\mathbb{Z}^{\,}_{2}$,
$V^{\,}_{d,r}$ is the union of two path-connected 
and compact topological subspaces 
that are indexed by the integers~(\ref{eq: Z2 index}) and
are of equal ``volume'' when the mean values for the Dirac masses vanish.
(c) When  
$\pi^{\,}_{0}(V^{\,}_{d,r})=\{0\}\equiv0$,
$V^{\,}_{d,r}$ is a path-connected and compact topological space. 
        }
\end{figure}

\begin{figure}[tb]
\begin{center}
\includegraphics[width=1.0\linewidth]{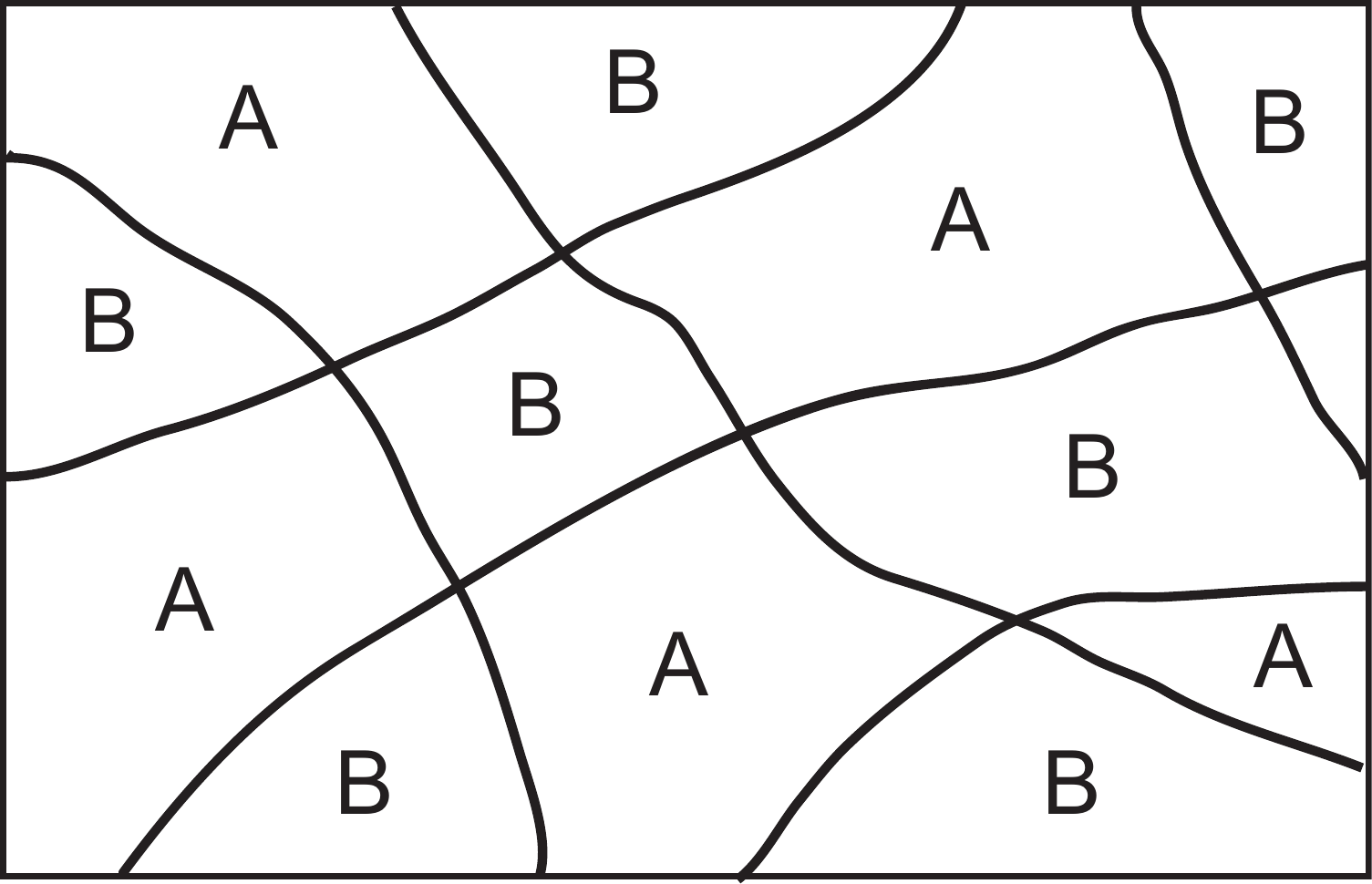}
\end{center}
\caption{\label{Fig: domains} 
For each realization of a random potential perturbing 
the massless Dirac Hamiltonian defined in $d$-dimensional 
Euclidean space $\mathbb{R}^{d}$, we may decompose 
$\mathbb{R}^{d}$ into open sets (domains)
of linear size $\xi^{\,}_{\mathrm{dis}}$.
In each of these domains,
the normalized Dirac masses correspond to a unique value of
their zeroth homotopy group. At the boundary between domains
differing by the values taken by their zeroth homotopy group,
the Dirac masses must vanish. Such boundaries support quasi-zero-energy
boundary states. When it is possible to classify the
elements of the zeroth homotopy group by the pair of indices
A and B, we may assign the letters A or B to any one of these
domains as is illustrated.
When the typical ``volume'' of a domain of type A equals
that of type B, quasi-zero modes undergo quantum percolation 
through the sample and thus establish either a critical or 
a metallic phase of quantum matter in $d$-dimensional space.
}
\end{figure}

\subsection{Existence and uniqueness of normalized Dirac masses}
\label{subsec: existence and uniqueness}

We summarize the properties of $d$-dimensional Dirac Hamiltonians
that follow from the topology of the classifying spaces $V^{\,}_{d}$ 
for the ten AZ symmetry classes.
\begin{enumerate}
\item
For each symmetry class, there exists a minimum rank 
$r^{\,}_{\mathrm{min}}(d)$, 
an even integer equal to or larger than the integer 2, 
for which the $d$-dimensional Dirac Hamiltonian supports a mass matrix
and below which either no mass matrix or no Dirac kinetic
contribution are allowed by symmetry.

\item
Suppose that the rank of the $d$-dimensional massive Dirac Hamiltonian is
$r=r^{\,}_{\mathrm{min}}(d)\,N$.
When there exists a mass matrix squaring to unity (up to a sign)
that commutes with all other symmetry-allowed mass matrices,
we call it the unique mass matrix.
There are the following three cases depending on
the entries in the $\pi^{\,}_{0}(V^{\,}_{d})$ column of
Table~\ref{table: AZ classes}.
\begin{enumerate}

\item
$\pi^{\,}_{0}(V^{\,}_{d})=\mathbb{Z}$:
there is always a unique mass matrix for $N\ge1$.

\item
$\pi^{\,}_{0}(V^{\,}_{d})=\mathbb{Z}^{\,}_{2}$:
there is a unique mass matrix $\beta^{\,}_{\min}$ only when $N=1$.
When $N$ is an even integer, for any given mass matrix,
there exists another matrix that anticommutes with it.
When $N$ is an odd integer larger than 1, the matrix
$\beta^{\,}_{\min}\tensor\openone^{\,}_{N}$ plays a role 
similar to that of the unique mass matrix 
in that the two matrices $\pm\beta^{\,}_{\min}\tensor\openone^{\,}_{N}$ 
belong to different connected components of $V^{\,}_{d}$,
even though there exists a normalized mass matrix that
neither commutes nor anticommutes with 
$\beta^{\,}_{\min}\tensor\openone^{\,}_{N}$.

\item 
$\pi^{\,}_{0}(V^{\,}_{d})=0$:
for any given mass matrix,
there exists another mass matrix that 
anticommutes with it.
That is, there is no unique mass matrix for $N\ge1$.

\end{enumerate}

When (i) $N=1$, (ii) the Dirac Hamiltonian has a unique mass matrix that
realizes topologically distinct ground states for different signs
of its mass, and (iii) the mass is varied smoothly in space, then
domain boundaries along which the mass vanishes are accompanied by
massless Dirac fermions, whose low-energy Hamiltonian is of rank
$r^{\,}_{\min}(d)/2$. It follows that $r^{\,}_{\min}(d-1)=r^{\,}_{\min}(d)$.
\end{enumerate}

We illustrate in Appendix~\ref{appsec: Lessons from ten ...}
these properties for 1D Dirac Hamiltonians in
the ten AZ symmetry classes.

\subsection{Relationship to higher homotopy groups}
\label{subsec: Relationship to higher homotopy groups}

The zeroth homotopy group of a topological space 
indicates if it is path connected and,
if not, how to index all its distinct subspaces that are path connected.
Equation~(\ref{eq: piD is related to pi0})
relates the zeroth to the higher homotopy groups of
the classifying spaces $V^{\,}_{d}$ defined in Table \ref{table: AZ classes}.
Equation~(\ref{eq: piD is related to pi0})
can be given the following interpretation.
 
We recall that the homotopy group $\pi^{\,}_{n}(X)$
is the set of homotopy classes of maps $f:S^{n}\to X$
between the unit sphere $S^{n}$ in $(n+1)$-dimensional Euclidean space
and the topological space $X$.
Following Teo and Kane in Ref.%
~\onlinecite{Teo10},
we identify $D$ in Eq.~(\ref{eq: piD is related to pi0}) 
as the dimensionality of the sphere $S^{D}$ 
that surrounds a defect in $d$-dimensional space. 
For point, line, and surface defects, $D=d-1$,
$D=d-2$, and $D=d-3$, respectively. In other words,
$D+1$ is the codimension of such defects in $d$-dimensional space. 
A homotopy group
$\pi^{\,}_{D}(V^{\,}_{d})$ 
with more than one element signals that 
defects of codimension $D+1$ in the normalized Dirac masses
can be indexed by a topological number. For given $D$ and $d$,
Eq.~(\ref{eq: piD is related to pi0}) 
[both $d$ and $d-D$ on the left- and right-hand sides of
$\pi^{\,}_{D}(V^{\,}_{d})=\pi^{\,}_{0}(V^{\,}_{d-D})$,
respectively, are defined either modulo 2 or modulo 8 depending on the AZ
symmetry class]
dictates through the homotopy reduction $D\to D-D=0$
and the dimensional reduction $d\to d-D$ 
which five of the ten AZ symmetry classes support topological
defects in their normalized Dirac masses. 
In particular, point defects inherit the
topological numbers from the zeroth homotopy group of $V^{\,}_{1}$.
Any Dirac Hamiltonian with a defective normalized Dirac mass
of topological character supports eigenstates with a vanishing
energy eigenvalue (zero modes) that are bound to the defect in
the directions transverse to it. These zero modes are robust
to any local perturbation as long as they respect the AZ symmetry class
and they are not too strong. 

\section{Anderson localization and the zeroth homotopy group of the classifying spaces}
\label{sec: Anderson localization and the zeroth homotopy group of ...}

We assume space to be $d$-dimensional.
Furthermore, we assume that a microscopic lattice model
with lattice spacing $\mathfrak{a}$ describing
noninteracting fermions propagating in a static random environment 
respecting one of the ten symmetry constraints from the AZ symmetry classes
is captured by the Dirac Hamiltonian
\begin{subequations}
\label{eq: def Dirac Hamiltonian with m(x)}
\begin{equation}
\mathcal{H}=
\sum_{i=1}^{d}
\alpha^{\,}_{i}\,
\frac{\partial}{\mathrm{i}\partial x^{\,}_{i}}
+
V(\bm{x})
+
\ldots,
\label{eq: def Dirac Hamiltonian with m(x) a}
\end{equation}
in the low-energy and long-wavelength limit.
The rank of the Dirac matrices is
\begin{equation}
r=
r^{\,}_{\mathrm{min}}\,N,
\label{eq: def Dirac Hamiltonian with m(x) b}
\end{equation}
where $r^{\,}_{\mathrm{min}}$ is the smallest rank that admits a Dirac mass
matrix and $N=1,2,\ldots$.
Hence, for any $\bm{x}\in\mathbb{R}^{d}$,
there exists a Dirac mass matrix $V(\bm{x})$ of rank $r$
that competes with the kinetic contribution
parametrized by the Dirac matrices 
$\bm{\alpha}=(\alpha^{\,}_{1},\ldots\alpha^{\,}_{d})$
[i.e., $V(\bm{x})$ anticommutes with $\bm{\alpha}$].
The matrix elements of the Dirac mass matrix $V(\bm{x})$
in Eq.\ (\ref{eq: def Dirac Hamiltonian with m(x) a}) 
are random functions of $\bm{x}\in\mathbb{R}^{d}$.
The dots represent all other
static random vector and scalar potentials allowed by the AZ symmetry class.
We fix the chemical potential $\mu$ to be $\mu=0$.

The matrix elements of the Dirac mass matrix $V(\bm{x})$ 
are assumed to be random functions that change smoothly in space
on the length scale of $\xi^{\,}_{\mathrm{dis}}$ ($\gg\mathfrak{a}$).
Their correlations are assumed local in that these matrix elements that
are not related by the AZ symmetries are
uncorrelated up to an exponential precision beyond the finite
length scale $\xi^{\,}_{\mathrm{dis}}\gg\mathfrak{a}$, 
e.g., (disorder averaging is denoted by an overline)
\begin{align}
\overline{V(\bm{x})}=: 
\mathsf{m}\,\beta^{\,}_{0},
\label{eq: def Dirac Hamiltonian with m(x) c}
\end{align}
and
\begin{align}
\frac{1}{r}
\overline{
\mathrm{tr}
[V(\bm{x})-\mathsf{m}\,\beta^{\,}_{0}]\,
[V(\bm{y})-\mathsf{m}\,\beta^{\,}_{0}]
         }=:
\mathsf{g}^{2}\,
e^{-|\bm{x}-\bm{y}|/\xi^{\,}_{\mathrm{dis}}},
\label{eq: def Dirac Hamiltonian with m(x) d}
\end{align}
\end{subequations} 
with all higher cumulants vanishing.
The choice of the normalized Dirac mass matrix $\beta^{\,}_{0}$
from the classifying space $V^{\,}_{d,r}$
will be done in Secs.%
~\ref{subsec: case of Z},
\ref{subsec: case of Z_{2}},
and
\ref{subsec: case of 0}
in such a way that the parameter space 
$(\mathsf{m},\mathsf{g})\in\mathbb{R}\times[0,\infty[$
captures the phase diagram representing the competition between 
delocalized and all topologically distinct localized phases 
of noninteracting fermions in a given $d$-dimensional AZ symmetry class.
The former phase is favored by the Dirac kinetic contribution. 
The latter phase is favored by the Dirac masses.
In other words,
localized (insulating) phases are favored by large $|\mathsf{m}|$,
whereas increasing $\mathsf{g}$ generates more density of states
in the band (mass) gap and, in doing so, helps delocalization.

The main result of this section is captured by Fig.%
~\ref{Fig: space of mass},
a consequence of Table \ref{table: AZ classes},
and Fig.~\ref{Fig: domains}.
Let $V^{\,}_{d,r}$ be the classifying space associated with
any typical realization of the random Dirac Hamiltonian%
~(\ref{eq: def Dirac Hamiltonian with m(x)}).
We then have the following phases
(at the band center $\varepsilon=0$ of the quasiparticle dispersion):
\begin{enumerate}
\item[(a)]
If
$\lim_{N \to \infinity}\pi^{\,}_{0}(V^{\,}_{d,r^{\,}_{\min}N})
=\mathbb{Z}$,
there are $N+1$ topologically distinct insulating phases
that are separated pairwise 
by either a critical point or a metallic phase. 
\item[(b)]
If 
$\pi^{\,}_{0}(V^{\,}_{d,r^{\,}_{\mathrm{min}}\,N})=\mathbb{Z}^{\,}_{2}$,
there are two topologically distinct insulating phases
that are separated by either a critical point 
or a metallic phase. 
\item[(c)]
If 
$\pi^{\,}_{0}(V^{\,}_{d,r^{\,}_{\mathrm{min}}\,N})=0$,
there is only a topologically trivial insulating phase.
\end{enumerate} 
When $d\ge2$ (where the equality holds for class AII, CII, D, and DIII),
the ground state is metallic for sufficiently large $\mathsf{g}$.

The key intuition to support items (a) and (b) is the following. 
Since the spatial variation of $V(\bm{x})$ is assumed to be smooth
for any realization of the disorder, $d$-dimensional
space can be decomposed into open sets (domains)
with the characteristic size
$\xi^{\,}_{\mathrm{dis}}$ such that
\begin{enumerate}
\item[(1)]
for each domain the values taken in it
by $V(\bm{x})$ can be assigned an index from
the zeroth homotopy group
$\pi^{\,}_{0}(V^{\,}_{d,r})$,
\item[(2)]
and $\mathrm{det}[V(\bm{x})]=0$
along the boundary of each domain.
\end{enumerate}
Figure \ref{Fig: domains} is an illustration of this decomposition
of $d$-dimensional space when either 
$\pi^{\,}_{0}(V^{\,}_{d,r})=\mathbb{Z}$ 
or $\pi^{\,}_{0}(V^{\,}_{d,r})=\mathbb{Z}^{\,}_{2}$.
There are then gapless modes bound to the boundaries 
of these domains.
We use a semiclassical picture in analogy with the Chalker-Coddington
network model of the IQHE.
When no connected boundaries
defined by the condition $\mathrm{det}[V(\bm{x})]=0$
percolate across 
$d$-dimensional space, we expect an insulating phase.
However, if a connected boundary along which
$\mathrm{det}[V(\bm{x})]=0$
percolates across $d$-dimensional space,
we expect departure from an insulating phase.
Quantum mechanics modifies this percolating picture by turning it into that
of a Chalker-Coddington-like (quantum) network model in dimension $d$, 
whereby the scattering matrix 
at each node of the network is fixed by the AZ symmetry class and 
$r=r^{\,}_\mathrm{min}\,N$. 

\subsection{Case of the zeroth homotopy group
$\mathbb{Z}$
\label{subsec: case of Z}
}

In each dimension $d$ of space, there are three AZ symmetry classes
whose classifying spaces $V^{\,}_{d,r^{\,}_{\min}\,N}$
are the unions of one of the three
Grassmannian manifolds, i.e., the classifying spaces are any one of
\begin{subequations}
\label{eq: Grassmanians finite N}
\begin{align}
&
\bigcup_{n=0,\cdots,N} \Big\{U(N)/[U(n)\times U(N-n)]\Big\},
\\
&
\bigcup_{n=0,\cdots,N} \Big\{O(N)/[O(n)\times O(N-n)]\Big\},
\\
&
\bigcup_{n=0,\cdots,N} \Big\{Sp(N)/[Sp(n)\times Sp(N-n)]\Big\}.
\end{align}
\end{subequations}
These unions of Grassmannian manifolds are realized by the
topological space of normalized Dirac masses whenever
there exists a unique (up to a sign) normalized mass matrix
that commutes with all other allowed normalized mass matrices. 
In the limit $N\to\infty$, the zeroth homotopy group
of either one of these Grassmannian manifolds is $\mathbb{Z}$.

When the rank~(\ref{eq: def Dirac Hamiltonian with m(x) b})
of the Dirac Hamiltonian~(\ref{eq: def Dirac Hamiltonian with m(x) a})
belonging to any one of these AZ symmetry classes
is the minimal one, $N=1$ and there is a unique (up to a sign)
normalized Dirac mass matrix 
$\beta^{\,}_{0}$ of rank $r^{\,}_{\mathrm{min}}$ such that
\begin{equation}
V(\bm{x})=m(\bm{x})\,\beta^{\,}_{0}
\end{equation} 
in Eq.~(\ref{eq: def Dirac Hamiltonian with m(x) a}).
The Dirac Hamiltonian
(\ref{eq: def Dirac Hamiltonian with m(x) a}) with the uniform
mass $m>0$ is topologically distinct from the 
Dirac Hamiltonian
(\ref{eq: def Dirac Hamiltonian with m(x) a}) with the uniform
mass $m<0$. Correspondingly, the classifying spaces
(\ref{eq: Grassmanians finite N})
reduce to
\begin{subequations}
\label{eq: Grassmanians N=1}
\begin{align}
&
\bigcup_{n=0,1}
U(1)/[U(n)\times U(1-n)]\simeq
\{-1,+1\},
\\
&
\bigcup_{n=0,1}
O(1)/[O(n)\times O(1-n)]\simeq
\{-1,+1\},
\\
&
\bigcup_{n=0,1}
Sp(1)/[Sp(n)\times Sp(1-n)]\simeq
\{-1,+1\}.
\end{align}
\end{subequations}
If the Dirac mass matrix $V(\bm{x})$ of minimal rank
is random with the statistical correlation%
~(\ref{eq: def Dirac Hamiltonian with m(x) c})
and
~(\ref{eq: def Dirac Hamiltonian with m(x) d}),
we may decompose $d$-dimensional space in disjoint domains
as depicted in Fig.~\ref{Fig: domains}.
A typical domain has the linear size
$\xi^{\,}_{\mathrm{dis}}$.
In it $m(\bm{x})\neq0$ with a given sign,
along its boundary $m(\bm{x})=0$, and a sign change is only permissible
across this boundary into another domain with opposite and
constant sign of the mass $m(\bm{x})\neq0$. 
Any boundary separating two domains with opposite
signs of $m(\bm{x})\neq0$ binds 
gapless boundary states.
Whenever two boundaries approach each other within a distance much
smaller than $\xi^{\,}_{\mathrm{dis}}$, boundary states undergo an
elastic quantum scattering process dictated by the AZ symmetry class.
In other words, Fig.~\ref{Fig: domains}
defines a quantum network model, whereby incoming plane waves 
along the boundaries defined by the condition $m(\bm{x})=0$
scatter off each other elastically at the nodes of this network
of boundaries.
The mean value $\mathsf{m}$
of the random Dirac mass $m(\bm{x})$
dictates the relative
volume occupied by the domains
with $\mathrm{sgn}[m(\bm{x})]=+1$
relative to the volume  occupied by the domains
with $\mathrm{sgn}[m(\bm{x})]=-1$.
When $\mathsf{m}=0$, 
both volume are typically equal, in which case
the domain boundaries percolate across the system and
the boundary states are delocalized and signal either a critical
or a metallic phase. 

The situation is different when $N=2$.
Indeed, the unions of Grassmannian manifolds 
(\ref{eq: Grassmanians finite N})
are now comprised of the pair of Grassmannians
\begin{subequations}
\label{eq: Grassmanian component 2/2x0}
\begin{align}
&
U(2)/[U(2)\times U(0)],
\qquad
U(2)/[U(0)\times U(2)],
\\
&
O(2)/[O(2)\times O(0)],
\qquad
O(2)/[O(0)\times O(2)],
\\
&
Sp(2)/[Sp(2)\times Sp(0)],
\quad
Sp(2)/[Sp(0)\times Sp(2)],
\end{align}
\end{subequations}
with the dimensions
0, 0, and 0, respectively, and the Grassmannians
\begin{subequations}
\label{eq: Grassmanian component 2/1x1}
\begin{align}
&
U(2)/[U(1)\times U(1)],
\\
&
O(2)/[O(1)\times O(1)],
\\
&
Sp(2)/[Sp(1)\times Sp(1)],
\end{align}
\end{subequations}
with the dimensions
2, 1, and 4, respectively. 
Accordingly, there are three topologically distinct insulating
phases when $N=2$.
To derive the dimensions of these Grassmannian manifolds,
we used the fact that $U(n)$, $O(n)$, and $Sp(n)$ have the dimensions
$n^{2}$, $n(n-1)/2$, and $n(2n+1)$, respectively.

Imagine that $d$-dimensional space is randomly decomposed
into three types of domains according to the rule that
the random Dirac mass matrix $V(\bm{x})$ is associated
with only one of the three path-connected Grassmannian manifolds
making up the classifying space (\ref{eq: Grassmanian component 2/1x1})
in each domain.
The domain boundaries bind gapless states,
percolation of which leads to delocalization
or criticality between localized phases as in the case of $N=1$.
The transitions can be induced by changing the parameter 
$\mathsf{m}$.
Unlike the $N=1$ case, however, 
the ground state at $\mathsf{m}=0$ 
and for any non-vanishing $\mathsf{g}$ not too strong
is most likely a localized phase.
Indeed, localization is most likely to occur because
$d$-dimensional space is randomly partitioned into domains
such that most of the
domains are characterized by a random Dirac mass matrix 
$V(\bm{x})$ associated with
the path-connected Grassmannian manifolds of largest dimension in
Eq.\ (\ref{eq: Grassmanian component 2/1x1}). 

This difference between the $N=1$ and $N=2$ cases
is not accidental. The same difference holds between the cases of 
odd $N$ and even $N$ integers, namely that the tuning $\mathsf{m}=0$
delivers typically a critical or metallic phase of quantum matter 
in $d$-dimensional space when $N$ is odd, 
while it delivers typically a localized phase of quantum matter 
in $d$-dimensional space when $N$ is even, as we now explain.

When the classifying space $V^{\,}_{d,r^{\,}_{\mathrm{min}}\,N}$
is any one of the three unions
of Grassmannian manifolds (\ref{eq: Grassmanians finite N}),
we can always choose to represent the Dirac Hamiltonian
(\ref{eq: def Dirac Hamiltonian with m(x) a})
with
\begin{subequations}
\label{eq: case pi0 when z}
\begin{equation}
\bm{\alpha}=
\bm{\alpha}^{\,}_{\mathrm{min}}\otimes \openone^{\,}_{N}
\end{equation}
for the Dirac kinetic contribution and
\begin{equation}
V(\bm{x})=
\beta^{\,}_{\mathrm{min}}\otimes M(\bm{x})
\end{equation}
\end{subequations}
for the Dirac mass contribution,
whereby $\bm{\alpha}^{\,}_{\mathrm{min}}$ and $\beta^{\,}_{\mathrm{min}}$
represent the Clifford algebra with rank $r^{\,}_{\mathrm{min}}$,
while $\openone^{\,}_{N}$ is a unit $N\times N$ matrix and
$M(\bm{x})$ is a random $N\times N$ Hermitian matrix
which is a smooth continuous function of $\bm{x}$.
For a given realization of the random matrix $M(\bm{x})$,
we may partition $d$-dimensional space into domains
whose boundaries are defined by $\mathrm{det}[M(\bm{x})]=0$.
Each domain can be assigned a topological index as follows.
We may index each element of
$\pi^{\,}_{0}(V^{\,}_{d,r^{\,}_{\mathrm{min}}\,N})$
with the values of $\nu$ defined by
\begin{subequations}
\label{eq: nu when z}
\begin{equation}
\nu:=
\frac{1}{2}\,
\mathrm{tr}
\big\{\mathrm{sgn}[M(\bm{x})]\big\}\in
\begin{cases}
\{-1/2,+1/2\},&\hbox{ if $N=1$},
\\
\{-1,0,+1\},&\hbox{ if $N=2$},
\\
\hbox{and so on},& \hbox{ if $N>2$},
\end{cases}
\end{equation}
where the Hermitian matrix $M(\bm{x})$ is diagonalized
by the unitary matrix $U(\bm{x})$,
\begin{equation}
M=:
U^{\dag}\,
\mathrm{diag}\,
\left(\lambda^{\,}_{1},\ldots,\lambda^{\,}_{N}\right)\,
U,
\end{equation}
and
\begin{equation}
\mathrm{sgn}[M(\bm{x})]:= 
U^{\dag}\,
\mathrm{diag}\,
\left(
\frac{\lambda^{\,}_{1}}{|\lambda^{\,}_{1}|},
\ldots,
\frac{\lambda^{\,}_{N}}{|\lambda^{\,}_{N}|}
\right)\,
U.
\label{eq: Z index}
\end{equation}
\end{subequations}
Each domain with $V(\bm{x})$ a smooth function
of $\bm{x}$ has thereby been assigned the topological index $\nu$.

We still need to \textit{choose} the parameter space 
$(\mathsf{m},\mathsf{g})\in\mathbb{R}\times[0,\infty[$
announced in Eqs.~(\ref{eq: def Dirac Hamiltonian with m(x) c}) 
and (\ref{eq: def Dirac Hamiltonian with m(x) d}),
when the zeroth homotopy group of the classifying space is $\mathbb{Z}$. 
We must distinguish two cases.

\textbf{Case when $\mathsf{g}>0$ is 
a relevant perturbation to the clean critical point:}
We select the normalized Dirac mass matrix
\begin{subequations}
\label{alternative definition 0}
\begin{equation}
\beta^{\,}_{0}:=
\beta^{\,}_{\mathrm{min}}\otimes\openone^{\,}_{N},
\label{alternative definition 0 a}
\end{equation}
that anticommutes with all the components of $\bm{\alpha}$ and
commutes with all Dirac mass matrices allowed for given $r$ and symmetry
constraints (see also Appendix \ref{appendix: interpretation of pi_{0}}).
The matrix $\beta^{\,}_{0}$ is the unique mass matrix
introduced in Sec.\ \ref{subsec: existence and uniqueness}.
We define the parameter space
$(\mathsf{m},\mathsf{g})\in\mathbb{R}\times[0,\infty[$
through the probability distribution of the Dirac mass
matrix $V(\bm{x})$ given by
\begin{align}
\overline{V(\bm{x})}=:
\mathsf{m}\,\beta^{\,}_{0},
\label{alternative definition 0 b}
\end{align}
and
\begin{align}
\frac{1}{r}\,
\overline{\mathrm{tr}
\left\{
[V(\bm{x})-\mathsf{m}\,\beta^{\,}_{0}]
[V(\bm{y})-\mathsf{m}\,\beta^{\,}_{0}]
\right\}}
=:\mathsf{g}^{2}\,e^{-|\bm{x}-\bm{y}|/\xi_\mathrm{dis}},
\label{alternative definition 0 c}
\end{align}
\end{subequations}
with all higher cumulants vanishing.
With the definitions
(\ref{alternative definition 0 b})
and
(\ref{alternative definition 0 c})
for the probability distribution of $V(\bm{x})$,
the point $\mathsf{m}=\mathsf{g}=0$ is a massless Dirac critical point.
The critical point $\mathsf{m}=\mathsf{g}=0$
separates two insulating phases with $\nu=\pm N/2$
along the horizontal axis $\mathsf{g}=0$ in parameter space.
By assumption, $\mathsf{g}>0$ is relevant in the vicinity of
the clean critical point at $\mathsf{m}=\mathsf{g}=0$.
This is the rule when $d=1$, in which case
there appears for $\mathsf{g}>0$ $N-1$ additional localized phases,
with $\nu=-(N/2)+1,-(N/2)+2,\ldots,(N/2)-1$ that are
pairwise separated by lines of critical points,
all of which emerge from the critical point $\mathsf{m}=\mathsf{g}=0$,
as was demonstrated in Ref.~\onlinecite{Brouwer98}.
(It is the dimerization denoted by $f$ in Ref.~\onlinecite{Brouwer98}
that plays the role of $\mathsf{m}$.)
For those symmetry classes in $d=2$ for which 
$\mathsf{g}>0$ is (marginally) relevant in the vicinity of
the clean critical point at $\mathsf{m}=\mathsf{g}=0$,
we conjecture that  $N-1$ additional localized phases
that we may again label with
$\nu=-(N/2)+1,-(N/2)+2,\ldots,(N/2)-1$
are stabilized when $\mathsf{g}>0$.
Consecutive localized phases are either separated by
a line of critical points or by a metallic phase.
The relevant symmetry classes are AIII, BDI, and CII in $d=1$
and A and C in $d=2$.

\textbf{Case when $\mathsf{g}>0$ is 
an irrelevant perturbation to a clean critical point:}
We define the parameter space
$(\mathsf{m},\mathsf{g})\in\mathbb{R}\times[0,\infty[$
through the probability distribution of the Dirac mass
matrix $V(\bm{x})$ given by
\begin{subequations}
\label{eq: definition of m for irrelevant g}
\begin{equation}
\overline{V(\bm{x})}=:
\mathsf{m}\,\beta^{\,}_{0}+V^{\,}_{0},
\label{eq: definition of m for irrelevant g a}
\end{equation}
where $\beta^{\,}_{0}$ again commutes with all other mass matrices
permitted by the symmetry class, and
\begin{align}
\frac{1}{r}\,
\overline{\mathrm{tr}
\left\{
[V(\bm{x})-\overline{V(\bm{x})}]
[V(\bm{y})-\overline{V(\bm{y})}]
\right\}}
=:\mathsf{g}^{2}\,e^{-|\bm{x}-\bm{y}|/\xi_\mathrm{dis}}.
\label{eq: definition of m for irrelevant g b}
\end{align}
Here,
$V^{\,}_{0}$ 
is any mass matrix permitted by the symmetry
that satisfies the condition
\begin{equation}
V^{\,}_{0}=\beta^{\,}_{\t{min}}\tensor M^{\,}_{0},
\label{eq: definition of m for irrelevant g c}
\end{equation}
\end{subequations}
where the $N\times N$ Hermitian matrix
$M^{\,}_{0}$ has $N$ nondegenerate eigenvalues. 
The existence of the $r\times r$ Hermitian matrix $V^{\,}_{0}$
is required to obtain $N+1$ distinct localized phases in the phase diagram
by changing the parameter $\mathsf{m}$ in the clean limit $\mathsf{g}=0$.
The prescription (\ref{eq: definition of m for irrelevant g}) 
applies to the symmetry class D in $d=2$ and all symmetry classes with 
$\pi^{\,}_{0}(V)=\mathbb{Z}$ in $d\ge3$.

More discussions on the relationship between the RG flow of 
$\mathsf{g}$ and the probability distribution of the Dirac mass matrix 
$V(x)$ are presented in 
Sec.~\ref{subsec: 3D Z} for 3D disordered systems and in
Appendix~\ref{appsubsec: OPE class D} 
for the 2D systems of the symmetry class D.

The zeroth homotopy group of the topological space
$V^{\,}_{d,r^{\,}_{\mathrm{min}}\,N}$
encodes the connectedness of 
$V^{\,}_{d,r^{\,}_{\mathrm{min}}\,N}$.
The ``volume'' of each path-connected component 
\begin{subequations}
\begin{align}
&
U(N)/\big[U(n)\times U(N-n)\big],
\\
&
O(N)/\big[O(n)\times O(N-n)\big],
\\
&
Sp(N)/\big[Sp(n)\times Sp(N-n)\big],
\end{align}
\end{subequations}
of $V^{\,}_{d,r^{\,}_{\mathrm{min}}\,N}$
is measured by the dimension
\begin{subequations}
\begin{align}
&
2n(N-n)=
N^{2}-n^{2}-(N-n)^{2},
\\
&
n(N-n)=
\frac{
N(N-1)
-
n(n-1)
     }
     {
2
     }
\nonumber\\
&\hphantom{n(N-n)=}
-
\frac{
(N-n)(N-n-1)
     }
     {
2
     },
\\
&
4n(N-n)=
N(2N+1)-(N-n)[2(N-n)+1]
\nonumber\\
&\hphantom{4n(N-n)=}
-n(2n+1),
\end{align}
\end{subequations}
respectively,
as is depicted schematically in Fig.~\ref{Fig: space of mass}(a).

When $N$ is odd, one can always write
\begin{subequations}
\label{eq: case Grassmanians with odd N}
\begin{equation}
V^{\,}_{d,r^{\,}_{\mathrm{min}}\,N}=
\mathrm{A}\cup\mathrm{B},
\end{equation}
where
\begin{equation}
\mathrm{A}:=
\bigcup_{n=0,\ldots,\frac{N-1}{2}} 
\big\{U(N)/[U(N-n)\times U(n)]\big\}
\end{equation}
and
\begin{equation}
\mathrm{B}:=
\bigcup_{n=\frac{N+1}{2},\ldots,N} 
\big\{U(N)/[U(n)\times U(N-n)]\big\},
\end{equation}
\end{subequations}
and similarly with the substitutions $U\to O,Sp$. 
The existence of a critical or metallic phase of quantum matter
when $\mathsf{m}=0$ in the parametrization (\ref{alternative definition 0})
follows from repeating the argumentation for
the $N=1$ case that is captured by Fig.~\ref{Fig: domains}.

When $N$ is even, one can always write
\begin{subequations}
\label{eq: case Grassmanians with even N}
\begin{equation}
V^{\,}_{d,r^{\,}_{\mathrm{min}}\,N}=
V^{\,}_{-}\cup V^{\,}_{0}\cup V^{\,}_{+},
\end{equation}
where
\begin{align}
&
V^{\,}_{-}:=
\bigcup_{n=0,\ldots,\frac{N}{2}-1} 
\big\{U(N)/[U(N-n)\times U(n)]\big\},
\\
&
V^{\,}_{0}:=
U(N)/[U(N/2)\times U(N/2)],
\end{align}
and
\begin{equation}
V^{\,}_{+}:=
\bigcup_{n=\frac{N}{2}+1,\ldots,N} 
\big\{U(N)/[U(n)\times U(N-n)]\big\},
\end{equation}
\end{subequations}
and similarly with the substitutions $U\to O,Sp$.
Since the index $\nu$ assigned to localized phases 
is an odd function of $\mathsf{m}$,
the ground state at $\mathsf{m}=0$ and $\mathsf{g}$ non-vanishing 
but not too large in the parametrization (\ref{alternative definition 0})
is expected to be in the localized phase,
for $d$-dimensional space is partitioned into
domains of Dirac mass matrices which are predominantly drawn
from $V^{\,}_{0}$.

\subsection{Case of the zeroth homotopy group $\mathbb{Z}^{\,}_{2}$}
\label{subsec: case of Z_{2}}

In each dimension $d$ of space, there are two AZ symmetry classes
whose classifying spaces $V^{\,}_{d,r^{\,}_{\mathrm{min}}\,N}$
are homeomorphic to either the orthogonal group $O(N)$
or the quotient space $O(2N)/U(N)$
and thus have the zeroth homotopy group $\mathbb{Z}^{\,}_{2}$.
The former case is called the first descendant $\mathbb{Z}^{\,}_{2}$.
The second case is called the second descendant $\mathbb{Z}^{\,}_{2}$.
If so, we can always choose to represent the Dirac Hamiltonian
(\ref{eq: def Dirac Hamiltonian with m(x) a})
with
\begin{subequations}
\label{eq: case pi0 when z2}
\begin{equation}
\bm{\alpha}=
\bm{\alpha}^{\,}_{\mathrm{min}}\otimes \openone^{\,}_{N}
\label{eq: case pi0 when z2 a}
\end{equation}
for the Dirac kinetic contribution and
\begin{equation}
V(\bm{x})=
\rho^{\,}_{\mathrm{min}}\otimes M(\bm{x})
\label{eq: case pi0 when z2 b}
\end{equation}
for the Dirac mass contribution
in each domain where the Dirac mass matrix 
$V(\bm{x})$ is continuous and invertible.
Here, $\rho^{\,}_{\mathrm{min}}$ is a 
$r^{\,}_{\mathrm{min}}/2 \times r^{\,}_{\mathrm{min}}/2$ 
matrix such that, when tensored with the antisymmetric Pauli matrix 
$\sigma^{\,}_{2}$,
$\bm{\alpha}^{\,}_{\mathrm{min}}$
and 
$\rho^{\,}_{\mathrm{min}}\tensor \sigma^{\,}_{2}$
deliver a representation of the Clifford algebra of
rank $r^{\,}_{\mathrm{min}}$. 
Finally, $\openone^{\,}_{N}$ is a unit $N\times N$ matrix and
$M(\bm{x})$ is a $2N\times 2N$ Hermitian matrix which is also antisymmetric, 
i.e.,
\begin{equation}
M(\bm{x})=M^{\dag}(\bm{x}),
\qquad
M(\bm{x})=-M^{\mathsf{T}}(\bm{x}).
\label{eq: case pi0 when z2 c}
\end{equation}

Each domain in the partition of $d$-dimensional space 
into domains defined by
the boundaries where $\mathrm{det}[M(\bm{x})]=0$
can be assigned a topological index as follows.
We may index each element of
$\pi^{\,}_{0}(V^{\,}_{d,r^{\,}_{\mathrm{min}}\,N})$
with the values of $\nu=0,1$ defined by
\begin{equation}
(-1)^{\nu}:=
\frac{\mathrm{Pf}\,[\mathrm{i}M(\bm{x})]}
     {\sqrt{\mathrm{det}\,[\mathrm{i}M(\bm{x})]}}.
\label{eq: Z2 index}
\end{equation}
\end{subequations}
This topological number $\nu$ is well-defined 
because $\mathrm{i}M(\bm{x})$ is a real-valued antisymmetric matrix and
$\mathrm{det}\,[\mathrm{i}M(\bm{x})]$ is positive.
Each domain has thereby been assigned the $\mathbb{Z}^{\,}_{2}$-valued 
topological index $\nu$.

When $N$ is odd, we choose the parameter space
$(\mathsf{m},\mathsf{g})\in\mathbb{R}\times[0,\infty[$
by selecting
\begin{subequations}
\label{alternative definition 0 bar if Z2}
\begin{equation}
\beta^{\,}_{0}:=
\rho^{\,}_{\mathrm{min}}\otimes\sigma^{\,}_{2}\otimes\openone^{\,}_{N}
\label{alternative definition 0 a bar if Z2}
\end{equation}
in the probability distribution of the Dirac mass
matrix $V(\bm{x})$ given by
\begin{align}
\overline{V(\bm{x})}=:
\mathsf{m}\,\beta^{\,}_{0},
\label{alternative definition 0 b bar if Z2}
\end{align}
and
\begin{align}
\frac{1}{r}\,
\overline{
\mathrm{tr}
\left\{
[V(\bm{x})-\mathsf{m}\,\beta^{\,}_{0}]
[V(\bm{y})-\mathsf{m}\,\beta^{\,}_{0}]
\right\}
         }
=:\mathsf{g}^{2}\,e^{-|\bm{x}-\bm{y}|/\xi_\mathrm{dis}},
\label{alternative definition 0 c bar if Z2}
\end{align}
\end{subequations}
with all higher cumulants vanishing.
In the clean limit $\mathsf{g}=0$, the point $\mathsf{m}=0$
is a massless Dirac critical point separating the two
insulating phases with $\nu=0$ ($\mathsf{m}>0$)
and $\nu=1$ ($\mathsf{m}<0$).
Given $\mathsf{g}>0$, the $d$-dimensional space is decomposed
into domains, and we may identify the domains labeled by A
in Fig.~\ref{Fig: domains}
with $\nu=0$ and the domains labeled by B in Fig.%
~\ref{Fig: domains}
with $\nu=1$ for any realization of the random Dirac mass matrix
$V(\bm{x})$. Hence tuning the mean value $\mathsf{m}$ to some critical
value (e.g., $\mathsf{m}=0$) realizes a situation
where the domains A and B appear 
with equal probability and the domain boundaries percolate.
This tuning stabilizes either a critical point or a metallic phase
in $d$-dimensional space.

When $N$ is even, we do not adopt the probability distribution
(\ref{alternative definition 0 bar if Z2}), for it leads to 
the massless Dirac point $\mathsf{m}=\mathsf{g}=0$ separating
two insulating phases belonging to the same topological phase with $\nu=0$ 
in the clean limit $\mathsf{g}=0$.
For example, when we consider $N=2$ and the first descendant 
$\mathbb{Z}^{\,}_{2}$ 
(see Appendix~\ref{appendisubsec: Interpretation of ...}a),
 we may choose
\begin{align}
M(\bm{x})=&\,
m^{\,}_{2,0}(\bm{x})\,
\sigma^{\,}_{2}\otimes\tau^{\,}_{0}
+
m^{\,}_{2,1}(\bm{x})\,
\sigma^{\,}_{2}\otimes\tau^{\,}_{1}
\nonumber\\
&\,
+
m^{\,}_{2,3}(\bm{x})\,
\sigma^{\,}_{2}\otimes\tau^{\,}_{3}
+
m^{\,}_{1,2}(\bm{x})\,
\sigma^{\,}_{1}\otimes\tau^{\,}_{2},
\end{align}
where the quadruplets 
$\sigma^{\,}_{\mu}$ 
and
$\tau^{\,}_{\mu}$, $\mu=0,1,2,3$,
are made of the unit $2\times2$ matrix and the three Pauli matrices,
respectively. In this case
\begin{align}
(-1)^\nu=\t{sgn}
(m_{2,0}^{2} + m_{1,2}^{2} - m_{2,1}^{2} - m_{2,3}^{2}).
\end{align}
If we define the controlling parameter $\mathsf{m}$ as in 
Eq.~(\ref{alternative definition 0 bar if Z2}),
we have $\overline{m^{\,}_{2,0}}=\mathsf{m}$ and  
$\overline{m^{\,}_{1,2}}=\overline{m^{\,}_{2,1}}=\overline{m^{\,}_{2,3}}=0$,
for which the localized phase with $\nu=0$ always appear for non-vanishing 
$\mathsf{m}$ with sufficiently small $\mathsf{g}$.
Instead, we choose a probability distribution such that
the massless Dirac point $\mathsf{m}=\mathsf{g}=0$ 
in the parameter space
$(\mathsf{m},\mathsf{g})\in\mathbb{R}\times[0,\infty[$
separates two insulating phases with $\nu=0$ and $\nu=1$
in the clean limit $\mathsf{g}=0$. 
For the above example of $N=2$,
such a choice is given by
\begin{subequations}
\begin{equation}
\overline{V(\bm{x})}=0, 
\end{equation}
\begin{equation}
\overline{m^{2}_{2,0}} 
+ 
\overline{m^{2}_{1,2}} 
- 
\overline{m^{2}_{2,1}} 
- 
\overline{m^{2}_{2,3}}
=:\mathsf{m},
\end{equation}
and 
\begin{align}
\frac{1}{r}\,
\overline{
\mathrm{tr}
\left[
V(\bm{x})\,V(\bm{y})
\right]}
=:\mathsf{g}^{2}\,e^{-|\bm{x}-\bm{y}|/\xi^{\,}_{\mathrm{dis}}}.
\end{align}
\end{subequations}
For general values of $N$, we may adopt
$\mathsf{m}:=\overline{\mathrm{Pf}\,[\mathrm{i}M(\bm{x})]}$
instead of Eq.~(\ref{alternative definition 0 bar if Z2}).

\subsection{Case of the zeroth homotopy group $\{0\}$}
\label{subsec: case of 0}

In each dimension $d$ of space, there are five AZ symmetry classes
whose classifying spaces $V^{\,}_{d,r}$
are compact and path-connected topological spaces
and have thus a vanishing zeroth homotopy groups.
We can always choose to represent the Dirac Hamiltonian
(\ref{eq: def Dirac Hamiltonian with m(x) a})
with
\begin{equation}
\bm{\alpha}=
\bm{\alpha}^{\,}_{\mathrm{min}}\otimes \openone^{\,}_{N}
\label{eq: case pi0 when 0 a}
\end{equation}
for the Dirac kinetic contribution, and then
we choose $\beta^{\,}_{\mathrm{min}}$ arbitrarily
from the allowed $r^{\,}_{\mathrm{min}}\times r^{\,}_{\mathrm{min}}$
normalized Dirac mass matrices
which anticommutes with all the components of $\bm{\alpha}_{\min}$.

For any of these five AZ symmetry classes,
we choose the parameter space
$(\mathsf{m},\mathsf{g})\in\mathbb{R}\times[0,\infty[$
by selecting
\begin{subequations}
\label{alternative definition 0 bar}
\begin{equation}
\beta^{\,}_{0}:=
\beta^{\,}_{\mathrm{min}}\otimes\openone^{\,}_{N}
\label{alternative definition 0 a bar}
\end{equation}
in the probability distribution of the Dirac mass
matrix $V(\bm{x})$ given by
\begin{align}
&
\overline{V(\bm{x})}=:
\mathsf{m}\,\beta^{\,}_{0},
\label{alternative definition 0 b bar}
\\
&
\frac{1}{r}\,
\overline{
\mathrm{tr}
\left\{
[V(\bm{x})-\mathsf{m}\,\beta^{\,}_{0}]
[V(\bm{y})-\mathsf{m}\,\beta^{\,}_{0}]
\right\}
         }
=:\mathsf{g}^{2}\,e^{-|\bm{x}-\bm{y}|/\xi_\mathrm{dis}},
\label{alternative definition 0 c bar}
\end{align}
\end{subequations}
with all higher cumulants vanishing.

In these five AZ symmetry classes,
the phase diagram has only a single localized phase that is
adiabatically connected to a topologically
trivial band insulator with reducing disorder.

Nevertheless, there can be anomalies of the conductivity and of 
the density of states at $\varepsilon=0$
when higher than the zeroth homotopy group of $V^{\,}_{d,r}$
have more than one elements, as we shall illustrate in Sec.%
~\ref{sec: Application to 2D space}.

\section{Application to 1D space}
\label{sec: Application to 1D space}

\begin{table*}[t]
\begin{center}
\caption{\label{table: 1D localization}
Interplay between the topology of the
normalized Dirac masses in a given AZ symmetry class
and Anderson localization when space is one-dimensional.
For each AZ symmetry class from the first column,
(i)
the second column gives the minimal rank $r^{\,}_{\mathrm{min}}$
of the Dirac matrices for which a Dirac mass is allowed by the symmetries,
(ii) the third column gives the corresponding topological space 
for the normalized Dirac masses
$V^{\,}_{d=1,r=r^{\,}_{\mathrm{min}}N}$,
(iii)
the fourth column gives the corresponding zeroth homotopy group 
$\pi^{\,}_{0}(V^{\,}_{d=1,r=r^{\,}_{\mathrm{min}}N})$,
and (iv) the fifth column gives the corresponding panel
from Fig.~\ref{Fig: 1D phase diagrams}.
Each panel from Fig.~\ref{Fig: 1D phase diagrams} is
a phase diagram for quasiparticles at the Fermi energy,
which is fixed to $\varepsilon=0$
in the cases of the three chiral classes (AIII, BDI, and CII)
and the four BdG classes (C, CI, D, and DIII).
Parameter space is two-dimensional with the mean value 
$\mathsf{m}\in\mathbb{R}$ of the characteristic
Dirac masses as horizontal axis and the characteristic
disorder strength $\mathsf{g}\geq0$ as vertical axis.  
The last column characterizes transport through the dependence of 
the mean conductance as a function of the length $L$ of the disordered region
when the random masses are identically distributed
with a vanishing mean value $\mathsf{m}=0$.
Insulating and critical phases are characterized by a mean conductance 
that decays exponentially and algebraically fast with $L/(N\ell)$, 
respectively, where $\ell$ is the mean free path.
The entries ``even-odd'' indicate that the conductance
decays algebraically for odd $N$ and exponentially fast
for even $N$.
In each of the zeroth homotopy column,
the three entries $\mathbb{Z}$ hold in the limit
$N\to\infty$, while the entry $0$ is a short hand for the group $\{0\}$
made of the single element $0$.
        }
\begin{tabular}[t]{cccccc}
\hline \hline
AZ symmetry class~ 
& 
~$r^{\,}_{\mathrm{min}}$~
&
~$V^{\,}_{d=1,r}$~
&
~$\pi^{\,}_{0}(V^{\,}_{d=1,r})$~
& 
~Phase diagram from Fig.~\ref{Fig: 1D phase diagrams}~
&
~Cut at $\mathsf{m}=0$
\\
\hline
A    
& 2 & $C^{\,}_{1}$ & 0 & (c) & ~insulating~
\\
~ AIII ~   
& 2 & $C^{\,}_{0}$ & $\mathbb{Z}$ & (a) & even-odd
\\
\hline
AI   
& 2 & $R^{\,}_{7}$ & 0 & (c) & insulating 
\\
BDI
& 2 & $R^{\,}_{0}$ & $\mathbb{Z}$ & (a) &  even-odd 
\\
D
& 2 & $R^{\,}_{1}$ & $\mathbb{Z}^{\,}_{2}$ & (b) & critical
\\
DIII 
& 4 & $R^{\,}_{2}$ & $\mathbb{Z}^{,}_{2}$ & (b) & critical
\\
AII
& 4 & $R^{\,}_{3}$ & 0 & (c) & insulating
\\
CII
& 4 & $R^{\,}_{4}$ & $\mathbb{Z}$ & (a) & even-odd
\\
C
& 4 & $R^{\,}_{5}$ & 0 & (c) & insulating
\\
CI
& 4 & $R^{\,}_{6}$ & 0 & (c) & insulating
\\
\hline \hline
\end{tabular}
\end{center}
\end{table*}

The effects of static and local disorder are always strong in 1D space.
The ballistic transport
for a finite number $N$ of 1D channels is unstable to disorder;
Anderson localization rules. 
However, there are exceptions to the rule, i.e.,
the localization length and the density of states are divergent at
the boundaries of insulating phases
in the phase diagrams of the symmetry classes AIII, BDI, CII, D, and DIII.
The main result of this section is that, according to Sec.%
~\ref{sec: Anderson localization and the zeroth homotopy group of ...},
these anomalies are caused by
the nature of the disconnectedness of the topological space 
parametrized by the normalized Dirac mass as 
encoded by the zeroth homotopy group of the relevant classifying
space in the third column of Table \ref{table: 1D localization}.

\begin{figure*}[tb]
\begin{center}
\includegraphics[width=1.0\linewidth]{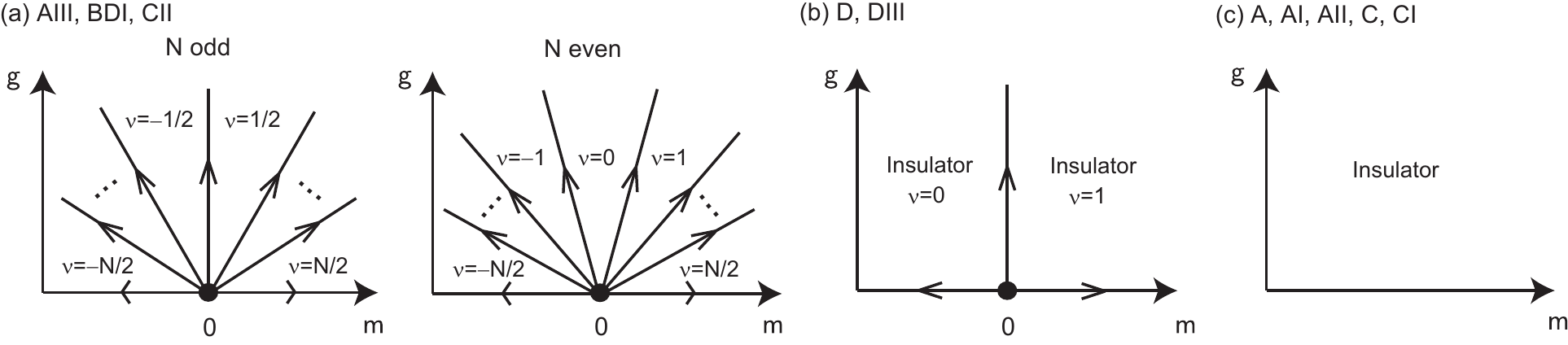}
\end{center}
\caption{\label{Fig: 1D phase diagrams} 
Qualitative quantum phase diagrams for 1D disordered wires
with the quasiparticle energy fixed at $\varepsilon=0$.
The horizontal axis is parametrized by the
characteristic value $\mathsf{m}\in\mathbb{R}$
of the disorder-averaged Dirac masses allowed by the symmetry class.
The vertical axis is parametrized by the characteristic value
$\mathsf{g}\geq0$ taken by the strength of the generic static and local
random mass disorder allowed by the symmetry class.  
Arrows on the phase boundaries and on the horizontal axis
indicate flows under renormalization group transformations.
        }
\end{figure*}

We can apply the lessons from 
Sec.~\ref{sec: Anderson localization and the zeroth homotopy group of ...}
to deduce the qualitative phase diagram for disordered
1D wires in any symmetry class.
The phase diagram is spanned by the control parameters
$\mathsf{m}$ and $\mathsf{g}$ defined in
Sec.~\ref{sec: Anderson localization and the zeroth homotopy group of ...},
as shown in Fig.~\ref{Fig: 1D phase diagrams}.

We note that, regarding transport of quasiparticles of any non-vanishing 
energy $\varepsilon$, 
there are only three relevant symmetry classes,
the standard symmetry classes A, AI, and AII.
The chiral symmetry classes AIII, BDI, and CII and
the BdG symmetry classes C, CI, D, and DIII
show their characteristic transport properties
at the band center $\varepsilon=0$
of their quasiparticle spectra.
In other words, a non-vanishing $\varepsilon$ induces a crossover
from one chiral or BdG class to 
one standard symmetry class.%
~\cite{mudry-crossover00}

Before presenting our results, we review the relevant literature.

\subsection{A brief review}

For one-dimensional lattice models with local random potentials,
analytical results for the conductance and density of states
have been obtained using the Fokker-Planck (FP) equation 
obeyed by the Lyapunov exponents of the 1D transfer matrix.%
~\cite{Dorokhov82,Mello88,Brouwer98,MBF-AIII-99,BMF-PRL00,BrouwerMudryFurusaki00,BFGM,mudry-crossover00,titov-ten-AZ-classes-01,Brouwer03,BFMRBUTSURI,Gruzberg05}

These results establish that the metallic phase 
[in the so-called quasi-one-dimensional (1D) limit defined by taking the number $N$ 
of one-dimensional channels to infinity in a suitable way~\cite{Efetov83}]
is always unstable to disorder away from the band center. 
Upon increasing the length of the 1D geometry,
disorder drives a crossover to an (Anderson) insulating phase.%
~\cite{Dorokhov82,Mello88} 
However, the band center $\varepsilon=0$ is anomalous
for seven of the ten AZ symmetry classes.%
~\cite{Brouwer98,MBF-AIII-99,BMF-PRL00,BrouwerMudryFurusaki00,BFGM,mudry-crossover00,titov-ten-AZ-classes-01,Brouwer03,BFMRBUTSURI,Gruzberg05}

On the one hand,
the chiral symmetry classes AIII, BDI, and CII display an even-odd effect
in the parity of the number $N$ of 1D transverse channels by which 
the disorder-averaged conductance at the band center alternates between 
exponential decay for even $N$ and algebraic decay for odd $N$.%
~\cite{Brouwer98}
On the other hand,
the disorder-averaged thermal conductance at the band center for any one of the
superconducting symmetry classes D and DIII shows a mean conductance 
with algebraic decay for any $N$.%
~\cite{BFGM}
These results hold for Gaussian disorder with vanishing mean.

The density of states in the neighborhood
of the band center also signals that the band center is a critical energy
for the symmetry classes AIII, BDI, CII, D and DIII separating 
two 1D Anderson insulating phases, 
as it fails to follow the dependence predicted
by random matrix theory (i.e., a suppression of the density of states
resulting from the enhanced level repulsion caused by the spectral symmetry
about the band center).%
~\cite{BMF-PRL00,titov-ten-AZ-classes-01,BFMRBUTSURI} 
In contrast, the disorder-averaged conductance is exponentially suppressed 
with the length of the wire in the symmetry classes C, CI, A, AI, and AII
for all energies while the density of states does not deviate from
its expected behavior at the diffusive (unstable) fixed point.

The Fokker-Planck equation (the DMPK equation)
obeyed by the Lyapunov exponents
describing the transfer matrix in 1D contains universal data
of geometric origin.%
~\cite{Huffmann90,BrouwerMudryFurusaki00} 
In each of the ten AZ symmetry classes,
the 1D transfer matrix defines a noncompact symmetric space,
of which the Lyapunov exponents are the radial coordinates.%
~\cite{Caselle96,Caselle04}
An infinitesimal increase in the length of the disordered region 
for one of the ten symmetry classes induces an
infinitesimal Brownian motion of the Lyapunov exponents 
that is solely controlled by the multiplicities of the
ordinary, long, and short roots of the corresponding
classical semi-simple Lie algebra under suitable assumptions
on the disorder (locality, weakness, and isotropy between all channels).
The long and short roots have a very special meaning
with regard to Anderson localization.
When the 1D transfer matrix describes the stability of the 
metallic phase of noninteracting fermions perturbed by
one-body local random potentials
in the bulk of a 1D lattice model, the multiplicity of the 
short root entering the Brownian motion of the Lyapunov exponents 
always vanish.%
~\cite{Dorokhov82,Mello88,Brouwer98,MBF-AIII-99,BrouwerMudryFurusaki00,BFGM,mudry-crossover00,BMF-PRL00,titov-ten-AZ-classes-01,Brouwer03,BFMRBUTSURI,Gruzberg05} 
Moreover, the multiplicities of the long roots also vanish
for the five 1D symmetry classes AIII, BDI, D, DIII, and CII.%
~\cite{Brouwer98,BFGM} 
This vanishing is the signature of the existence of
critical points separating topological insulating phases in 1D wires.
However, when the 1D transfer matrix describes the 1D 
boundary of a 2D topological band insulator moderately perturbed by
local random potentials, the multiplicities
of the short roots is non-vanishing in the Brownian motions of the 
Lyapunov exponents in the five AZ symmetry classes A, AII,%
~\cite{Takane04} 
D, DIII, and C.
Correspondingly, the conductance is of order one
along the infinitely long boundary, i.e., 
the insulating bulk supports extended edge states.
These extended edge states can be thought of as
realizing a 1D ballistic phase robust to disorder.
The interpretation of a DMPK equation 
with non-vanishing multiplicity of the short root 
is that it describes transport
along the boundary of a 2D topological insulator.

The same conclusions for the stability of a bulk 1D metallic 
phase,%
~\cite{Zirnbauer92,Altland01,Lamacraft04,altland14-prl}
and for the stability of a 1D ballistic phase along the 
boundary of a 2D band insulator perturbed by moderate disorder%
~\cite{schnyder-ryu08,schnyder-aip-proc09,ryu-njp10} 
also follow from describing these disordered noninteracting fermionic models
with NLSMs. The role played
by the multiplicities of the short roots in order to evade Anderson 
localization within the 1D transfer matrix approach
is played by the presence of a topological term in the 
NLSM with the target spaces corresponding to the symmetry classes
A, AII, D, DIII, and C.%
~\cite{schnyder-ryu08,schnyder-aip-proc09,ryu-njp10}  

\subsection{$\pi^{\,}_{0}(V_{d=1,r})=\mathbb{Z}$}

According to the third column of Table \ref{table: 1D localization},
the topological space $V^{\,}_{d=1,r=r^{\,}_{\mathrm{min}}\,N}$ 
that parametrizes the space of Dirac masses
is the union of the Grassmannian manifolds indexed by the
number of 1D channels $N$ for the symmetry classes AIII, BDI, and CII.

Figure~\ref{Fig: 1D phase diagrams}(a)
is the schematic phase diagram for
the chiral symmetry classes AIII, BDI, and CII
at the band center $\varepsilon=0$.
A 1D disordered wire with $N$ channels
is characterized by $N+1$ distinct insulating phases 
that are indexed by the set of half-integers and integers
with $N$ odd and even, respectively.
The flow of $\mathsf{g}$ along the phase
boundary separating two neighboring insulating
phases in Fig.~\ref{Fig: 1D phase diagrams}(a)
is to the infinitely strong disorder fix points
$\mathsf{g}\to\infty$.\cite{Dyson53,Eggarter78}
The topological argument for this property is the following.
The topological space
$V^{\,}_{d=1,r=r^{\,}_{\min}\,N}$
parametrized by the Dirac masses allowed in any one of the three chiral
symmetry classes is disconnected.
One Dirac mass is singled out,
the Dirac mass whose Dirac matrix commutes with all other Dirac masses 
present when $N>1$:
$\beta^{\,}_{0}$ defined in Eq.~(\ref{alternative definition 0 a}).
The value $\mathsf{m}$ taken by averaging this mass
over its probability distribution is the control parameter
$\mathsf{m}$ that triggers
a quantum phase transition between two distinct neighboring insulating 
phases in Fig.~\ref{Fig: 1D phase diagrams}(a).
At $\mathsf{m}=0$, a disordered wire is in the $\nu=0$ localized phase
when $N$ is even and on the phase boundary between the $\nu=\pm\frac12$
insulating phases when $N$ is odd.
This qualitative
even-odd effect in $N$ that originates with the parity of the number of 
disconnected components $N+1$
of $V^{\,}_{d=1,r=r^{\,}_{\mathrm{min}}\,N}$ is captured quantitatively
by the Fokker-Planck approach,%
~\cite{Brouwer98,MBF-AIII-99,BMF-PRL00,BrouwerMudryFurusaki00,BFGM,mudry-crossover00,titov-ten-AZ-classes-01,Brouwer03,BFMRBUTSURI}
the NLSM approach,
~\cite{Altland01,Lamacraft04,altland14-prl,Altland15}
or the application of noncommutative geometry%
~\cite{Mondragon-Shem14a,Prodan14a,Prodan14b,Song14a}
and quantum-entanglement techniques.%
~\cite{Mondragon-Shem14b}
One-parameter scaling is lost at the phase boundaries
in the vicinity of the band center.%
~\cite{Brouwer98,MBF-AIII-99,BMF-PRL00,BrouwerMudryFurusaki00}
Recently, 
analytic results encoding a two-parameter scaling similar 
to that characterizing the integer quantum Hall effect were obtained 
for the symmetry classes AIII and BDI from the supersymmetric NLSM approach.%
~\cite{altland14-prl}
Moreover, the fan-like phase diagram depicted in
Fig.~\ref{Fig: 2D phase diagrams}(a),
in which $N+1$ localized phases emerge from $\mathsf{m}=\mathsf{g}=0$, 
has been confirmed for the symmetry class BDI from
a numerical study of the Lyapunov exponents of transfer matrices.%
~\cite{altland14-prl}
Note that a nonlinearity in the band dispersion makes the
fan-like diagram asymmetric about $\mathsf{m}=0$ in tight-binding models.

The anomalous behavior at the phase boundaries can be deduced from
the results obtained previously by the Fokker-Planck approach.%
~\cite{Brouwer98,MBF-AIII-99,BMF-PRL00,BrouwerMudryFurusaki00,BFGM,mudry-crossover00,titov-ten-AZ-classes-01,Brouwer03,BFMRBUTSURI}
At any phase boundary the dimensionless Landauer conductance $g$ shows
anomalous dependence on the wire length $L$ ($\gg N\ell$),
\begin{subequations}
\label{eq: critical behavior in 1D}
\begin{align}
\overline{\ln g} \propto
-\sqrt{\frac{L}{N\,\ell}},
\qquad
\ln\overline{g} \propto
-\frac{1}{2}\ln\!\left(\frac{L}{N\,\ell}\right),
\label{eq: anomalous g}
\end{align}
and the density of states near $\varepsilon=0$
has the strongest divergence, the Dyson singularity,
\begin{align}
\rho(\varepsilon)\propto
\frac{1}
     {\left|\varepsilon\,\tau\left[\ln(|\varepsilon|\,\tau)\right]^3\right|},
\label{eq: Dyson singularity}
\end{align}
\end{subequations}%
where $\tau=N^{2}\,\ell$, $\ell$ is the mean free path, and
the Fermi velocity has been set to unity.
The universal critical behavior~(\ref{eq: critical behavior in 1D})
is a manifestation of an (unstable) infinite-disorder 
quantum critical point controlled by Griffiths effects.%
~\cite{Fisher95,Shelton98,Balents97,Hastings01,Motrunich-1D-01}

In an insulating phase the dimensionless conductance has
exponential dependence on $L$,
\begin{align}
\overline{\ln g}\propto
-\frac{L}{\xi},
\label{eq: g in a localized phase}
\end{align}
where the localization length $\xi$ is of order $N\,\ell$ and
diverges at phase boundaries.
The exponent in the power-law $\varepsilon$ dependence of
$\rho(\varepsilon)$ increases from $-1$
as $\mathsf{m}$ is changed from its value at a phase boundary.
Half way between two consecutive transitions,
the density of states is
the closest to the prediction from random matrix theory,
\begin{align}
\rho(\varepsilon)\propto
(\varepsilon\,\tau)^{m^{\,}_{\mathrm{o}}-1}\,|\ln(\varepsilon\,\tau)|,
\label{eq: D1 DoS chital classes N even}
\end{align}
where $m^{\,}_{\mathrm{o}}=1,2,4$ are the multiplicities of the ordinary roots
for the chiral symmetry classes BDI, AIII, and CII, respectively.%
~\cite{BMF-PRL00}
However, the density of states (\ref{eq: D1 DoS chital classes N even})
is not quite the one expected from random
matrix theory, for it acquires a multiplicative logarithmic correction,
as shown in Ref.~\onlinecite{BMF-PRL00}.
The density of states interpolates between these two limiting functions
as a function of the control parameter $\mathsf{m}$.%
~\cite{rieder14}

\subsection{$\pi^{\,}_{0}(V^{\,}_{d=1,r})=\mathbb{Z}^{\,}_{2}$}

According to the third column of Table~\ref{table: 1D localization},
the topological space $V^{\,}_{d=1,r=r^{\,}_{\mathrm{min}}\,N}$ 
that parametrizes the normalized random Dirac mass
in 1D has two disconnected components
for the BdG symmetry classes D and DIII. 
Accordingly, there should be two topologically distinct insulating phases
separated by a phase-boundary line, at which the localization length
and the density of states diverge.
Tuning the control parameter $\mathsf{m}$ defined
in Sec.~\ref{subsec: case of Z_{2}} to zero
selects the phase-boundary line,%
~\cite{BFGM,titov-ten-AZ-classes-01,BFMRBUTSURI}
around which one-parameter scaling is broken.%
~\cite{Motrunich-1D-01,Brouwer03,Gruzberg05}

Figure~\ref{Fig: 1D phase diagrams}(b) is the phase diagram
for the BdG symmetry classes D and DIII
at the Fermi level $\varepsilon=0$.
A 1D disordered wire is characterized by
two insulating phases that are separated by a phase boundary along
which the disorder strength $\mathsf{g}$ flows to the infinitely
strong disorder fix point, the Dyson fix point.
This phase boundary is located at $\mathsf{m}=0$. 
The topological argument for this property is the following.
The topological space $V^{\,}_{d=1,r=r^{\,}_{\mathrm{min}}\,N}$
parametrized by the Dirac masses allowed in any one of these two
BdG symmetry classes is made of two disconnected components.
For any realization of the random potential
these two components are indexed by the $\mathbb{Z}^{\,}_{2}$
index defined in Eq.\ (\ref{eq: nu when z}).
Hence the control parameter $\mathsf{m}$
that drives the quantum phase transition between
the two distinct insulating phases can be chosen to be the disorder-averaged 
value over this Pfaffian.
On the other hand, had we started from the symmetry class BDI
and weakly broken the time-reversal and chiral symmetries by the addition 
of weak perturbations that bring the Dirac Hamiltonian to one in 
the symmetry class D, we may then keep the control parameter 
$\mathsf{m}$ of the symmetry class BDI along the horizontal axis in
parameter space. There follows the same phase diagram 
as in Fig.~\ref{Fig: 2D phase diagrams}(a)
with the caveat that the $N+1$ unperturbed localized phases with the 
$\mathbb{Z}$ topological numbers $\nu$ 
become $N+1$ localized phases with alternating 
$\mathbb{Z}^{\,}_{2}$ topological numbers $\nu$ after perturbation.%
~\cite{rieder14}

The anomalous behavior at the phase boundary can be deduced from
the results obtained previously by the Fokker-Planck approach.%
~\cite{BFGM,titov-ten-AZ-classes-01,Brouwer03,BFMRBUTSURI}
The anomalous dependence on the wire length $L$ of the dimensionless
Landauer conductance $g$ is the same as Eq.\ (\ref{eq: anomalous g}),
and the density of states $\rho(\varepsilon)$ exhibits
the Dyson singularity (\ref{eq: Dyson singularity}).
In the insulating phases the conductance has the exponential dependence
on $L$, Eq.\ (\ref{eq: g in a localized phase}).
The density of states $\rho(\varepsilon)$ has the power-law dependence
on the excitation energy $\varepsilon$ with the exponent continuously
varying from $-1$ of the Dyson singularity to the exponent of
the random matrix theory\cite{AZ-classes}
(0 and 1 for class D and DIII, respectively),
as $|\mathsf{m}|$ is increased from the critical point $\mathsf{m}=0$.
The fact that the symmetry classes D and DIII share with the 
chiral symmetry classes the same Dyson singularity
is known as ``superuniversality''.%
~\cite{Gruzberg05,rieder14}

\subsection{$\pi^{\,}_{0}(V^{\,}_{d=1,r})=\{0\}$}

According to the third column of Table 
\ref{table: 1D localization},
the topological space $V^{\,}_{d=1,r=r^{\,}_{\mathrm{min}}\,N}$ 
that parametrizes the normalized random Dirac masses
is path-connected for the symmetry classes A, AI, AII, C, and CI.
Figure~\ref{Fig: 1D phase diagrams}(c)
is the schematic phase diagram for 
the symmetry classes A, AI, AII, C, and CI
at $\varepsilon=0$ 
and to all symmetry classes away from $\varepsilon=0$.
A 1D disordered wire is always localized in these symmetry classes.
The topological argument for this property is the following.
The topological space $V^{\,}_{d=1,r}$
parametrized by the Dirac masses allowed in any one of these
five symmetry classes is path connected.
Consequently, the sign of $\mathsf{m}$ has no topological meaning.
The allowed Dirac masses select a unique insulating phase.

The density of states in the vicinity of 
the quasiparticle energy $\varepsilon=0$
is affected by the enhanced level repulsion in the
BdG symmetry classes C and CI,
but this effect is captured by random matrix theory%
~\cite{AZ-classes}
since the localization length is never divergent;%
~\cite{BFGM,titov-ten-AZ-classes-01}
$\rho^{~}_{\mathrm{C}}(\varepsilon)\propto\varepsilon^{2}$
and
$\rho^{~}_{\mathrm{CI}}(\varepsilon)\propto|\varepsilon|$.

\section{Application to 2D space}
\label{sec: Application to 2D space}

\begin{table*}[tb]
\begin{center}
\caption{\label{table: 2D localization}
Interplay between the topology of the
normalized Dirac masses in a given AZ symmetry class
and Anderson localization when space is two-dimensional.
For each AZ symmetry class from the first column,
(i) the second column gives the minimal rank $r^{\,}_{\mathrm{min}}$
of the Dirac matrices for which a Dirac mass is allowed by the symmetries,
(ii) the third column gives the corresponding topological space 
for the normalized Dirac masses $V^{\,}_{d=2,r=r^{\,}_{\mathrm{min}}N}$,
(iii) the fourth column gives the corresponding zeroth homotopy group 
$\pi^{\,}_{0}(V^{\,}_{d=2,r=r^{\,}_{\mathrm{min}}N})$, 
(iv) the fifth column gives the corresponding first homotopy group 
$\pi^{\,}_{1}(V^{\,}_{d=2,r=r^{\,}_{\mathrm{min}}N})$, 
and (v) the sixth column gives the corresponding panel
from Fig.~\ref{Fig: 2D phase diagrams}.
Each panel from Fig.~\ref{Fig: 2D phase diagrams} is
a phase diagram for quasiparticles at the Fermi energy,
which is fixed to $\varepsilon=0$.
Parameter space is two-dimensional with the mean value
$\mathsf{m}\in\mathbb{R}$ for the characteristic
Dirac masses as horizontal axis and the characteristic
disorder strength $\mathsf{g}\geq0$ as vertical axis.  
The last column characterizes transport through the dependence of 
the mean conductivity 
as a function of the linear length $L$ of the disordered region
when $\mathsf{m}=0$.
Insulating phases are characterized by a mean conductivity 
that decays exponentially with $L$.
A metallic phase has a mean conductivity that grows with $L$.
A critical phase has a mean conductivity of the order of $e^{2}/h$.
The entries ``even-odd'' indicate a critical or metallic phase
for odd $N$ and an insulating phase for even $N$.
The entries ``Gade singularity'' indicate a diverging density of state
associated with
$\pi^{\,}_{1}(V^{\,}_{d=2,r})=\mathbb{Z}$. 
In each homotopy column, the three entries $\mathbb{Z}$ 
hold for $N$ larger than an integer (infinity included)
that depends on the order
of the homotopy group and the classifying space, 
the two entries $\mathbb{Z}^{\,}_{2}$ 
hold for $N$ larger than an integer that also depends on the order
of the homotopy group and the classifying space.
The entry $0$ is a short hand for the group $\{0\}$
made of the single element $0$. 
        }
\begin{tabular}[t]{ccccccc}
\hline \hline
AZ symmetry class~ 
&
~$r^{\,}_{\mathrm{min}}$~
&
~$V^{\,}_{d=2,r}$~
&
~$\pi^{\,}_{0}(V^{\,}_{d=2,r})$~
&
~$\pi^{\,}_{1}(V^{\,}_{d=2,r})$~
&  
~Phase diagram from Fig.~\ref{Fig: 2D phase diagrams}~
& 
~Cut at $\mathsf{m}=0$
\\
\hline
A    & 2 & $C^{\,}_{0}$ & $\mathbb{Z}$ & 0 & (a,b) & even-odd 
\\
AIII & 4 & $C^{\,}_{1}$ & 0 & $\mathbb{Z}$  & (i) & Gade singularity 
\\
\hline
AI   & 4 & $R^{\,}_{6}$ & 0 & 0 & (h) & insulating 
\\
BDI  & 4 & $R^{\,}_{7}$ & 0 & $\mathbb{Z}$  & (i) & Gade singularity 
\\
D    & 2 & $R^{\,}_{0}$ & $\mathbb{Z}$ & $\mathbb{Z}^{\,}_{2}$ & (c,d)
 &  even-odd 
\\
DIII & 4 & $R^{\,}_{1}$ & $\mathbb{Z}^{\,}_{2}$ & $\mathbb{Z}^{\,}_{2}$  &
(e,f) & metallic
\\
AII  & 4 & $R^{\,}_{2}$ & $\mathbb{Z}^{\,}_{2}$ & 0 & (g) & metallic
\\
CII  & 8 & $R^{\,}_{3}$ & 0 & $\mathbb{Z}$  & (i) & Gade singularity
\\
C    & 4 & $R^{\,}_{4}$ & $\mathbb{Z}$ & 0  & (a,b) & even-odd 
\\
CI   & 8 & $R^{\,}_{5}$ & 0 & 0 & (h) & insulating 
\\
\hline \hline
\end{tabular}
\end{center}
\end{table*}

This section is dedicated to Anderson localization
in the ten 2D AZ symmetry classes when a local disorder is present.  
First, we review known facts about these 2D disordered systems.
Unlike the Fokker-Planck approach for which we only
know how to extract controlled analytical results in 1D,%
~\cite{Beenakker97}
the NLSM approach delivers perturbative renormalization-group (RG)
results in any dimensions close to 2D.%
~\cite{Friedan85} 
We then show how the topology of classifying spaces 
combined with perturbative RG flows
allow to deduce the phase diagram that delimits the 
topologically distinct localized and metallic phases 
for each 2D AZ symmetry class.  
This analysis delivers a unifying explanation for many
past analytical and numerical results regarding
Anderson localization in the 2D symmetry classes
when the disorder is local.  

\subsection{A brief review}

In the symmetry classes AI,
all the single-particle states are exponentially localized 
upon perturbation by a random potential.
This result is understood perturbatively as the phenomenon of
weak localization by which the first non-vanishing quantum correction 
to the longitudinal conductivity caused by a disorder is negative.%
~\cite{abrahams79}
This result is also understood 
nonperturbatively as a consequence of the fact that the 2D NLSM 
that encodes the physics of Anderson localization
in the symmetry class AI has a coupling, the inverse of the value of the
longitudinal conductivity in the diffusive regime, 
that flows to strong coupling.%
~\cite{Efetov80,Hikami80}

In the symmetry class A,
the phase diagram of single-particle states consists of
multiple of localized phases that are distinguished 
by their quantized Hall conductivity
and are separated by phase boundaries at which the localization
length diverges.%
~\cite{Laughlin81,Halperin82,Khmelnitskii83,Levine83,Pruisken84,Laughlin84}
The existence of critical points separating neighboring localized phases
can be understood as a result of the topological term which can be
added to the 2D NLSM for class A.%
~\cite{Pruisken84}
These critical states levitate 
with increasing disorder strength until they ``annihilate'' upon merging.%
~\cite{Khmelnitskii84,Laughlin84}

In the 2D symmetry class AII, only the single-particle states 
close to the edges of the Bloch bands are exponentially localized 
in the presence of a weak disorder.%
~\cite{Hikami80,Efetov80}
As the disorder strength is increased, 
the lower and upper mobility edges separating in 
energy the Lifshitz tails from the extended single-particle states approach 
each other until they merge and all single-particle states are exponentially 
localized for sufficiently strong disorder. 

Consequently, the metallic phase for noninteracting electrons
propagating in 2D space is unstable to the presence of any 
local disorder in the symmetry classes A and AI, 
while it is stable for a sufficiently weak disorder 
in the symmetry class AII, 
as long as the chemical potential is 
sufficiently far from the unperturbed band edges.

These three outcomes for the competition between the kinetic energy and 
the local random potential
in 2D space are not exhaustive in the presence of a particle-hole or 
a chiral symmetry, as it does in the symmetry classes
AIII, BDI, D, DIII, CII, C, and CI.
Among the BdG symmetry classes,
symmetry class CI is insulating very much in the same way
as symmetry class AI is.%
~\cite{Senthil98}
Symmetry class C can display a spin IQHE
very much in the same way as symmetry class A can display the charge IQHE.%
~\cite{Senthil98}
The metallic phases in the symmetry classes D and DIII are stable
to weak disorder.%
~\cite{Senthil99,Senthil00,Bocquet00}
Symmetry class D can display a thermal IQHE
very much in the same way as symmetry class A can display the charge IQHE.%
~\cite{Senthil00,read-green00,chalker-classD01,Mildenberger07}
Symmetry class DIII can realize a thermal $\mathbb{Z}^{\,}_{2}$
topological insulator very much in the same way as symmetry class AII can.%
~\cite{schnyder-ryu08,Fulga-DIII-12}
Finally, the chiral classes AIII, BDI, and CII 
are anomalous at the band center
with diverging localization length and density of states,
while they crossover to the
symmetry classes A, AI, and AII, respectively,
for any non-vanishing value of the chemical potential.%
~\cite{gade-wegner,gade1993anderson}

All of these features that are of topological origin
can be understood qualitatively from
the zeroth and first homotopy groups from Table 
\ref{table: 2D localization},
as we now explain below.

\subsection{Implications of the topology of the classifying spaces}

Equation (\ref{eq: def Dirac Hamiltonian with m(x)}) 
with $d=2$ and with the rank $r=r^{\,}_{\mathrm{min}}N$
is our starting point. 
It is assumed that the local random disorder 
entering the Dirac Hamiltonian is
the most general $r\times r$ Hermitian matrix with random
identically independently distributed (iid)
matrix elements up to the constraints imposed by the 
selected AZ symmetry classes.
We fix the chemical potential $\mu=0$ and discuss
transport properties of quasiparticles of energy $\varepsilon=0$
for all ten symmetry classes.

\begin{figure*}[tb]
\begin{center}
\includegraphics[width=1.0\linewidth]{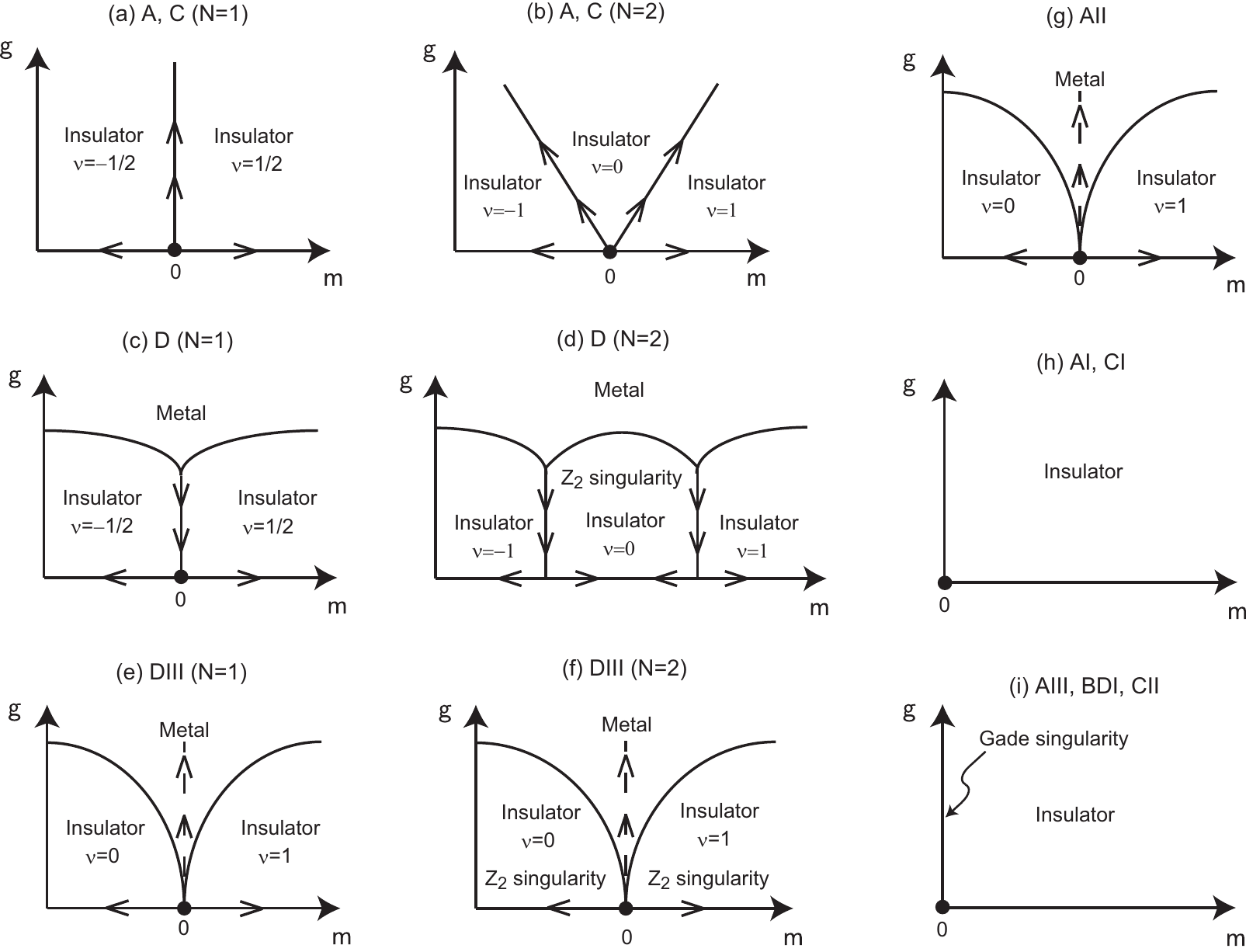}
\end{center}
\caption{\label{Fig: 2D phase diagrams}
Qualitative quantum phase diagrams for 2D random Dirac Hamiltonians
with the quasiparticle energy fixed at $\varepsilon=0$.
The horizontal axis is parametrized by the
characteristic value $\mathsf{m}\in\mathbb{R}$
of the disorder-averaged Dirac masses allowed by the symmetry class.
We choose the probability distribution%
~(\ref{eq: definition of m for irrelevant g}) 
for the case of $N=2$ in the symmetry class D.
The vertical axis is parametrized by the characteristic value
$\mathsf{g}\geq0$ taken by the strength of the generic local
disorder allowed by the 2D AZ symmetry class.
The zeroth homotopy group associated with the space of 
normalized Dirac mass fixes the indexing 
of the topologically distinct insulating phases. 
The first homotopy group of the normalized Dirac masses determines if the
density of states is regular or singular at the band center.
The existence of a metallic phase, the flows along phase boundaries,
and the dimensionality of the phase boundaries all follow from perturbative
renormalization-group calculations.
        }
\end{figure*}

For each AZ symmetry class in the first column of
Table \ref{table: 2D localization},
we give the minimum rank $r^{\,}_{\mathrm{min}}$
for which the Dirac Hamiltonian admits a Dirac mass,
the classifying space $V^{\,}_{d=2,r}$,
its zeroth homotopy group 
$\pi^{\,}_{0}(V^{\,}_{d=2,r^{\,}_{\mathrm{min}}\,N})$,
its first homotopy group 
$\pi^{\,}_{1}(V^{\,}_{d=2,r^{\,}_{\mathrm{min}}\,N})$,
and its phase diagram for $N=1$ and $N=2$.
All other entries hold for any $N$.
Each panel from Fig.~\ref{Fig: 2D phase diagrams}
is a phase diagram in a two-dimensional parameter space.
The horizontal and vertical axes are the characteristic mean value
$\mathsf{m}$ of the Dirac masses and the characteristic disorder
strength $\mathsf{g}$, respectively, introduced
in Sec.~\ref{sec: Anderson localization and the zeroth homotopy group of ...}.
The last column characterizes transport through the dependence of 
the mean conductivity 
as a function of the linear length $L$
of the disordered region when $\mathsf{m}=0$
and $\varepsilon=0$.
Insulating phases are characterized by a mean conductivity 
that decays exponentially with $L$.
A metallic phase has a mean conductivity that grows with $L$.
Critical phases have a mean conductivity 
that is non-vanishing but finite for $L\to\infty$.
The entry ``even-odd'' indicates that the parity of
$N$ selects either an insulating or a critical phase,
depending on whether $N$ is even or odd, respectively. 
The new entry ``Gade singularity'' 
compared to the 1D Table \ref{table: 1D localization}
signals a conductivity that is independent of
the disorder strength and a divergence of the density of states 
upon approaching the band center. This new entry is a signature
of $\pi^{\,}_{1}(V^{\,}_{d=2,r})=\mathbb{Z}$, as we are going to explain.

The main result of this section is that, according to Sec.%
~\ref{sec: Anderson localization and the zeroth homotopy group of ...},
there are deviations in Table \ref{table: 2D localization}
away from the insulating behavior found in the 2D symmetry classes AI and CI.
These deviations are attributed to the zeroth and first 
homotopy groups of the normalized Dirac masses being either
$\mathbb{Z}^{\,}_{2}$ or $\mathbb{Z}$.

\subsection{$\pi^{\,}_{0}(V^{\,}_{d=2,r})=\mathbb{Z}$}

According to Table \ref{table: 2D localization},
the 2D AZ symmetry classes A, D, and C
are associated with Grassmannian manifolds.
Hence the zeroth homotopy group of
the topological spaces associated with the normalized Dirac masses 
in these AZ symmetry classes is $\mathbb{Z}$.
According to Sec.%
~\ref{sec: Anderson localization and the zeroth homotopy group of ...},
their phase diagrams must host $N+1$ topologically distinct
insulating phases and $N$ transition lines separating them.

The arguments of Sec.%
~\ref{sec: Anderson localization and the zeroth homotopy group of ...}
follow from the zeroth homotopy group of the
Grassmannian manifolds. The 2D symmetry classes A and C share
the same trivial first homotopy group. In view of the fact
that the 2D metallic phase is perturbatively unstable
(due to weak localization)
in both symmetry classes (a fact not encoded in the homotopy groups
of Table \ref{table: 2D localization}),
we expect 2D phase diagrams
in the 2D symmetry classes A and C that are similar to the one
in the 1D chiral symmetry classes, the one shown in 
Figs.~\ref{Fig: 2D phase diagrams}(a)
and \ref{Fig: 2D phase diagrams}(b)
for $N=1$ and $N=2$, respectively.

Contrary to the 2D symmetry classes A and C,
the metallic phase of the 2D symmetry class D is stable
(due to weak antilocalization) to weak disorder.
Hence the 2D symmetry class D
is expected to host a (thermal) metallic phase 
when the characteristic disorder strength
$\mathsf{g}$ is sufficiently large,
in addition to $N+1$ topologically distinct insulating phases.
When the metallic phase is robust to weak disorder,
topologically distinct localized phases can be separated by 
a metallic phase rather than by a mere phase boundary line.
It is the RG flow of 
the random Dirac mass in the Dirac Hamiltonian that
determines which of the metallic phase or 
the critical boundary separates two topologically distinct
insulating phases.
Now, the characteristic coupling $\mathsf{g}$ for the disorder
is marginally irrelevant in the vicinity of the critical point
$\mathsf{m}=0$,
$\mathsf{g}=0$ 
for the symmetry class D with $N=1$, 
as is reviewed in Appendix~\ref{app: OPE}.
Consequently, there is a RG flow along the
phase boundary between topologically distinct insulating phases
into the critical point 
$\mathsf{m}=0$,
$\mathsf{g}=0$
when $N=1$.
More generally, phase boundaries separate 
topologically distinct insulating phases
at small $\mathsf{g}$.
Thus, starting from any of the $N+1$ insulating phase
separated by phase boundaries, we have a transition driven by increasing 
$\mathsf{g}$ into the metallic phase of the 2D symmetry class D,
as shown in Figs.~\ref{Fig: 2D phase diagrams}(c) ($N=1$)
and \ref{Fig: 2D phase diagrams}(d) ($N=2$).
With regard to the case of $N>1$, say $N=2$ as depicted 
in Fig.~\ref{Fig: 2D phase diagrams}(d), 
we are choosing the probability distribution%
~(\ref{eq: definition of m for irrelevant g})
to parametrize the horizontal axis of the phase diagram
instead of the  probability distribution%
~(\ref{alternative definition 0})
used to parametrize the horizontal axis of the phase diagram%
~\ref{Fig: 2D phase diagrams}(c) when $N=1$.
(See Appendix~\ref{appsubsec: OPE class D}
for a more detailed discussion of the relationship between 
the irrelevant RG flow of $\mathsf{g}$ 
and the expectation values of mass terms.) 
Moreover, the Grassmannian manifold for the 2D symmetry class
D also differs from that for the symmetry classes A and C
in that its first homotopy group is nontrivial and given
by $\mathbb{Z}^{\,}_{2}$.
In the $N=2$ case, it is the path-connected component $O(2)/O(1)\times O(1)$ 
from the space of Dirac masses (the $\nu=0$ insulating phase) that has 
a nontrivial first homotopy group and can thus host $\mathbb{Z}^{\,}_{2}$
vortices, each supporting a Majorana zero mode.
Such zero modes would lead to a Griffiths-like singularity at $\varepsilon=0$
in the density of states.
However, $\mathbb{Z}^{\,}_{2}$ vortices are not expected to exist in any
insulating phase whose space of Dirac masses has a trivial
first homotopy group. For example, in the $N=2$ case, the point-like components
$O(2)/O(2)\times O(0)$
or
$O(2)/O(0)\times O(2)$
from the space of Dirac masses (the $\nu=\pm1$ insulating phases) 
have a trivial first homotopy group. Hence, if it is the
$\nu=\pm1$ insulating phases of Dirac fermions in the symmetry class D with
$N=2$ that are realized in the vicinity of
$\varepsilon=0$, then no Griffiths-like singularity
is expected at $\varepsilon=0$ in the density of states.
In other words, we expect a density of states that deviates from the
one predicted by random matrix theory in the symmetry class D for
any insulating phase supporting Majorana zero modes 
bound to $\mathbb{Z}^{\,}_{2}$ Dirac mass vortices.%
~\cite{footnote-classD}

These predictions are consistent with the following numerical studies of
2D quantum network models.
The cases $N=1$ and $N=2$
in the symmetry class A are the best known applications of
quantum network models.
Chalker and Coddington studied numerically 
in Ref.~\onlinecite{Chalker88}
the 2D quantum network model corresponding to the case $N=1$ in the symmetry
class A.~\cite{ludwig94,Ho96} 
Lee and Chalker studied numerically 
in Ref.~\onlinecite{lee-chalker94}
the 2D quantum network model corresponding to the 
case $N=2$ in the symmetry class A
with their investigation of spin-degenerate Landau levels
and found two quantum phase transitions separating three insulating
phases of 2D quantum matter. 
Two-dimensional quantum network models for the symmetry class C 
have also been studied analytically and numerically.%
~\cite{Gruzberg99,Kagalovsky99,Beamond02,Ortuno09}
They deliver the same phase diagrams as for the 2D symmetry class A.
The case of $N=1$ in the symmetry class
D can also be regularized by a 2D quantum network model with nodal
scattering matrices of appropriate symmetry and rank,%
~\cite{chalker-classD01}
or by two-band lattice models for Majorana fermions.%
~\cite{kraus-stern-11} 
The former model shows a critical boundary with the
attractive fixed point corresponding to
the Dirac Hamiltonian at $\mathsf{m}=\mathsf{g}=0$
that separates two insulating
phases of 2D quantum matter for not too strong $\mathsf{g}$,
as shown in Fig.~\ref{Fig: 2D phase diagrams}(c)
when $N=1$.
The two insulating phases are unstable 
to the metallic phase upon increasing $\mathsf{g}$,
as shown in Fig.~\ref{Fig: 2D phase diagrams}(c).
There have been numerical studies on
the critical properties around the multicritical point where the
two insulating and one metallic phases meet at $\mathsf{g}>0$.%
~\cite{Mildenberger07,Fulga-DIII-12}
For $N=2$, there are three insulating phases on top of which
sits a metallic phase,
as shown in Fig.~\ref{Fig: 2D phase diagrams}(d).
Moreover, quasi-zero modes bound to $\mathbb{Z}^{\,}_{2}$
vortices in the Dirac masses are present in any insulating phase
for which the first homotopy group is $\mathbb{Z}^{\,}_{2}$,
the $\nu=0$ phase in Fig.~\ref{Fig: 2D phase diagrams}(d)
when $N=2$.%

\subsection{$\pi^{\,}_{0}(V^{\,}_{d=2,r})=\mathbb{Z}^{\,}_{2}$}

According to Table \ref{table: 2D localization},
the 2D AZ symmetry classes DIII and AII are associated with
disconnected topological spaces with two path-connected components, 
i.e., the zeroth homotopy group of 
the topological spaces associated with the normalized Dirac masses 
in these AZ symmetry classes is $\mathbb{Z}^{\,}_{2}$.
According to Sec.%
~\ref{sec: Anderson localization and the zeroth homotopy group of ...},
their phase diagrams must host two topologically distinct
insulating phases and one critical boundary or one metallic phase 
separating them at $\mathsf{m}=0$.

From the point of view of topology, the normalized Dirac masses
in the 2D symmetry class DIII differ from those 
in the 2D symmetry class AII in that
the former has the first homotopy group $\mathbb{Z}^{\,}_{2}$ 
when $N\geq2$, while the latter has a trivial first homotopy group
for all $N=1,2,\ldots$.
This difference manifests itself through a density of states that
deviates from the one predicted from random matrix theory
in the 2D insulating phases of the symmetry class DIII when $N=2,3,\ldots$
[Fig.~\ref{Fig: 2D phase diagrams}(f)].
The metallic phase is stable (due to weak antilocalization)
to the presence of weak disorder in both 2D symmetry classes. 
Now, the characteristic coupling $\mathsf{g}$ for the disorder
is marginally relevant in the vicinity of the critical point 
$\mathsf{m}=\mathsf{g}=0$  
for symmetry classes DIII and AII, 
as is reviewed in Appendix~\ref{app: OPE}.
Consequently, the metallic phase separates two
topologically distinct insulating phases 
for any non-vanishing $\mathsf{g}>0$
all the way down to the critical point
$\mathsf{m}=\mathsf{g}=0$ 
and sits on top of the insulating phases as depicted
in Figs.~\ref{Fig: 2D phase diagrams}(e), 
\ref{Fig: 2D phase diagrams}(f), and \ref{Fig: 2D phase diagrams}(g).

These predictions are consistent with the following published works.
Studies of random Dirac Hamiltonians%
~\cite{Ringel12,Mong12,Morimoto-weak14} 
and a network model%
~\cite{Obuse-weakTI13}
in the 2D symmetry class AII 
have confirmed the robustness of the
metallic phase all the way down to the critical point
$\mathsf{m}=\mathsf{g}=0$ 
in Fig.~\ref{Fig: 2D phase diagrams}(g).
The numerical study in Ref.~\onlinecite{Fulga-DIII-12} of a 
2D quantum network model for the symmetry 
class DIII is also consistent with Fig.%
~\ref{Fig: 2D phase diagrams}(e).

\subsection{$\pi^{\,}_{0}(V^{\,}_{d=2,r})=\{0\}$}

According to Table \ref{table: 2D localization},
the 2D AZ symmetry classes AIII, AI, BDI, CII, and CI are associated with
path-connected topological spaces, i.e., the zeroth homotopy group of 
the topological spaces associated with the normalized Dirac masses 
in these AZ symmetry classes is the trivial group $\{0\}$.
According to Sec.%
~\ref{sec: Anderson localization and the zeroth homotopy group of ...},
their phase diagrams must host no more than one
insulating phases.

The symmetry classes AI and CI do not support a 2D metallic phase,
as it is unstable due to weak localization.
For this reason, their phase diagrams in Fig.%
~\ref{Fig: 2D phase diagrams}(h)
consist of a single insulating phase.
An example of a localized phase in class CI is 
a dirty $d^{\,}_{x^{2}-y^{2}}$ superconductor 
with TRS and spin $SU(2)$ rotation symmetry.%
\cite{Senthil98,Senthil99}

The chiral symmetry classes AIII, BDI, and CII 
stand out in 2D space because
(i) 
the quantum corrections to their longitudinal Drude conductivities
vanish to any order in perturbation theory precisely at the band center,
and 
(ii) their density of states diverges at
the band center $\varepsilon=0$, as was shown by Gade and Wegner
for a sublattice model perturbed by disorder with chiral symmetry
in Refs.~\onlinecite{gade-wegner} and \onlinecite{gade1993anderson}.
Items (i) and (ii) are also true for 
random Dirac Hamiltonians in the 2D symmetry classes
AIII, BDI, and CII at the band center.%
~\cite{guruswamy2000gl,mudry-gade-singularity03,ryu-global-phase12}

On the one hand,
the topology of the classifying spaces of the 2D chiral symmetry classes 
AIII, BDI, and CII is not predictive regarding item (i). 
On the other hand, the fact that the first homotopy group of
the normalized Dirac masses in the 2D symmetry classes
AIII, BDI, and CII is $\mathbb{Z}$ provides an explanation
for the diverging density of states upon approaching the band center
alternative to the one based on the RG calculations
from Refs.%
~\onlinecite{gade-wegner,gade1993anderson,guruswamy2000gl,mudry-gade-singularity03,ryu-global-phase12}
or to the one based on Griffith effects from
Ref.~\onlinecite{Motrunich-2D-02}.
The physical interpretation of 
$\pi^{\,}_{1}(V^{\mathrm{AIII}}_{d=2,r=4})=
\pi^{\,}_{1}(V^{\mathrm{BDI}}_{d=2,r=4})=
\pi^{\,}_{1}(V^{\mathrm{CII}}_{d=2,r=8})=\mathbb{Z}$
is that one finds the (noncontractible) unit circle $S^{1}$ 
in the topological space of normalized Dirac
masses from the 2D symmetry classes AIII, BDI, and CII.
This subspace is generated by two anticommuting mass matrices
as was pointed out in the studies of charge fractionalization
in graphene.%
~\cite{Hou07,Jackiw07,Chamon08a,Chamon08b,Ryu09,Santos10,Santos11}
Such a unit circle, corresponding as it is to
a pair of normalized masses, supports point-like defects,
$U(1)$ vortices, which can bind an integer number of zero modes if 
isolated.%
~\cite{jackiw-rossi81}
A finite but dilute density of vortices binds midgap states
resulting in the singular density of states at the band center.
We show the phase diagrams of the 2D chiral symmetry classes at 
$\epsilon=0$ in Fig.~\ref{Fig: 2D phase diagrams}(i).
In the insulating phase the density of states exhibits a Griffiths
singularity $\rho(\varepsilon)\propto|\varepsilon|^{\alpha}$
at the band center $\varepsilon=0$ with the exponent $\alpha$ varying
continuously with $|\mathsf{m}|$
from $\alpha=-1$ at $\mathsf{m}=0$ (Gade singularity)
to the value expected from random matrix theory $\alpha=1,0,3$ for
class AIII, BDI, and CII, respectively.
We note that the 2D symmetry class CII differs from the 2D symmetry classes
AIII and BDI when $\varepsilon \neq 0$ and the disorder is weak.
Indeed, any deviation of the chemical potential away from the band
center is a relevant symmetry-breaking perturbation that drives a crossover
between the symmetry class CII to the symmetry class AII.
As spin-orbit coupling favors antiweak localization,
the metallic phase is stable to weak disorder in the symmetry class AII.

\section{Application to 3D space}
\label{sec: Application to 3D space}

\begin{table*}[tb]
\begin{center}
\caption{\label{table: 3D localization}
Interplay between the topology of the
normalized Dirac masses in a given AZ symmetry class
and Anderson localization when space is three-dimensional.
For each AZ symmetry class from the first column,
(i) the second column gives the minimal rank $r^{\,}_{\mathrm{min}}$
of the Dirac matrices for which a Dirac mass is allowed by the symmetries,
(ii) the third column gives the corresponding topological space 
for the normalized Dirac masses $V^{\,}_{d=3,r=r^{\,}_{\mathrm{min}}N}$,
(iii) the fourth column gives the corresponding zeroth homotopy group 
$\pi^{\,}_{0}(V^{\,}_{d=3,r=r^{\,}_{\mathrm{min}}N})$, 
(iv) the fifth column gives the corresponding first homotopy group 
$\pi^{\,}_{1}(V^{\,}_{d=3,r=r^{\,}_{\mathrm{min}}N})$, 
(v) the sixth column gives the corresponding second homotopy group 
$\pi^{\,}_{2}(V^{\,}_{d=3,r=r^{\,}_{\mathrm{min}}N})$,
(vi) the seventh column gives the corresponding panel
from Fig.~\ref{Fig: 3D phase diagrams}, and
(vii) the last column signals if the density of states per unit energy
and per unit volume is singular at the Fermi energy,
which is fixed to $\varepsilon=0$.
A singular density of states 
is associated with nontrivial point defects characterized by
$\pi^{\,}_{2}(V^{\,}_{d=3,r=r^{\,}_{\mathrm{min}}\,N})$.
In each homotopy column, the three entries $\mathbb{Z}$ 
hold for $N$ larger than an integer (infinity included)
that depends on the order
of the homotopy group and the classifying space, 
the two entries $\mathbb{Z}^{\,}_{2}$ 
hold for $N$ larger than an integer that also depends on the order
of the homotopy group and the classifying space.
The entry $0$ is a short hand for the group $\{0\}$
made of the single element $0$.
        }
\begin{tabular}[t]{cccccccc}
\hline \hline
AZ symmetry class~ 
&
~$r^{\,}_{\mathrm{min}}$~
&
~$V^{\,}_{d=3,r}$~
&
~$\pi^{\,}_{0}(V^{\,}_{d=3,r})$~
&
~$\pi^{\,}_{1}(V^{\,}_{d=3,r})$~
&
~$\pi^{\,}_{2}(V^{\,}_{d=3,r})$~
&  
~Phase diagram from Fig.~\ref{Fig: 3D phase diagrams}~
& 
~Density of states
\\
\hline
A    & 4 & $C^{\,}_{1}$ & 0 & $\mathbb{Z}$ & 0 & (c) & \hbox{Nonsingular}
\\
AIII & 4 & $C^{\,}_{0}$ & $\mathbb{Z}$ & 0 & $\mathbb{Z}$  & (a) & \hbox{Singular}
\\
\hline
AI & 8 & $R^{\,}_{5}$ & 0 & 0 & 0 & (c) & 
\hbox{Nonsingular}
\\
BDI& 8 & $R^{\,}_{6}$ & 0 & 0 & $\mathbb{Z}$  & (c) & 
\hbox{Singular}
\\
D  & 4 & $R^{\,}_{7}$ & 0 & $\mathbb{Z}$ & $\mathbb{Z}^{\,}_{2}$ & (c) & 
\hbox{Singular}
\\
DIII& 4 & $R^{\,}_{0}$ & $\mathbb{Z}$ & $\mathbb{Z}^{\,}_{2}$ & $\mathbb{Z}^{\,}_{2}$ & (a) & 
\hbox{Singular}
\\
AII & 4 & $R^{\,}_{1}$ & $\mathbb{Z}^{\,}_{2}$ & $\mathbb{Z}^{\,}_{2}$ & 0 & (b) & 
\hbox{Nonsingular}
\\
CII & 8 & $R^{\,}_{2}$ &  $\mathbb{Z}^{\,}_{2}$ & 0 & $\mathbb{Z}$  & (b) & 
\hbox{Singular}
\\
C   & 8 & $R^{\,}_{3}$ & 0 &  $\mathbb{Z}$ &  0 & (c) & 
\hbox{Nonsingular}
\\
CI  & 8 & $R^{\,}_{4}$ & $\mathbb{Z}$ & 0 & 0 & (a) & 
\hbox{Nonsingular}
\\
\hline \hline
\end{tabular}
\end{center}
\end{table*}

\begin{figure*}[tb]
\begin{center}
\includegraphics[width=1.0\linewidth]{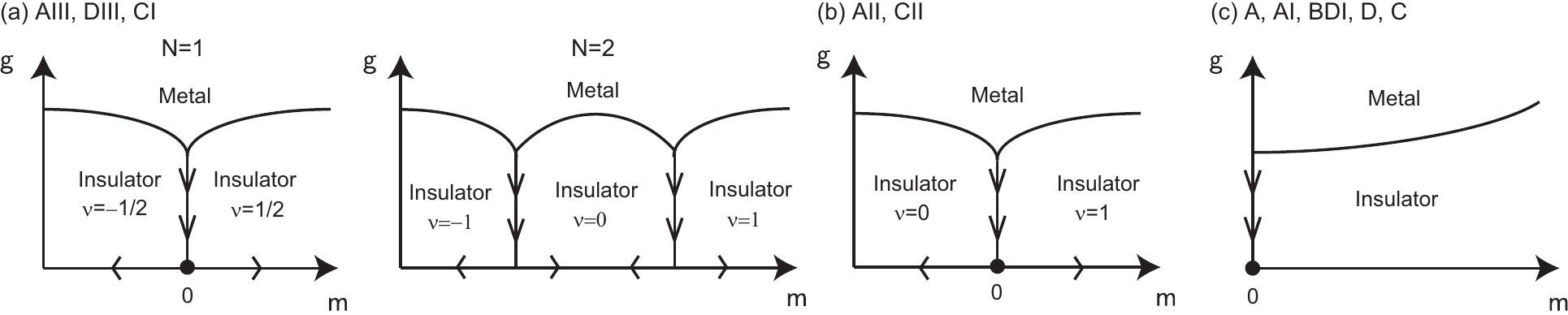}
\end{center}
\caption{\label{Fig: 3D phase diagrams}
Qualitative quantum phase diagrams for 3D random Dirac Hamiltonians
with the quasiparticle energy fixed at $\varepsilon=0$.
The horizontal axis is parametrized by the
characteristic value $\mathsf{m}\in\mathbb{R}$
of the disorder-averaged Dirac masses allowed by the symmetry class.
We choose the probability distribution%
~(\ref{eq: definition of m for irrelevant g}) 
for the case of $N=2$ in the symmetry classes AIII, DIII, and CI.
The vertical axis is parametrized by the characteristic value
$\mathsf{g}\geq0$ taken by the strength of the generic local
disorder allowed by the 3D AZ symmetry class. 
The zeroth homotopy group of the topological space associated with
the normalized Dirac mass fixes the indexing 
of the topologically distinct insulating phases. 
The existence of a metallic phase, the flows along phase boundaries,
and the dimensionality of the phase boundaries all follow from perturbative
renormalization-group calculations.
        }
\end{figure*}

In this section, we are going to employ the machinery of
classifying spaces in order to deduce the qualitative phase diagrams
capturing the physics of Anderson localization when the effects
of local disorder are not always strong, or 
marginally strong as they were in 1D and 2D, respectively. 
The lowest dimension in which
such disorder can be treated perturbatively 
in all ten AZ symmetry classes is $d=3$.
We shall show that the machinery of the classifying spaces
applied to the normalized Dirac masses 
is predictive regarding the phase diagrams when combined
with the fact that the characteristic coupling $\mathsf{g}$ 
for the disorder is irrelevant at the critical Dirac point
for any $d\geq3$. Before doing so however, 
we briefly review salient past results on
Anderson localization in 3D.

The metallic phase is stable in 3D space for the standard
symmetry classes A, AI, and AII.
A sufficiently strong disorder is required to
stabilize the insulating phase.%
~\cite{Slevin97,Slevin99,Kobayashi14}
The same is true for all chiral and symmetry classes at the band center.

The insulating phases in 3D space for the standard
symmetry classes A and AI are topologically trivial.
There are two topologically distinct insulating phases
in the 3D symmetry class AII.%
~\cite{roy-3dTI09,moore-balents07,fu-kane-mele07}
At the band center,
four more nonstandard symmetry classes support 
topologically distinct insulating phases.%
~\cite{schnyder-ryu08,schnyder-aip-proc09,ryu-njp10}

\subsection{Higher homotopy groups and topological defects}

More interestingly from the point of view of this paper,
we predict that there are 3D symmetry phases with
topologically trivial insulating phases
that support a singular density of states at the band center
due to the fact that their normalized Dirac masses support
topological defects, as we are going to explain now.

To investigate this possibility,
we start from the Dirac Hamiltonian%
~(\ref{eq: def Dirac Hamiltonian with m(x)})
with $d=3$ and the rank $r=r^{\,}_{\mathrm{min}}\,N$.
It is assumed that the local disorder 
entering the Dirac Hamiltonian is
the most general $r\times r$ Hermitian matrix with random
iid matrix elements up to the constraints imposed by the 
selected AZ symmetry classes. 
We fix the chemical potential $\mu=0$ and discuss
transport properties of quasiparticles of energy $\varepsilon=0$
for all ten symmetry classes.

According to Sec.~\ref{subsec: Relationship to higher homotopy groups},
when $\pi^{\,}_{d-1}(V^{\,}_{d})$ has a nontrivial entry,
point defects in the normalized Dirac masses may bound zero modes.
If so, they contribute to a singular density of states.
From the transition rule in Eq.~(\ref{eq: piD is related to pi0}),
in any dimension $d$ of space,
$\pi^{\,}_{d-1}(V^{\,}_{d})= \mathbb{Z}$ 
for the chiral symmetry classes AIII, BDI, and CII, while 
$\pi^{\,}_{d-1}(V^{\,}_{d})= \mathbb{Z}^{\,}_{2}$ 
in the BdG symmetry classes D and DIII.
Hence, in any dimension $d$ of space, 
a random Dirac Hamiltonian of the form
(\ref{eq: def Dirac Hamiltonian with m(x)})
displays a singular density of states at the
band center in the chiral symmetry classes AIII, BDI, and CII
and in the BdG symmetry classes D and DIII
due to the proliferation of midgap states bound to
$\mathbb{Z}$ and $\mathbb{Z}^{\,}_{2}$ point defects, respectively.
Topological defects of higher dimensions 
$d^{\,}_{\mathrm{def}}=1,\ldots,d-1$
also bind midgap states that contribute
to the density of states around $\varepsilon=0$.
However, they are not expected to cause
a singularity of the density of states at $\varepsilon=0$. 
Indeed, either the density of extended defects 
scales with the linear size $L$ of space as $L^{-d^{\,}_{\mathrm{def}}}$ 
and is thus subextensive, or the midgap states bound 
to compact extended defects with the characteristic
linear length $L^{\,}_{\mathrm{def}}$ are a level spacing of order
$1/L^{\,}_{\mathrm{def}}$ away from  $\varepsilon=0$.
Above the dimensionality of space $d=3$, all AZ symmetry classes
support at least one nontrivial homotopy group.
The form of the singularity of the density of states at the band center 
depends parametrically on the statistical distribution
of the Dirac masses. It is generically nonuniversal, 
except at the critical points that govern
the transition between distinct phases.

\subsection{$\pi^{\,}_{0}(V^{\,}_{d=3,r})=\mathbb{Z}$ \label{subsec: 3D Z}}

The normalized Dirac masses realize the Grassmannian manifolds
in the 3D symmetry classes AIII, DIII, and CI
according to Table \ref{table: 3D localization}. 
We deduce that these 3D symmetry classes support $N+1$ 
topologically distinct insulating phases separated by $N$ boundaries
by combining the arguments of
Sec.~\ref{sec: Anderson localization and the zeroth homotopy group of ...}
with the fact that the characteristic coupling $\mathsf{g}$ 
for the disorder is irrelevant at the critical Dirac point
for any $d\geq3$, see Fig.~\ref{Fig: 3D phase diagrams}(a).
Since the characteristic coupling $\mathsf{g}$ 
is irrelevant, we must distinguish the case when $N=1$ from the case when
$N>1$. When $N=1$, we choose the probability distribution%
~(\ref{alternative definition 0}).
When $N>1$, we choose the probability distribution%
~(\ref{eq: definition of m for irrelevant g}).
Indeed, we recall that if we label the horizontal axis 
(the clean limit) of the phase diagram
with $\mathsf{m}$ 
defined by Eq.~(\ref{alternative definition 0 b}), then
the horizontal axis supports no more than two insulating phases 
with $\nu=\pm \frac{N}{2}$, respectively. Given that $\mathsf{g}$ 
is irrelevant in the vicinity of $\mathsf{m}=\mathsf{g}=0$,
the choice~(\ref{alternative definition 0 b}) delivers two localized
phases below the metallic dome. In other words, the phase diagram
when $N>1$ with the choice~(\ref{alternative definition 0})
is identical to that of Fig.~\ref{Fig: 3D phase diagrams}(a) 
with the caveat that the index $\nu=\pm\frac{1}{2}$ 
is to be replaced with the index $\nu=\pm\frac{N}{2}$
for the two localized phases. To display $N+1>2$ localized phases 
in the phase diagram, it is necessary to have $N+1>2$ insulating phases
along the horizontal axis (the clean limit) of the phase diagram.
This is achieved with the choice~(\ref{eq: definition of m for irrelevant g})
that insures that the eigenvalues of the Dirac mass matrix $\beta^{\,}_{0}$ 
in Eq.~(\ref{eq: definition of m for irrelevant g a})
competes with $N$ nondegenerate eigenvalues of the matrix $M^{\,}_{0}$
in Eq.~(\ref{eq: definition of m for irrelevant g c})
The second homotopy group of the normalized Dirac masses
is $\mathbb{Z}$  in the 3D symmetry class AIII 
according to Table \ref{table: 3D localization}. 
The first and second homotopy groups of the
normalized Dirac masses are 
$\mathbb{Z}^{\,}_{2}$
in the 3D symmetry class DIII
according to Table \ref{table: 3D localization}. 
Hence the density of states 
(the number of states per unit energy and per unit volume)
is singular at the band center
due to the proliferations of midgap states that are bound to
point-defects of the normalized Dirac masses for the 3D symmetry
classes AIII and DIII. The 3D symmetry class CI has
trivial first and second homotopy groups so that the
density of states is nonsingular at the band center.

\subsection{$\pi^{\,}_{0}(V^{\,}_{d=3,r})=\mathbb{Z}^{\,}_{2}$}

The normalized Dirac masses are the union of two path-connected
compact topological spaces in the 3D symmetry classes AII and CII
according to Table \ref{table: 3D localization}.  We deduce that these
3D symmetry classes support two topologically distinct insulating
phases separated by one boundary by combining the arguments of
Sec.~\ref{sec: Anderson localization and the zeroth homotopy group of
  ...}  with the fact that the characteristic coupling $\mathsf{g}$
for the disorder is irrelevant at the critical Dirac point for any
$d\geq3$.  A metallic phase sits on top of these two topologically
distinct insulating phases. Thus, we deduce the phase diagram for
classes AII and CII shown in 
Fig.~\ref{Fig: 3D phase diagrams}(b).
This phase diagram is consistent with the self-consistent Born
approximation%
~\cite{Shindou-murakami09} and nonperturbative
numerical studies in the symmetry class AII.%
~\cite{Kobayashi14}
On the other hand, had we started from the symmetry class DIII and
weakly broken the particle-hole and chiral symmetries by
the addition of weak perturbations that bring the Dirac
Hamiltonian to one in the symmetry class AII, we may
then keep the control parameter $\mathsf{m}$ of the symmetry class
DIII along the horizontal axis in parameter space.
This would give the same phase diagram as in
Fig.~\ref{Fig: 3D phase diagrams}(a) for $N>1$ with the
caveat that the $N+1$ unperturbed localized phases with
the $\mathbb{Z}$ topological numbers become $N + 1$ localized phases
with alternating $\mathbb{Z}_2$ topological numbers
after perturbation.

The first and second homotopy groups of the normalized
Dirac masses from the 3D symmetry classes AII and CII differ.
The second homotopy group is trivial in the 3D symmetry classes AII
according to Table \ref{table: 3D localization}.
Hence the density of states is not singular in the 3D symmetry classes AII.
The second homotopy group is $\mathbb{Z}$ in the 3D symmetry classes CII
according to Table \ref{table: 3D localization}.
Hence the density of states is singular in the 3D symmetry classes CII.
The first homotopy groups of the 3D symmetry classes AII and CII are
$\mathbb{Z}^{\,}_{2}$ and trivial, respectively.
In the former case, the density of states receives 
a contribution from midgap states bound to the line defects.
This contribution need not be singular at $\varepsilon=0$.

\subsection{$\pi^{\,}_{0}(V^{\,}_{d=3,r})=\{0\}$}

The normalized Dirac masses are path-connected compact topological
spaces in the 3D symmetry classes A, AI, BDI, D, and C according to
Table \ref{table: 3D localization}.  These symmetry classes support a
single insulating phase, see Fig.~\ref{Fig: 3D phase diagrams}(c).  
Of these, only the 3D symmetry class AI has
the trivial group for its three homotopy groups. The normalized Dirac
masses in the 3D symmetry classes BDI and D have the second homotopy
groups $\mathbb{Z}$ and $\mathbb{Z}^{\,}_{2}$ for $N$ sufficiently
large, respectively. Thus, the 3D symmetry classes BDI and D have
densities of states that are singular at the band center because of
the proliferation of midgap states bound to $\mathbb{Z}$ and
$\mathbb{Z}^{\,}_{2}$ point-like defects in the normalized Dirac
masses, respectively.  Line defects of the normalized Dirac masses are
also allowed in the 3D symmetry classes A, D, and C.  They also
contribute to the density of states, but they need not give rise
to a singularity at $\varepsilon=0$.  We note that the density of
states computed for a network model from the symmetry class C was
found to be nonsingular in Ref.~\onlinecite{Ortuno09}.

\section{Boundaries of topological insulators}
\label{sec: Boundaries of topological insulators}

So far, we have been concerned with the interplay between
topology, local symmetries, and Anderson localization 
for $d$-dimensional random massive Dirac Hamiltonians 
that capture the effects
of local, and smooth disorder of $d$-dimensional lattice models
such as quantum network models at low energies and long wavelengths.
We have explained in Sec.%
~\ref{subsec: The tenfold way for the classifying spaces V}
how Table%
~\ref{table: AZ classes}
can be used to identify in each dimension $d$ of space the five
AZ symmetry classes that support topologically distinct 
insulating phases of $d$-dimensional quantum matter.
There is an alternative road to 
the classification of TIs and TSCs that was explained in Sec.%
~\ref{subsec: The tenfold way for the classifying spaces V}, 
which we now review for the sake of completeness.

Any $(d-1)$-dimensional boundary of a $d$-dimensional
topological insulators or superconductors is immune to Anderson localization.%
~\cite{schnyder-ryu08}
This immunity is captured by the presence
of a topological term in the NLSM that captures the physics of 
Anderson localization on the $(d-1)$-dimensional boundary
subject to the symmetry constraint imposed by one of the ten
AZ symmetry classes.%
~\cite{schnyder-ryu08} 
Alternatively, this property holds for
those AZ symmetry constraints that prohibit the existence of 
any Dirac mass matrix of rank
$r=\widetilde{r}^{\,}_{\mathrm{min}}<r^{\,}_{\mathrm{min}}$
entering the random Dirac Hamiltonian that
encodes single-particle transport 
at low energies and long wavelength on any
$(d-1)$-dimensional boundary.
In Appendix~\ref{appendixsubsec: Existence condition of ...},
we formulate the conditions for the presence
(or absence) of Dirac mass matrices. 
We show that the conditions for the absence of normalized Dirac masses
in $(d-1)$-dimensions are equivalent to the conditions for
a nontrivial zeroth homotopy group of the topological spaces associated
with the normalized Dirac masses in $d$-dimensions.
This alternative derivation of the tenfold way for TI/TSCs 
is captured by Table~\ref{table: surface localization}
for 2D and 3D spaces.

\begin{table}[tb]
\begin{center}
\caption{\label{table: surface localization}
Immunity to Anderson localization in five out of the ten
AZ symmetry classes along any boundary
of a topological insulator (TI) or a 
topological superconductors (TSC). 
This immunity is a consequence of the absence of any Dirac mass matrix 
with the rank
$r=\widetilde{r}^{\,}_{\mathrm{min}}<r^{\,}_{\mathrm{min}}$
(see Appendix \ref{appendixsubsec: Existence condition of ...})
for a random Dirac Hamiltonian capturing the low-energy and 
long-wavelength effects of local
disorder on the boundary in the corresponding AZ symmetry class.
        }
\begin{tabular}[t]{ccc}
\hline \hline
class &  ~edge of 2D TI/TSC~  & ~surface of 3D TI/TSC~  \\
\hline
A    
& ~ballistic~ & -
\\
~ AIII ~   
&  - & ~metallic~
\\
\hline
AI   
& - & -
\\
BDI
& - &  -
\\
D
& ballistic & -
\\
DIII 
& ballistic & metallic
\\
AII
& ballistic & metallic
\\
CII
& - & metallic
\\
C
& ballistic & -
\\
CI
& - & metallic
\\
\hline \hline
\end{tabular}

\end{center}
\end{table}

\section{Surface stability of topological insulators with reflection symmetry}
\label{sec: Surface stability of reflection topological insulators}

\subsection{General discussions}
\label{subsec: STCI general discussions}

This section is devoted to the stability 
of topological crystalline insulators (TCIs) 
with a reflection symmetry
when perturbed by a local random potential that retains
the reflection symmetry on average. 
The disorder is assumed weak in the bulk 
in that its characteristic strength is
much smaller than the band gap from the $d$-dimensional bulk. 
In this limit, the $(d-1)$-dimensional boundary may be treated in isolation
and represented in terms of massless Dirac fermions 
perturbed by local random potentials obeying 
appropriate symmetries. The material SnTe  is an example of a TCI
and the alloying with Pb in Sn${}^{\,}_{1-x}$Pb${}^{\,}_{x\vphantom{1}}$Te
can be modeled by a weak local random potential.

By definition, a $d$-dimensional TCI is a band insulator that
supports on any one of its $(d-1)$-dimensional boundaries 
that is invariant under the crystalline symmetry, 
here a reflection symmetry, boundary states that
disperse across the bulk gap with a linear (Dirac)
dispersion relation in the close vicinity to their band crossing.%
~\cite{fu-tci11}
In other words, the reflection symmetry protects 
$(d-1)$-dimensional boundary massless Dirac cones.

\begin{table*}[tb]
\begin{center}
\caption{\label{table: reflection TCI 2D}
Topological classification of crystalline insulators (TCIs)
in 2D with reflection symmetry and
their edge stability when the reflection symmetry is retained on 
average, a situation for which we use the acronym STCI
(statistical topological crystalline insulator).
For each AZ symmetry class, we give in the second column the
topological classification of insulators (TIs).
Columns three to six are the TCIs for the
symmetry classes BDI, DIII, CII, and CI
with the reflection symmetry $R^{\eta^{\,}_{T},\eta^{\,}_{C}}$, where the
superscripts $\eta^{\,}_{T}=\pm$ and $\eta^{\,}_{C}=\pm$
specify if the operation for reflection
commutes ($+$ sign) or anticommutes ($-$ sign) 
with reversal of time $T$ and exchange of particles and holes $C$,
respectively. Here, we choose the convention 
$(R^{\eta^{\,}_{T},\eta^{\,}_{C}})^{2}=+1$.
The third and fourth columns for the symmetry
class AIII/(AI, AII)/(D, C) are the topological
classification of the TCIs with the reflection symmetry $R^{\eta,\eta}$,
where the superscript $\eta$ specifies if the operation for reflection
commutes ($+$) or anticommutes ($-$) with chiral/time-reversal/particle-hole
transformation, respectively.
The quotation marks for the entry ``$\mathbb{Z}^{\,}_{2}$''
corresponding to a
TCI with the $R^{-,-}$ symmetry in class CII indicates instability
to interavalley scattering.%
~\protect{\cite{Chiu13,morimoto-clifford13}}
The last column gives the STCIs. The superscript
$\dag$ over the trivial zeroth homotopy groups in the STCI column
signals when the edge states of a TCI are unstable to the presence of 
disorder that respects the reflection symmetry on average.
        }
\begin{tabular}[t]{ccccccc}
\hline \hline
\multicolumn{1}{c}{AZ symmetry class~} 
&
\multicolumn{1}{c}{~2D TI~} 
&
\multicolumn{4}{c}{~2D TCI~} 
& 
\multicolumn{1}{c}{~2D STCI} \\
& $-$ & $~R^{+,+}~$ & $~R^{-,-}~$ & $~R^{+,-}~$ & $~R^{-,+}~$ &   \\
\hline
A    
& $\mathbb{Z}$ & 0 & $-$ & $-$ & $-$ & 0 
\\
~ AIII ~   
& 0 & $\mathbb{Z}$ & 0 & $-$ & $-$ & $\mathbb{Z}$ 
\\
\hline
AI
& 0 & 0 & 0 & $-$ & $-$ & 0 
\\
BDI
& 0 & $\mathbb{Z}$ &  0 & 0 & 0 & $\mathbb{Z}$ 
\\
D  
& $\mathbb{Z}$ & $\mathbb{Z}^{\,}_{2}$ & 0 & $-$ & $-$ & $\mathbb{Z}^{\,}_{2}$ 
\\
DIII
& $\mathbb{Z}^{\,}_{2}$ & $\mathbb{Z}^{\,}_{2}$ & $\mathbb{Z}$ & 0 & $\mathbb{Z}^{\,}_{2}$ & $\mathbb{Z}^{\,}_{2}$ 
\\
AII
& $\mathbb{Z}^{\,}_{2}$ & 0 & $\mathbb{Z}^{\,}_{2}$ & $-$ & $-$ & 0$^{\dag}$
\\
CII 
& 0 & $\mathbb{Z}$ & ``$\mathbb{Z}^{\,}_{2}$'' & 0 & 0 & $\mathbb{Z}$ 
\\
C
& $\mathbb{Z}$ & 0 & 0 & $-$ & $-$ & 0 
\\
CI
& 0 & 0 & $\mathbb{Z}$ & 0 & 0 & 0$^{\dag}$ 
\\
\hline \hline
\end{tabular}
\end{center}
\end{table*}

\begin{table*}[tb]
\begin{center}
\caption{\label{table: reflection TCI 3D}
Topological classification of crystalline insulators (TCIs)
in 3D with reflection symmetry and
their surface stability when the reflection symmetry is retained on 
the average, a situation for which we use the acronym STCI
(statistical topological crystalline insulator).
For each AZ symmetry class, we give in the second column the
topological classification of insulators (TIs).
Columns three to six are the TCIs for
the symmetry classes BDI, DIII, CII, and CI
with the reflection symmetry $R^{\eta^{\,}_{T},\eta^{\,}_{C}}$, where the
superscripts $\eta^{\,}_{T}=\pm$ and $\eta^{\,}_{C}=\pm$
specify if the operation for reflection
commutes ($+$ sign) or anticommutes ($-$ sign) 
with reversal of time $T$ and exchange of particles and holes $C$,
respectively. 
Here, we choose the convention $(R^{\eta^{\,}_{T},\eta^{\,}_{C}})^{2}=+1$.
The third column for symmetry class A indicates 
the $\mathbb{Z}$ classification
of TCIs with reflection symmetry.
The third and fourth columns for the symmetry
class AIII/(AI, AII)/(D, C) are the topological
classification of the TCIs with the reflection symmetry $R^{\eta,\eta}$,
where the superscript $\eta$ specifies if the operation for reflection
commutes ($+$) or anticommutes ($-$) with the 
chiral/time-reversal/particle-hole transformation, respectively.
The quotation marks for the entry ``$\mathbb{Z}^{\,}_{2}$''
corresponding to a
TCI with the $R^{-,-}$ symmetry in class C indicates instability
to intervalley scattering.%
~\protect{\cite{Chiu13,morimoto-clifford13}}
The last column gives the STCIs. The superscript
$\dag$ over the trivial zeroth homotopy groups in the STCI column
signals when the surface states of a TCI are unstable to the presence of 
disorder that respects the reflection symmetry on the average.    
\\
}
\begin{tabular}[t]{ccccccc}
\hline\hline
\multirow{2}{*}{AZ symmetry class~} 
&
\multicolumn{1}{c}{~3D TI~} 
&
\multicolumn{4}{c}{~3D TCI~} 
& 
\multicolumn{1}{c}{~3D STCI} 
\\
& $-$ &~$R^{+,+}$~&~$R^{-,-}$~&~$R^{+,-}$~&~$R^{-,+}$~&   
\\
\hline
A   
& 0 & $\mathbb{Z}$ & $-$ & $-$ & $-$ & $\mathbb{Z}$ 
\\
~ AIII ~    
& $\mathbb{Z}$ & 0 & $\mathbb{Z}$ & $-$ & $-$ & $0^\dagger$ 
\\
\hline
AI
& 0 & 0 & $\mathbb{Z}$ & $-$ & $-$ & 0$^{\dag}$ 
\\
BDI
& 0 & 0 & 0 & 0 & $\mathbb{Z}$ & 0$^{\dag}$ 
\\
D
& 0 & $\mathbb{Z}$ &  0 & $-$ & $-$ & $\mathbb{Z}$ 
\\
DIII  
& $\mathbb{Z}$ & $\mathbb{Z}^{\,}_{2}$ & 0 & $\mathbb{Z}$ & $\mathbb{Z}$ & $\mathbb{Z}^{\,}_{2}$ 
\\
AII
& $\mathbb{Z}^{\,}_{2}$ & $\mathbb{Z}^{\,}_{2}$ & $\mathbb{Z}$ & $-$ & $-$ & $\mathbb{Z}^{\,}_{2}$ 
\\
CII
& $\mathbb{Z}^{\,}_{2}$ & 0 & $\mathbb{Z}^{\,}_{2}$ & $\mathbb{Z}^{\,}_{2}$ & $\mathbb{Z}$ & 0$^{\dag}$
\\
C
& 0 & $\mathbb{Z}$ & ``$\mathbb{Z}^{\,}_{2}$'' & $-$ & $-$ & $\mathbb{Z}$ 
\\
CI
& $\mathbb{Z}$ & 0 & 0 & $\mathbb{Z}$ & $\mathbb{Z}$ & 0$^{\dag}$
\\
\hline \hline
\end{tabular}
\end{center}
\end{table*}

A classification of $d$-dimensional 
TCI with reflection symmetry has been obtained
in Ref.~\onlinecite{morimoto-clifford13} (see also Ref.~\onlinecite{Chiu13})
from studying the zeroth homotopy groups of 
the topological spaces associated with the normalized
Dirac masses entering $d$-dimensional massive Dirac Hamiltonians
obeying nonlocal reflection symmetries in addition to the AZ local symmetries.
For each of the AZ symmetry classes ordered as in the first columns 
of Tables \ref{table: reflection TCI 2D} and \ref{table: reflection TCI 3D},
the results from Ref.~\onlinecite{morimoto-clifford13}
are summarized in columns three to six when space is two- and 
three-dimensional, respectively. For completeness, the second
columns of Tables \ref{table: reflection TCI 2D} and
\ref{table: reflection TCI 3D}
are the same as in Table~\ref{table: AZ classes}
when $d=2$ and $d=3$, respectively.
Hence, for the AZ symmetry classes BDI, DIII, CII, and CI, 
the zeroth homotopy group of the topological space associated with
the normalized Dirac masses when reflection symmetry is not imposed 
(column two)
is supplemented by the zeroth homotopy group of 
the topological space associated with the normalized Dirac masses
when the reflection symmetry $R^{\eta^{\,}_{T},\eta^{\,}_{C}}$ holds
(columns three to six).
We choose the convention $(R^{\eta^{\,}_{T},\eta^{\,}_{C}})^{2}=+1$.
Here, any one of the superscripts $\eta^{\,}_{T}=\pm$ and $\eta^{\,}_{C}=\pm$ 
takes the value $+$ ($-$) if the operation of reflection commutes
(anticommutes) with the operation $T$ for the reversal of time 
and the operation $C$ for the exchange of particles and holes, respectively.
For the AZ symmetry classes AIII/(AI, AII)/(D,C), 
the third and fourth columns indicate the topological classification with
the reflection symmetry $R^{\eta,\eta}$,
where the superscript $\eta$ takes the value $+$ ($-$) if the reflection
operator commutes (anticommutes) with the operation for
chiral/time-reversal/particle-hole transformation, respectively.
For the symmetry class A, the third column shows the classification
in the presence of a reflection symmetry.
Any entry $``-''$ indicates the absence of the symmetries under 
reversal of time $T$ or under exchange of particles and holes $C$.

The question we want to address is what is the fate of 
those boundary massless Dirac cones protected by a combination of
AZ and reflection symmetries
if a boundary is perturbed by a local random potential 
belonging to a statistical ensemble such that 
(i) there exist realizations of the random potential
that break the reflection symmetry, 
(ii) even though the reflection symmetry holds on average.
The acronym STCI for statistical topological crystalline insulator
is used when items (i) and (ii) are met.

The answer is found in the last columns
of Tables \ref{table: reflection TCI 2D} and \ref{table: reflection TCI 3D}
in 2D and 3D space, respectively.
Each entry of the column STCI from the
Tables \ref{table: reflection TCI 2D} and \ref{table: reflection TCI 3D}
gives the zeroth homotopy group of the topological space associated with
the normalized Dirac masses in $(d-1)$-dimensional space
that are allowed by imposing no other symmetries than
the ones from the AZ symmetry class 
to which the $d$-dimensional STCI belongs. 
The intuition for this result is that a $d$-dimensional
STCI supports either a metallic or critical phase
on any of its $(d-1)$-dimensional boundary that is invariant 
under the crystalline operation if Anderson localization
in the corresponding $(d-1)$-dimensional AZ symmetry class is preempted
for topological reasons.

The interpretation of the non-vanishing zeroth homotopy groups in 
the columns three to six from
Tables \ref{table: reflection TCI 2D} and \ref{table: reflection TCI 3D}
is the following. Whenever the representation of the $d$-dimensional
TCI in terms of a Dirac Hamiltonian with
Dirac matrices of rank $r$ is the minimal one,
\begin{subequations}
\label{eq: def TCI with r=rmin}
\begin{equation}
r=r^{\,}_{\mathrm{min}},
\end{equation} 
i.e., no mass matrix is compatible with the symmetries
for any smaller rank of the Dirac matrices
(the actual value of $r^{\,}_{\min}$ depends on the AZ class
and the type of the $R$ symmetry), then there exists
a normalized mass matrix $\beta$ with the following two properties.
First, $\beta$ is unique up to a sign.
Second, $\beta$ is invariant under
the reflection represented by $R$,
\begin{equation}
R\,\beta\,R^{-1}=\beta.
\label{eq: R commutes with beta}
\end{equation}
\end{subequations}
Two Dirac Hamiltonians that differ through the sign 
$\pm\beta$ with which $\beta$ enters select two topologically 
distinct crystalline insulating phases in $d$-dimensional space.
The topologically nontrivial TCI phase is the one that supports a
massless Dirac cone on any one of the $(d-1)$-dimensional boundaries
that are invariant under the reflection symmetry. 
Such a massless Dirac cone is derived from the Dirac Hamiltonian 
$\widetilde{\bm{\alpha}}\cdot\widetilde{\bm{p}}$ with the Dirac matrices
$\widetilde{\bm{\alpha}}\equiv
(\widetilde{\alpha}^{\,}_{1},\ldots,\widetilde{\alpha}^{\,}_{d-1})$
of rank $r^{\,}_{\mathrm{min}}/2$ and the 
boundary momentum $\widetilde{\bm{p}}\in\mathbb{R}^{d-1}$.

Assume that a topologically nontrivial phase of a $d$-dimensional 
TCI is selected by the sign with which the mass matrix $\beta$
enters a Dirac Hamiltonian of minimal rank.
Furthermore, assume that,
without the protection arising from the reflection symmetry,
this TCI is a topologically trivial band insulator.
This implies that a normalized mass matrix $\beta'$ must exist such that
$\beta'$ anticommutes with $\beta$,
\begin{subequations}
\label{eq: properties beta'}
\begin{equation}
\{\beta,\beta'\}=0,
\label{eq: properties beta' a}
\end{equation} 
and $\beta'$ is odd under the transformation
\begin{equation}
R\,\beta'\,R^{-1}=-\beta'
\label{eq: properties beta' b}
\end{equation}
\end{subequations}
by the operation $R$.

Consider a $d$-dimensional Dirac Hamiltonian
with the pair of mass terms 
$m\,\beta$
and $m'\,\beta'$
that satisfies
Eqs.~(\ref{eq: def TCI with r=rmin}) 
and (\ref{eq: properties beta'}).
Assume that the breaking of the reflection symmetry by
$m'\,\beta'$ is small, 
\begin{equation}
|m'|\ll|m|.
\label{eq: def m' weak wrt m}
\end{equation}
If the sign of $m$ selects a topologically nontrivial crystalline insulating 
phase in the limit $m'=0$, then any $(d-1)$-dimensional boundary left 
invariant under the reflection symmetry must
support a massive Dirac cone with the gap proportional to $|m'|$.
Such a massive Dirac cone is derived from the Dirac Hamiltonian 
$\widetilde{\bm{\alpha}}\cdot\widetilde{\bm{p}}
+\widetilde{m}'\,\widetilde{\beta}'$ 
with the Dirac matrices
$\widetilde{\bm{\alpha}}\equiv
(\widetilde{\alpha}^{\,}_{1},\ldots,\widetilde{\alpha}^{\,}_{d-1})$
and $\widetilde{\beta}'$
of rank $r^{\,}_{\mathrm{min}}/2$ and the 
boundary momentum $\widetilde{\bm{p}}\in\mathbb{R}^{d-1}$.
The Dirac mass matrix $\widetilde{\beta}'$ at the boundary 
originates from the Dirac mass matrix $\beta'$ in the bulk.

Assume that $\beta'$ is multiplied by a random function
\begin{subequations}
\label{eq: def weak random m'}
\begin{equation}
m':\mathbb{R}^{d}\to\mathbb{R},\bm{x}\mapsto m'(\bm{x})
\end{equation}
that satisfies the bound 
\begin{equation}
\sup_{\bm{x}\in\mathbb{R}^{d}}|m'(\bm{x})|<|m|,
\end{equation}
\end{subequations}
where $m$ selects a topologically nontrivial crystalline insulating 
phase in the limit for which the reflection symmetry is recovered.
The fate of the boundary states that are protected by the
reflection symmetry in the presence of the weak 
reflection-symmetry-breaking perturbation
(\ref{eq: def weak random m'})
is the same as the fate of massless Dirac fermions in $(d-1)$ dimensions
perturbed by a generic local random potential with vanishing mean
[by item (ii)]
that obeys no other symmetries than the ones from the
AZ symmetry class to which the $d$-dimensional TCI belongs. 
The latter stability analysis can be done with
the methods from Secs.%
~\ref{sec: Classifying spaces of normalized Dirac masses}%
--\ref{sec: Boundaries of topological insulators}.
When the classifying space of the normalized Dirac masses 
in $(d-1)$-dimensions has nontrivial zeroth homotopy group 
(listed in the last columns of 
Tables \ref{table: reflection TCI 2D} and \ref{table: reflection TCI 3D}),
it has two path-connected compact subspaces indexed
by the ambiguity to assign a sign to the unique normalized Dirac mass matrix 
$\widetilde{\beta}'$ of the minimal rank 
(half the minimal rank of the bulk TCI, i.e., $r^{\,}_{\mathrm{min}}/2$). 
The anticommutation relation (\ref{eq: properties beta' b}) 
implies that this pair of path-connected compact subspaces are homeomorphic
to each other by the reflection transformation $R$.

Assume now that we are given a statistical ensemble
of random Dirac Hamiltonians of rank $r=r^{\,}_{\mathrm{min}}$,
each realization of which would be the same $d$-dimensional TCI,
perturbed by reflection-symmetry-breaking random potentials,
with the reflection symmetry $R$ obeying 
Eq.~(\ref{eq: R commutes with beta}).
We may then decompose any $(d-1)$-dimensional 
boundary that is invariant under the reflection symmetry
into patches labeled by the sign of 
the Dirac mass $\widetilde{m}'(\widetilde{\bm{x}})$ multiplying
the $\widetilde{\beta}^{\prime}$ matrix that smoothly varies
along this boundary, say A if the sign is $+$
and B if the sign is $-$, as was done in Fig.~\ref{Fig: domains}.
The relative size of all the regions labeled A compared to all
the regions labeled B in Fig.~\ref{Fig: domains}
is fixed by the mean value of the mass term
with the mass matrix $\widetilde{\beta}^{\prime}$.
This difference in sizes vanishes
if the reflection symmetry holds on average.
The existence of
a $(d-2$)-dimensional submanifold between regions A and regions B that 
(i) percolates across the $(d-1)$-dimensional boundary and 
(ii) binds a midgap state signals a metallic or critical phase of
$(d-1)$-dimensional boundary.

The argument of the previous paragraph breaks down if we
relax the assumption $r=r^{\,}_{\mathrm{min}}$
to $r=r^{\,}_{\mathrm{min}}\, N$ with $N=2,3,\ldots$.
Indeed, when the rank of the Dirac matrices is the integer
multiple $N=2,3,\ldots$ of the minimum rank 
$r^{\,}_{\mathrm{min}}$,
the operation for reflection does not need to interchange
regions A and B from Fig.~\ref{Fig: domains}
anymore. Imposing the reflection symmetry on average
is not sufficient anymore to ensure that the difference in the sizes
of all regions labeled A and all regions labeled B
in Fig.~\ref{Fig: domains} 
vanishes. Stronger assumptions obeyed by the statistical
ensemble of random masses must be fulfilled to guarantee the existence
of a $(d-2)$-dimensional path between all regions A and B
that percolates across the $(d-1)$-dimensional boundary
depicted in Fig.~\ref{Fig: domains} and binds a midgap state.
To appreciate this point, we consider the following two cases.

Consider first the case of the zeroth homotopy group
$\mathbb{Z}^{\,}_{2}$ and $N=2$.
We can then make the choice
(\ref{eq: case pi0 when z2})
with $M$ a $4\times4$ matrix.
The reflection transformation $R$ still anticommutes with
the mass matrix~(\ref{eq: case pi0 when z2 b}),
as assumed in Eq.\ (\ref{eq: properties beta' b}).
However, it leaves the Pfaffian
(\ref{eq: Z2 index})
used to index the topological sector of the mass term unchanged.
Hence $R$ cannot be used to construct a homeomorphism between
the subspaces belonging to distinct topological sectors when
$N=2$. Imposing that the randomness preserves the reflection symmetry
on average is not sufficient to enforce a metallic or
critical phase of $(d-1)$-dimensional boundary.

We can generalize this argument to arbitrary $N$.
For odd $N$, preserving the reflection symmetry on average leads to
a vanishing value for the disorder average
of the $\mathbb{Z}^{\,}_{2}$ topological number 
(\ref{eq: Z2 index}).
Either a metallic or critical phase follows at the boundary of a STCI.
For even $N$, preserving the reflection symmetry on average
does not constrain the disorder average of the topological number
(\ref{eq: Z2 index}).
We thus expect an even-odd effect
(a sensitivity to the parity of $N$) with regard to
the stability of surface states of a STCI 
when the zeroth homotopy group is $\mathbb{Z}^{\,}_{2}$.

In the case of $\mathbb{Z}$ for the zeroth homotopy group of 
the topological space associated with
the normalized Dirac masses in $(d-1)$ dimensions,
there exists a unique normalized Dirac mass matrix, up to a sign, 
that commutes with all other permissible mass matrices.
This Dirac matrix corresponds to 
$\beta^{\,}_{0}$ in Eq.\ (\ref{alternative definition 0 a}).
The disorder average value of this mass term
vanishes (i.e., $\mathsf{m}=0$)
due to statistical (average) reflection symmetry of the STCI.
According to Tables \ref{table: 1D localization} and 
\ref{table: 2D localization},
we have the following even-odd effect.
The $(d-1)$-dimensional boundary of a STCI
is metallic or critical for odd $N$, 
while it is localized for even $N$,
as depicted in Figs.\ 3(a) and 4(a)-4(d).

Next, we consider two examples of a three-dimensional TCI.

\subsection{Example in the 3D symmetry class AII}

First, we consider a 3D TI in the symmetry class AII.
We then impose a crystalline symmetry by demanding
invariance under the reflection
\begin{subequations}
\begin{align}
&
\bm{x}\equiv(x,y,z)\mapsto
(-x,y,z)\\
\label{eq: def inversion about 0 of x}
&
\bm{s}\equiv(s^{\,}_{x},s^{\,}_{y},s^{\,}_{z})
\mapsto(s^{\,}_{x},-s^{\,}_{y},-s^{\,}_{z}),
\end{align}
\end{subequations}
for the space and spin degrees of freedom, respectively.

To this end, we choose the $4\times4$ massive Dirac Hamiltonian
\begin{equation}
\begin{split}
\mathcal{H}(\bm{k}):=&\,
\left(
k^{\,}_{x}\, 
s^{\,}_{y}
-
k^{\,}_{y}\,
s^{\,}_{x}
\right)
\otimes
\sigma^{\,}_{x}
+
k^{\,}_{z}\,
s^{\,}_{0}
\otimes
\sigma^{\,}_{y}
+
m\, 
s^{\,}_{0}
\otimes
\sigma^{\,}_{z}.
\label{eq: def snte}
\end{split}
\end{equation}
The $2\times2$ unit matrix $s^{\,}_{0}$ and the Pauli matrices
$\bm{s}$ act on the spin-1/2 degrees of freedom. They enter
the kinetic energy through a spin-orbit coupling.
The $2\times2$ unit matrix $\sigma^{\,}_{0}$ and the Pauli matrices
$\bm{\sigma}$ act on the orbital degrees of freedom.
Reversal of time is represented by conjugation of
$\mathcal{H}(\bm{k})$
with 
\begin{equation}
T:=\mathrm{i}s^{\,}_{y}\otimes\sigma^{\,}_{0}\,\mathsf{K},
\end{equation} 
where complex conjugation is denoted by $\mathsf{K}$. 
The massive Dirac Hamiltonian (\ref{eq: def snte})
realizes a 3D insulator in the symmetry class AII
with the Dirac matrices
\begin{equation}
\bm{\alpha}\equiv
\left(
 s^{\,}_{y}\otimes\sigma^{\,}_{x},
-s^{\,}_{x}\otimes\sigma^{\,}_{x},
s^{\,}_{0}\otimes\sigma^{\,}_{y}
\right),
\quad
\beta\equiv
s^{\,}_{0}\otimes\sigma^{\,}_{z},
\end{equation}
as
\begin{equation}
T\,\mathcal{H}(-\bm{k})\,T^{-1}=\mathcal{H}(\bm{k}).
\end{equation}
The rank $r=4$ of the Dirac matrices is the minimal one
for 3D topological insulators from the symmetry class AII
represented by single-particle massive Dirac Hamiltonians,
\begin{equation}
r^{\,}_{\mathrm{min}}=4.
\end{equation}

According to Ref.~\onlinecite{hsieh12},
the 3D Dirac Hamiltonian~(\ref{eq: def snte})
of rank $r=r^{\,}_{\mathrm{min}}=4$
captures the low-energy and long-wavelength limit 
in the vicinity of \textit{one out of four inequivalent} 
L points from the face-centered-cubic (fcc) Brillouin zone
of the band insulator SnTe for one sign of $m$
or the band insulator PbTe for the opposite sign of $m$.
The rank $r=r^{\,}_{\mathrm{min}}=4$ arises because 
(1) both SnTe and PbTe
have a simple rocksalt structure
(two atoms in the unit cell),
(2) both display a spin-orbit coupling,
and
(3)
because the splitting in energy between the conduction and 
valence bands is minimum at the center
of the eight hexagonal faces of the fcc Brillouin zone,
the so-called L points.
SnTe and PbTe belong to the same trivial insulating phase
in the entry AII from the second column of
Table \ref{table: reflection TCI 3D},
because this phase supports an even number of gapless surface states.
However, the plane 
\begin{equation}
(k^{\,}_{x},k^{\,}_{y},k^{\,}_{z})=
(0,k^{\,}_{y},k^{\,}_{z})
\end{equation}
from the face centered cubic Brillouin zone
is invariant under the reflection 
(\ref{eq: def inversion about 0 of x}).
On this so-called mirror plane, the massive 3D Dirac Hamiltonian
(\ref{eq: def snte})
simplifies to the direct sum of two massive
2D Dirac Hamiltonians of rank $r=2$ from the 
2D symmetry class D whose relative Chern numbers differ by one.

The topological space of the normalized Dirac masses in
the 3D symmetry class AII follows here from the extension problem
\begin{subequations}
\label{eq: extension problem for 4 by 4 example in AII}
\begin{equation}
\begin{split}
Cl^{\,}_{2,3}\equiv&
\left\{
T,TJ;\alpha^{\,}_{x},\alpha^{\,}_{y},\alpha^{\,}_{z}
\right\}
\\
&
\to
Cl^{\,}_{3,3}\equiv
\left\{
J\beta,T,TJ;\alpha^{\,}_{x},\alpha^{\,}_{y},\alpha^{\,}_{z}
\right\}
\end{split}
\label{eq: extension problem for 4 by 4 example in AII a}
\end{equation} 
with the classifying space
\begin{equation}
R^{\,}_{1}= O(N)
\label{eq: extension problem for 4 by 4 example in AII b}
\end{equation}
[see Eq.~(\ref{eq: extension problem with solution R p-q+2})
and Table \ref{table: all homotopies classifying spaces}],
whose zeroth homotopy group is~\cite{3Footnote-on-exampleSnTe}
\begin{equation}
\pi^{\,}_{0}(R^{\,}_{1})=
\mathbb{Z}^{\,}_{2},
\label{eq: extension problem for 4 by 4 example in AII c}
\end{equation}
\end{subequations}
as the solution.
We recall
that the topological number is $\mathbb{Z}^{\,}_{2}$
rather than $\mathbb{Z}$ from the fact that the tensoring
\begin{subequations}
\label{eq: def 8 by 8 example in AII}
\begin{equation}
\mathcal{H}(\bm{k})\to
\mathcal{H}(\bm{k})
\otimes
\tau^{\,}_{0},
\label{eq: def 8 by 8 example in AII a}
\end{equation}
where $\tau^{\,}_{0}$ is a $2\times2$ unit matrix and $\bm{\tau}$
are Pauli matrices,
admits the normalized mass term 
\begin{equation}
\beta'\equiv
s^{\,}_{z}\otimes\sigma^{\,}_{x}\otimes\tau^{\,}_{y}
\label{eq: def 8 by 8 example in AII b}
\end{equation}
that anticommutes with
\begin{equation}
\beta\equiv
s^{\,}_{0}\otimes\sigma^{\,}_{z}\otimes\tau^{\,}_{0}
\label{eq: def 8 by 8 example in AII c}
\end{equation}
\end{subequations}
and is invariant under the transformation
$T=\mathrm{i}s^{\,}_{y}\otimes\sigma^{\,}_{0}\otimes\tau^{\,}_{0}\,\mathsf{K}$
that implements reversal of time.

We now impose the crystalline symmetry originating from the operation
(\ref{eq: def inversion about 0 of x})
on the 3D massive Dirac Hamiltonian 
(\ref{eq: def 8 by 8 example in AII a}). 
To this end, we define
\begin{subequations}
\label{eq: def R-x symmetry for 8 by 8} 
\begin{equation}
R^{-}_{x}:=
s^{\,}_{x}\otimes\sigma^{\,}_{0},
\quad
\left(R^{-}_{x}\right)^{2}=
s^{\,}_{0}\otimes\sigma^{\,}_{0},
\quad
\{
R^{-}_{x},
T
\}=0,
\label{eq: def R-x symmetry for 8 by 8 a}  
\end{equation}
and demand that
\begin{equation}
R^{-}_{x}\,
\mathcal{H}(-k^{\,}_{x},k^{\,}_{y},k^{\,}_{z})\,
R^{-}_{x}\,=
\mathcal{H}(k^{\,}_{x},k^{\,}_{y},k^{\,}_{z}).
\label{eq: def R-x symmetry for 8 by 8 b} 
\end{equation}
\end{subequations}
One observes that the mass matrix $\beta$ is compatible with
the crystalline symmetry originating from the operation
(\ref{eq: def inversion about 0 of x}).
However, the crystalline symmetry originating from the operation
(\ref{eq: def inversion about 0 of x})
is not compatible with the mass matrix $\beta'$ defined by
Eq.~(\ref{eq: def 8 by 8 example in AII b}) if 
\begin{subequations}
\begin{equation}
R^{-}_{x}:=
s^{\,}_{x}\otimes\sigma^{\,}_{0}\otimes\tau^{\,}_{0},
\qquad
T:=
\mathrm{i} s^{\,}_{y}\otimes
\sigma^{\,}_{0}\otimes
\tau^{\,}_{0}\,
\mathsf{K},
\end{equation}
is combined with the symmetry condition
\begin{equation}
R^{-}_{x}\,
\mathcal{H}(-k^{\,}_{x},k^{\,}_{y},k^{\,}_{z})\otimes\tau^{\,}_{0}\,
R^{-}_{x}\,=
\mathcal{H}(k^{\,}_{x},k^{\,}_{y},k^{\,}_{z})\otimes\tau^{\,}_{0}.
\end{equation}
\end{subequations}
This conclusion can be generalized in the following way.

If we impose the symmetry under the reflection
(\ref{eq: def inversion about 0 of x})
that anticommutes with reversal of time
to any single-particle 3D Dirac Hamiltonian belonging to the symmetry class
AII, we must supplement the Clifford algebra for class AII 
[recall Eq.~(\ref{eq: real Clifford algebra})]
with a new generator $J\,\alpha^{\,}_{x}\,R^{-}_{x}$ 
that squares to $+1$ times the identity operator. 
The topological space of the normalized Dirac masses in
the 3D symmetry class AII obeying this reflection symmetry
follows here from the extension problem
\begin{subequations}
\label{eq: extension problem AII with reflection x to -x}
\begin{equation}
\begin{split}
Cl^{\,}_{2,4}\equiv&
\left\{
T,TJ;\alpha^{\,}_{x},\alpha^{\,}_{y},\alpha^{\,}_{z},J\alpha^{\,}_{x}R^{-}_{x}
\right\}
\\
&
\to
Cl^{\,}_{3,4}\equiv
\left\{
J\beta,T,TJ;
\alpha^{\,}_{x},\alpha^{\,}_{y},\alpha^{\,}_{z},J\alpha^{\,}_{x}R^{-}_{x}
\right\}
\end{split}
\label{eq: extension problem AII with reflection x to -x a}
\end{equation} 
with the classifying space 
\begin{equation}
R^{\,}_{0}=
\lim_{N\to\infty}
\cup_{n=0}^{N}\big\{O(N)/\big[O(n)\times O(N-n)\big]\big\}
\label{eq: extension problem AII with reflection x to -x b}
\end{equation}
[see Eq.~(\ref{eq: extension problem with solution R p-q+2})
and Table \ref{table: all homotopies classifying spaces}],
whose zeroth homotopy group is
\begin{equation}
\pi^{\,}_{0}(R^{\,}_{0})=
\mathbb{Z},
\label{eq: extension problem AII with reflection x to -x c}
\end{equation}
\end{subequations}
as the solution.

The zeroth homotopy group $\mathbb{Z}$ for the 
normalized Dirac masses in the 3D symmetry class AII obeying
the crystalline symmetry under the operation 
(\ref{eq: def inversion about 0 of x})
is the reason why an even number of surface massless Dirac cones
represented by rank two Dirac matrices is protected,
when no such protection is operative if we relax this crystalline symmetry
and deal with the 3D symmetry class AII.
In other words, an insulator from the 3D symmetry class AII that would support
an even number of surface Dirac cones in the clean limit
is topologically trivial because a generic local random
potential from the symmetry class AII gaps
out any even number of surface  Dirac cones. 
Since the clean limit of SnTe can be approximated at low energies 
by the massive Dirac Hamiltonian (\ref{eq: def snte}) 
tensored with the $N\times N$ unit matrix
where $N=4$ is the even number of Dirac cones 
centered around the $N=4$ inequivalent L points 
in the fcc Brillouin zone,%
~\cite{4Footnote-on-L-points}
we conclude that SnTe is a topologically trivial insulator
under a generic single-particle local
perturbation belonging to the symmetry class AII,
while SnTe is a topologically nontrivial insulator
whose topologically distinct phases are indexed in $\mathbb{Z}$,
if the crystalline symmetry under the operation 
(\ref{eq: def inversion about 0 of x})
is imposed. Accordingly, the corresponding entries in 
Table~\ref{table: reflection TCI 3D}
are $\mathbb{Z}^{\,}_{2}$ and $\mathbb{Z}$ in the second and
fourth columns, respectively. It remains to explain the
entry $\mathbb{Z}^{\,}_{2}$ in the last column.

We now assume that the $8\times8$
Dirac Hamiltonian (\ref{eq: def 8 by 8 example in AII a})
is perturbed by a local random potential that 
(1) belongs to the symmetry class AII,
(2) may break the crystalline symmetry
(\ref{eq: def R-x symmetry for 8 by 8})
for any given realization, 
(3) but preserves the crystalline symmetry
(\ref{eq: def R-x symmetry for 8 by 8})
on the average. Hence the mass matrix
$\beta'$ defined in Eq.~(\ref{eq: def 8 by 8 example in AII b})
enters the random potential multiplied 
by a random function 
$m':\mathbb{R}^{3}\to\mathbb{R},\bm{x}\mapsto m'(\bm{x})$
that varies smoothly in space
but averages to zero everywhere in 3D space.
This mass matrix anticommutes with the mass matrix
(\ref{eq: def 8 by 8 example in AII c}).
Any surface of the TI that is invariant 
under the operation~(\ref{eq: def inversion about 0 of x})
can be divided, for any realization of the random function $m'$,
into the regions A  in which $\mathrm{sgn}\,m'(\bm{x})$
is $+$ and the regions B in which $\mathrm{sgn}\,m'(\bm{x})$
is $-$, as is done in Fig.~\ref{Fig: domains}.
Along the one-dimensional
boundary between two path-connected regions A and B, $m'(\bm{x})=0$.
Such an edge binds a pair of midgap helical edge states. 
The typical difference between the area of all the regions A and the area of
the regions B is fixed by the mean value of the random function $m'$.
This typical difference is zero if the mean value of the random function $m'$ 
vanishes everywhere on the surface.
If so, a pair of helical midgap states are bound to
a one-dimensional path that percolates across the surface.
Hence they are extended.

The parameter $x$ in the alloy Sn${}^{\,}_{1-x}$Pb${}^{\,}_{x}$Te 
is associated with a local random potential as the selection of the
sites on which Sn is to be substituted by Pb is essentially a random
Poisson process to a first approximation. 
In the clean limit $x\to0$, 
the projection of the massive 3D Dirac cones (\ref{eq: def snte})
at the four inequivalent L points
onto one of the mirror symmetric surface Brillouin zone
delivers an even number of 2D massless Dirac cones.
Specifically, four Dirac cones appear on the [001] surface ($N=2$),%
~\cite{4Footnote-on-L-points}
while six Dirac cones appear on the [111] surface ($N=3$).
The disorder in the alloy Sn${}^{\,}_{1-x}$Pb${}^{\,}_{x}$Te
belongs to the symmetry class AII.
It is also believed to preserve, on average, the reflection
symmetry present in the clean limit (\ref{eq: def snte}).
Thus, the alloy Sn${}^{\,}_{1-x}$Pb${}^{\,}_{x}$Te is an example of the STCIs.%
~\cite{hsieh12,Diez14}
According to the even-odd effect in the case of
the zeroth homotopy group $\mathbb{Z}^{\,}_{2}$
discussed in Sec.~\ref{subsec: STCI general discussions}, 
the gapless surface states of Sn${}^{\,}_{1-x}$Pb${}^{\,}_{x}$Te
are stable (extended) on the [111] surface,
but they are unstable (localized) on the [001] surface
for generic disorder that preserves the reflection symmetry on average.
Nevertheless, the four Dirac cones at the [001] surface can also be stable 
if the $C^{\,}_{4}$ rotation symmetry interchanging four Dirac cones is preserved
on average. Alternatively, if one of the two reflection symmetries, which
are present at 
the [001] surface, is broken by a lattice distortion,
only two out of the four Dirac cones remain massless.
This pair of surviving massless Dirac cones is immune
to Anderson localization in the presence of weak disorder
that preserves one of the two reflection symmetries
at the [001] surface on average.

\subsection{Example in the 3D symmetry class AI}

Second, we consider a 3D TI in the symmetry class AI
upon which we impose a crystalline symmetry under
\begin{equation}
\bm{x}\equiv(x,y,z)\mapsto
(-x,y,z)
\label{eq: def inversion about 0 of x for AI}
\end{equation}
that anticommutes with reversal of time.

To this end, we choose the $8\times8$ massive Dirac Hamiltonian
($r^{\,}_{\mathrm{min}}=8$ in the 3D symmetry class AI)
\begin{subequations}
\label{eq: def 8  by 8  for AI}
\begin{equation}
\mathcal{H}(\bm{k}):=
k^{\,}_{x}\, 
X^{\,}_{233}
+
k^{\,}_{y}\,
X^{\,}_{023}
+
k^{\,}_{z}\,
X^{\,}_{002}
+
\sum_{I}
m^{\,}_{I}\, 
X^{\,}_{I}.
\label{eq: def 8  by 8  for AI a}
\end{equation}
Here, 
\begin{equation}
X^{\,}_{\mu\nu\rho}:=
s^{\,}_{\mu}\otimes
\sigma^{\,}_{\nu}\otimes
\tau^{\,}_{\rho},
\qquad
\mu,\nu,\rho=0,1,2,3,
\label{eq: def 8  by 8  for AI b}
\end{equation}
where each quartet 
$(s^{\,}_{0},s^{\,}_{1},s^{\,}_{2},s^{\,}_{3})$,
$(\sigma^{\,}_{0},\sigma^{\,}_{1},\sigma^{\,}_{2},\sigma^{\,}_{3})$,
and
$(\tau^{\,}_{0},\tau^{\,}_{1},\tau^{\,}_{2},\tau^{\,}_{3})$
of $2\times 2$ Hermitian matrices
is built out of the $2\times2$ 
unit matrix together with the three Pauli matrices, 
and $I$ denotes the collective index taking the values
\begin{equation}
I=001,013,133,333.
\label{eq: def 8  by 8  for AI c}
\end{equation}
\end{subequations}

Represent reversal of time $T$ by complex conjugation
$\mathsf{K}$, 
\begin{equation}
T:=
X^{\,}_{000}\,\mathsf{K}.
\end{equation}
Represent the reflection (\ref{eq: def inversion about 0 of x for AI})
that anticommutes with $T$ by
\begin{equation}
R^{-}_{x}:=
X^{\,}_{020}.
\end{equation}
One verifies that
\begin{equation}
T\,\mathcal{H}(-\bm{k})\,T=
\mathcal{H}(\bm{k}).
\end{equation}
Furthermore, one verifies that imposing 
\begin{equation}
R^{-}_{x}\,\mathcal{H}(-k^{\,}_{x},k^{\,}_{y},k^{\,}_{z})\, R^{-}_{x}=
\mathcal{H}(k^{\,}_{x},k^{\,}_{y},k^{\,}_{z})
\label{eq: imposing R-x on 8 by 8 AI}
\end{equation}
demands of the four masses entering
the massive $8\times8$ Dirac Hamiltonian
(\ref{eq: def 8  by 8  for AI a}) that three of them vanish,
\begin{equation}
m^{\,}_{013}=m^{\,}_{133}=m^{\,}_{333}=0.
\end{equation}
Since only one normalized mass matrix is compatible
with the AZ symmetry class AI and the reflection symmetry
(\ref{eq: imposing R-x on 8 by 8 AI}) in 3D space,
it follows that the topological space of normalized Dirac masses
in the minimal representation has two compact and path-connected components
$\{\pm X^{\,}_{001}\}$. Consequently, a boundary that is left invariant by
the reflection symmetry (\ref{eq: def inversion about 0 of x for AI})
must host boundary states that are described by  a massless
$4\times4$ Dirac equation in the low-energy and long-wavelength limit.
Accordingly, we must find a non-vanishing entry in the corresponding entry
of the $R^{--}$ column of Table~\ref{table: reflection TCI 3D}. In order
to conclude that this entry is $\mathbb{Z}$, we must double
the rank of the Dirac Hamiltonian (\ref{eq: def 8  by 8  for AI a})
and verify that one mass commutes with all other allowed masses.
Alternatively, we may reach the same conclusion from the fact that,
without the reflection symmetry, 
the extension problem $Cl^{\,}_{0,5}\to Cl^{\,}_{1,5}$ has 
the classifying space $R^{\,}_{5}$ with $\pi^{\,}_{0}(R^{\,}_{5})=0$
for solution 
according to Eq.~(\ref{eq: extension problem with solution R p-q+2}),  
while, with the reflection symmetry,
the extension problem $Cl^{\,}_{0,6}\to Cl^{\,}_{1,6}$ has  
the classifying space $R^{\,}_{4}$ with $\pi^{\,}_{0}(R^{\,}_{4})=\mathbb{Z}$
for solution
according to Eq.~(\ref{eq: extension problem with solution R p-q+2}).
The fact that the reflection symmetry
(\ref{eq: imposing R-x on 8 by 8 AI}) in the 3D symmetry class AI
preempts more than one mass matrix is the reason for which
a 3D STCI in the symmetry class AI has no protected surface states
when the reflection symmetry holds only on average. 

\subsection{Relation to surface states of weak topological insulators}
\label{subsec: Relation to surface states of weak topological insulators}

Surface states of weak topological insulators (WTIs) are robust 
to local disorder that preserves the translation symmetry 
on average.%
~\cite{Ringel12,Mong12,fulga2012statistical,Morimoto-weak14}
This is understood in the same manner as with STCIs.
In a STCI, the average reflection symmetry 
tunes the surface states to be on the critical boundary
or in the metallic phase at $\mathsf{m}=0$ for odd $N$.
In a WTI, the average translation symmetry 
tunes the surface states to be on the critical boundary
or in the metallic phase at $\mathsf{m}=0$ for odd $N$,
as was shown
for $\mathbb{Z}^{\,}_{2}$ WTI of $N=1$
in Ref.~\onlinecite{Morimoto-weak14}.
When $N$ is an even integer, the average translation symmetry
is not sufficient for the surface states of a WTI to
evade Anderson localization, as in STCIs.

\section{Summary}
\label{sec: Summary}

The problem of Anderson localization for noninteracting fermions
realizing random Dirac Hamiltonians was revisited
from the perspective of the topologies of the spaces
associated with Dirac mass matrices allowed by symmetries.

It was first shown that the topological space $V$ of normalized 
Dirac masses can be characterized in a systematic fashion 
with the help of the homotopy groups
$\pi^{\,}_{0}(V)$,
$\pi^{\,}_{1}(V)$,
...,
$\pi^{\,}_{d-1}(V)$
by imposing on $V$ the Altland-Zirnbauer (AZ)
symmetries obeyed by all realizations of the local disorder.
(The AZ symmetries are local.)
For any spatial dimension $d$ and for any $n=0,\ldots,d-1$, one finds
three $d$-dimensional AZ symmetry classes
for which $\pi^{\,}_{n}(V)=\mathbb{Z}$
and two $d$-dimensional AZ symmetry classes for which
$\pi^{\,}_{n}(V)=\mathbb{Z}^{\,}_{2}$.
In particular, the zeroth homotopy group is nontrivial whenever 
the topological space $V$ is disconnected. 
The zeroth homotopy group $\pi^{\,}_{0}(V)$ indexes
the path-connected subspaces of $V$ and enumerates the topologically distinct
insulating phases in $d$ dimensions.
By considering a mapping of conducting channels, formed
along boundaries between topologically distinct insulating regions,
to a $d$-dimensional random quantum network model, we showed 
(i) which pairs of
topologically distinct insulating phases are separated by either a metallic
or a critical phase and
(ii) that the control parameter across this intervening
metallic phase or critical boundary involves either the disorder-averaged
value taken by the Dirac mass that commutes with all other Dirac masses
when $\pi^{\,}_{0}(V)=\mathbb{Z}$,
or the disorder-averaged value over the Pfaffian of the Dirac mass
when $\pi^{\,}_{0}(V)=\mathbb{Z}^{\,}_{2}$. 
Qualitative phase diagrams
in a two-dimensional parameter space spanned by this control parameter
and the characteristic disorder strength were deduced
for each $d$-dimensional AZ symmetry when the chemical potential is
at the band center 
by combining $\pi^{\,}_{0}(V)$ with a one-loop RG calculation.

We have also explained under what conditions
the self-averaging density of states per unit energy and 
per unit volume is singular in the vicinity of the band center
and when this singularity becomes universal. A sufficient condition 
for a singular density of states
is that the homotopy group $\pi^{\,}_{d-1}(V)$ is nontrivial
and point defects bounding zero-energy states may appear.
However, topology alone
is not sufficient to predict the nature of the singularity.

Finally, we have extended the study of the interplay between the AZ
symmetry classes, topology, and Anderson localization by allowing
crystalline (reflection) symmetries to hold on average,
a situation for which we reserve the acronym STCI
(statistical topological crystalline insulator).
Thereto, we find protected boundary states in five out of
ten AZ symmetry classes by demanding that it also
obeys a nonlocal reflection symmetry on average. 
We have shown that the stability against disorder of $(d-1)$-dimensional
boundary states of $d$-dimensional TCIs is understood from
the phase diagram of $(d-1)$-dimensional Dirac Hamiltonians with random mass.
We have also pointed out the existence of an even-odd dependence 
of the stability on the parameter $N$ that specifies the matrix dimension 
of Dirac Hamiltonians.
Finally, we have shown that the alloy Sn${}^{\,}_{1-x}$Pb${}^{\,}_{x}$Te 
is an example of a STCI in the 3D symmetry class AII
with an average reflection symmetry  
such that its surface states are protected by the
zeroth homotopy group $\pi^{\,}_{0}(V)=\mathbb{Z}^{\,}_{2}$ 
for surface normalized Dirac masses.

Before closing this section, we comment on two limitations of
our approach.
First, the random Dirac Hamiltonians that
we have studied are low-energy effective
models that should be derived from tight-binding models on regular lattices.
In the context of topological insulators, Dirac Hamiltonians are only
good effective models at energy scales below the band gap.
Our phase diagrams for Dirac Hamiltonians with random masses indicate that
the metallic phases are stable even in the limit of strong disorder,
whereas strong disorder should generically result in localization 
in lattice models.
There is no contradiction here, 
since there is an implicit assumption in our analysis of Dirac Hamiltonians 
that the disorder strength is always smaller than the energy scale below 
which Dirac Hamiltonians
are valid approximations to the underlying lattice models.
We note that several attempts were published
to study within a NLSM approach the
localization transitions in the strong-disorder
regime (of lattice models) in the chiral and AII symmetry classes.%
~\cite{fu-kane-average12,Koenig12,ostrovsky14}

Second, our approach was based on K-theory.
K-theory aims at identifying 
stable homotopy groups of topological spaces. From the perspective
of band theory, stability is to be interpreted as a topological attribute
that is robust against any addition of trivial bands.
Thus, there are cases to which our theory cannot be applied.
An example is the localization problem of a band insulator characterized
by a nontrivial Hopf map.
A band insulator characterized by a Hopf map%
~\cite{MoorePRL08,schnyder-ryu08,Kennedy2014}
[a nontrivial element of $\pi^{\,}_{3}(S^{2})$] is unstable to
the addition of trivial bands.
Hence such a Hopf map is not 
expected to be captured by the zeroth homotopy groups of the topological
spaces of normalized Dirac masses $\pi^{\,}_{0}(V)$.
Indeed, one cannot write down a Dirac Hamiltonian with a constant mass
term supporting a non-trivial Hopf map,
i.e., no Dirac Hamiltonian with a constant mass term
corresponds to a band insulator with a non-trivial Hopf map.

\section*{Acknowledgment}
This work was supported by Grants-in-Aid from the Japan Society for
Promotion of Science (Grants No.~24840047 and No.~24540338)
and by the RIKEN iTHES Project.
CMM and AF are grateful for the hospitality of the
Advanced Study Group 
``Topological Band Structures and Their Instabilities''
at the MPI-PKS (Dresden).
CM would like to acknowledge useful discussions
with A. Altland, D. Bagrets, and M. Zirnbauer.

\appendix

\section{Explicit constructions of the compact topological
spaces $V$ in 1D}
\label{appsec: Lessons from ten ...}

The properties of Sec.~\ref{sec: Classifying spaces of normalized Dirac masses}
are central to this paper. They are applications of deep results in
algebraic topology. They can be verified through explicit constructions
of massive Dirac Hamiltonians for any given dimension $d$ of space, 
as we are going to do when $d=1$ in Secs.~\ref{subsec: Examples 1} 
and~\ref{subsec: Examples 2}. We shall then motivate why
the properties of Sec.~\ref{sec: Classifying spaces of normalized Dirac masses}
hold in any $d=1,2,\ldots$ 
in Sec.~\ref{appendisubsec: Interpretation of ...}.
We close Appendix~\ref{appsec: Lessons from ten ...}
with a discussion of massless Dirac Hamiltonians
in Sec.~\ref{appendixsubsec: Existence condition of ...}.

\subsection{Examples of Dirac Hamiltonian of rank 2} 
\label{subsec: Examples 1} 

For simplicity, we represent
a generic Dirac Hamiltonian of rank $r=2$
that obeys periodic boundary (ring geometry)
conditions
by
\begin{equation}
\mathcal{H}(k):=
\tau^{\,}_{3}\,k
+
\tau^{\,}_{3}\, A^{\,}_{1}
+
\tau^{\,}_{2}\,M^{\,}_{2}
+
\tau^{\,}_{1}\,M^{\,}_{1}
+
\tau^{\,}_{0}\,A^{\,}_{0}.
\label{eq: generic 2 by 2 Dirac Hamiltonian d=1}
\end{equation}
The Fermi velocity and the Planck constants have been set to unity.
The unit $2\times2$ matrix is denoted by $\tau^{\,}_{0}$ and
$\bm{\tau}=(\tau^{\,}_{1},\tau^{\,}_{2},\tau^{\,}_{3})$ are three Pauli matrices.
The real number $A^{\,}_{1}$ couples as a vector potential induced
by a magnetic flux at the center of a ring would do.
The real number $A^{\,}_{0}$ couples as a scalar potential would do,
i.e., as the chemical potential.
The real parameters $M^{\,}_{1}$ and $M^{\,}_{2}$ couple
to the two available Pauli matrices that anticommute with the kinetic
energy $\tau^{\,}_{3}k$.
The four real parameters $A^{\,}_{1}$, $M^{\,}_{1}$, $M^{\,}_{2}$,
and $A^{\,}_{0}$ furnish an exhaustive parametrization of a 
$2\times2$ Hermitian matrix.  

When $A^{\,}_{1}=A^{\,}_{0}=0$, the Dirac Hamiltonian 
(\ref{eq: generic 2 by 2 Dirac Hamiltonian d=1})
has the eigenvalues
\begin{equation}
\varepsilon(k)=\pm\sqrt{k^{2}+M^{2}},
\qquad
M^{2}:=M^{2}_{1}+M^{2}_{2}.
\end{equation}
The real parameters $M^{\,}_{1}$ and $M^{\,}_{2}$
thus parametrize a Dirac mass.
More precisely, we may write
\begin{subequations}
\begin{equation}
\tau^{\,}_{2}\,M^{\,}_{2}
+
\tau^{\,}_{1}\,M^{\,}_{1}=
M\,\beta(\theta)
\end{equation}
where
\begin{equation}
\begin{split}
&
M:=\sqrt{M^{2}_{1}+M^{2}_{2}},
\quad
\theta:=
\arctan\frac{M^{\,}_{2}}{M^{\,}_{1}},
\\
&
\beta(\theta):=
\tau^{\,}_{1}\,
\cos\theta
+
\tau^{\,}_{2}\,
\sin\theta,
\end{split}
\end{equation}
\end{subequations}
in which case ($\alpha\equiv\tau^{\,}_{3}$)
\begin{equation}
\beta^{2}(\theta)=1,
\qquad
\{\beta(\theta),\alpha\}=0.
\end{equation} 
Thus, the normalized Dirac mass $\beta(\theta)$ 
is parametrized by the angle $0\leq\theta<2\pi$.
As a topological set, 
the normalized Dirac masses $\{\beta(\theta)|0\leq\theta<2\pi\}$ 
are homeomorphic to the circle $S^{1}$. They are also
homeomorphic to $U(1)$ through the map 
\begin{equation}
\beta^{\,}_{12}:[0,2\pi[\to U(1),
\theta\mapsto [\beta(\theta)]^{\,}_{12}.
\end{equation}

\textbf{Symmetry class A:}
The Dirac Hamiltonian~(\ref{eq: generic 2 by 2 Dirac Hamiltonian d=1})
with $A^{\,}_{1}$, $M^{\,}_{1}$, $M^{\,}_{2}$, $A^{\,}_{0}$ smooth 
and non-vanishing functions
of the position $x$ along the ring is said to belong to
the AZ symmetry class A.
As we have seen, the set
\begin{equation}
V^{\mathrm{A}}_{d=1,r=2}:=
\{\beta(\theta)|0\leq\theta<2\pi\}=: 
S^{1},
\label{eq: Vd=1 for A}
\end{equation}
a circle, is the topological space of normalized Dirac masses
associated with the Dirac Hamiltonian
by adding to the kinetic contribution 
with the Dirac matrix $\alpha$
the normalized mass matrix $\beta(\theta)$ 
for $0\leq\theta<2\pi$.
We note that $V^{\mathrm{A}}_{d=1,r=2}$ and $U(1)$ 
are homeomorphic as topological spaces.
Thus, they share the same homotopy groups.
On the other hand,
$V^{\mathrm{A}}_{d=1,r=2}$ is not a group under matrix multiplication
while $U(1)$ is.

In addition to the conservation of the fermion number, we may
impose TRS on the Dirac Hamiltonian%
~(\ref{eq: generic 2 by 2 Dirac Hamiltonian d=1})
There are two possibilities to do so.

\textbf{Symmetry class AII:}
If charge conservation holds and TRS is imposed through
\begin{subequations}
\label{eq: AII 2x2}
\begin{equation}
\mathcal{H}(k)=
+
\tau^{\,}_{2}\,
\mathcal{H}^{*}(-k)\,
\tau^{\,}_{2},
\label{eq: AII 2x2 a}
\end{equation}
then
\begin{equation}
\mathcal{H}(k)=
\tau^{\,}_{3}\,k
+
\tau^{\,}_{0}\,A^{\,}_{0}.
\label{eq: AII 2x2 b}
\end{equation}
No mass matrix is permissible if TRS squares to minus the identity.
The topological space of normalized Dirac masses in the symmetry class
AII is the empty set
\begin{equation}
V^{\mathrm{AII}}_{d=1,r=2}=\emptyset.
\end{equation}
\end{subequations}
Because of the fermion-doubling problem,%
~\cite{Nielsen81}
the only way to realize (\ref{eq: AII 2x2 b})
as the low-energy and long wavelength limit of a lattice model
is on the boundary of a two-dimensional topological insulator
in the symmetry class AII.

\textbf{Symmetry class AI:}
If charge conservation holds and TRS is imposed through
\begin{subequations}
\label{eq: AI 2x2}
\begin{equation}
\mathcal{H}(k)=
+
\tau^{\,}_{1}\,
\mathcal{H}^{*}(-k)\,
\tau^{\,}_{1},
\label{eq: AI 2x2 a}
\end{equation}
then
\begin{equation}
\mathcal{H}(k)=
\tau^{\,}_{3}\,k
+
\tau^{\,}_{2}\,M^{\,}_{2}
+
\tau^{\,}_{1}\,M^{\,}_{1}
+
\tau^{\,}_{0}\,A^{\,}_{0}.
\label{eq: AI 2x2 b}
\end{equation}
The same mass matrix as in the symmetry class
A is permissible if TRS squares to the identity.
The homeomorphy between the allowed masses in the symmetry classes
A and AI is accidental. It does not hold for larger representations
of the Dirac matrix as we shall verify explicity when considering
a rank 4 Dirac matrix below. 
The topological space of normalized Dirac masses obtained by
augmenting the Dirac kinetic contribution by adding
a mass matrix squaring to unity and obeying this TRS is
\begin{equation}
V^{\mathrm{AI}}_{d=1,r=2}:=
\{\beta(\theta)|0\leq\theta<2\pi\}
=: 
S^{1}.
\label{eq: Vd=1 for AI}
\end{equation}
\end{subequations}
The topological spaces
$V^{\mathrm{AI}}_{d=1,r=2}$ and $U(1)$ are homeomorphic.
(This homeomorphism is not a group homomorphism, for
$V^{\mathrm{A}}_{d=1,r=2}$ is not a group while $U(1)$ is.)
Consequently, they share the same homotopy groups.

The standard symmetry classes A, AII, and AI can be further constrained
by imposing the CHS. This gives the following three possibilities.

\textbf{Symmetry class AIII:}
If charge conservation holds together with the CHS
\begin{subequations}
\label{eq: AIII 2x2}
\begin{equation}
\mathcal{H}(k)=
-
\tau^{\,}_{1}\,
\mathcal{H}(k)\,
\tau^{\,}_{1},
\label{eq: AIII 2x2 a}
\end{equation}
then
\begin{equation}
\mathcal{H}(k)=
\tau^{\,}_{3}\,k
+
\tau^{\,}_{3}\, A^{\,}_{1}
+
\tau^{\,}_{2}\,M^{\,}_{2}.
\label{eq: AIII 2x2 b}
\end{equation}
There is a unique mass matrix.
The topological space of normalized Dirac masses obtained by adding to
the Dirac kinetic contribution a mass matrix 
squaring to unity and obeying the CHS is
\begin{equation}
V^{\mathrm{AIII}}_{d=1,r=2}=
\{\pm\tau^{\,}_{2}\}.
\label{eq: AIII 2x2 c}
\end{equation}
\end{subequations}

\textbf{Symmetry class CII:}
It is not possible to write down a $2\times2$ Dirac equation
in the symmetry class CII. For example, imposing
\begin{equation}
\mathcal{H}(k)=
-
\tau^{\,}_{1}\,
\mathcal{H}(k)\,
\tau^{\,}_{1},
\qquad
\mathcal{H}(k)=
+
\tau^{\,}_{2}\,
\mathcal{H}^{*}(-k)\,
\tau^{\,}_{2},
\end{equation}
enforces the symmetry class DIII, for
composing the CHS with the TRS delivers a PHS that squares to the
unity and not minus the unity. In order to implement 
the symmetry constraints of class CII,
we need to consider a $4\times4$ Dirac equation
(see Sec.~\ref{subsec: Examples 2}).

\textbf{Symmetry class BDI:}
If charge conservation holds together with
\begin{subequations}
\label{eq: BDI 2x2}
\begin{equation}
\mathcal{H}(k)=
-
\tau^{\,}_{1}\,
\mathcal{H}(k)\,
\tau^{\,}_{1},
\qquad
\mathcal{H}(k)=
+
\tau^{\,}_{1}\,
\mathcal{H}^{*}(-k)\,
\tau^{\,}_{1},
\label{eq: BDI 2x2 a}
\end{equation}
then
\begin{equation}
\mathcal{H}(k)=
\tau^{\,}_{3}\,k
+
\tau^{\,}_{2}\,M^{\,}_{2}.
\label{eq: BDI 2x2 b}
\end{equation}
There is a unique mass matrix.
The topological space of normalized Dirac masses obtained by adding to
the Dirac kinetic contribution a mass matrix 
squaring to unity while preserving TRS and PHS 
(a product of TRS and CHS), both of which square to unity, is
\begin{equation}
V^{\mathrm{BDI}}_{d=1,r=2}=
\{\pm\tau^{\,}_{2}\}.
\end{equation}
\end{subequations}

Now, we move on to the four Bogoliubov-de-Gennes symmetry classes
with PHS.

\textbf{Symmetry class D:}
If we impose PHS through
\begin{subequations}
\label{eq: D 2x2}
\begin{equation}
\mathcal{H}(k)=
-
\mathcal{H}^{*}(-k),
\label{eq: D 2x2 a}
\end{equation}
then
\begin{equation}
\mathcal{H}(k)=
\tau^{\,}_{3}\,k
+
\tau^{\,}_{2}\,M^{\,}_{2}.
\label{eq: D 2x2 b}
\end{equation}
There is a unique mass matrix.
The topological space of normalized Dirac masses obtained by adding to
the Dirac kinetic contribution a mass matrix 
squaring to unity and preserving the
PHS squaring to unity is
\begin{equation}
V^{\mathrm{D}}_{d=1,r=2}=
\{\pm\tau^{\,}_{1}\}.
\end{equation}
\end{subequations}

\textbf{Symmetry class DIII:}
If we impose PHS and TRS through
\begin{subequations}
\label{eq: DIII 2x2}
\begin{equation}
\mathcal{H}(k)=
-
\mathcal{H}^{*}(-k),
\qquad
\mathcal{H}(k)=
+
\tau^{\,}_{2}\,
\mathcal{H}^{*}(-k)\,
\tau^{\,}_{2},
\label{eq: DIII 2x2 a}
\end{equation}
respectively, then
\begin{equation}
\mathcal{H}(k)=
\tau^{\,}_{3}\,k.
\label{eq: DIII 2x2 b}
\end{equation}
No mass matrix is permissible if TRS squares to minus the identity.
The topological space of normalized Dirac masses 
in the symmetry class DIII is the empty set
\begin{equation}
V^{\mathrm{DIII}}_{d=1,r=2}=\emptyset.
\end{equation}
\end{subequations}
Because of the fermion-doubling problem,%
~\cite{Nielsen81}
the only way to realize (\ref{eq: DIII 2x2 b})
as the low-energy and long wavelength limit of a lattice model
is on the boundary of a two-dimensional topological 
superconductor in the symmetry class DIII.

\textbf{Symmetry class C:}
If we impose PHS through
\begin{subequations}
\label{eq: C 2x2}
\begin{equation}
\mathcal{H}(k)=
-
\tau^{\,}_{2}\,
\mathcal{H}^{*}(-k)\,
\tau^{\,}_{2},
\label{eq: C 2x2 a}
\end{equation}
then
\begin{equation}
\mathcal{H}(k)=
\tau^{\,}_{3}\,A^{\,}_{1}
+
\tau^{\,}_{2}\,M^{\,}_{2}
+
\tau^{\,}_{1}\,M^{\,}_{1}.
\label{eq: C 2x2 b}
\end{equation}
\end{subequations}
PHS squaring to minus unity prohibits a kinetic energy in
any Dirac Hamiltonian of rank 2 in the symmetry
class C.

\textbf{Symmetry class CI:}
If we impose PHS and TRS through
\begin{subequations}
\label{eq: CI 2x2}
\begin{equation}
\mathcal{H}(k)=
-
\tau^{\,}_{2}\,
\mathcal{H}^{*}(-k)\,
\tau^{\,}_{2},
\qquad
\mathcal{H}(k)=
+
\tau^{\,}_{1}\,
\mathcal{H}^{*}(-k)\,
\tau^{\,}_{1},
\label{eq: CI 2x2 a}
\end{equation}
respectively,
then
\begin{equation}
\mathcal{H}(k)=
\tau^{\,}_{2}\,M^{\,}_{2}
+
\tau^{\,}_{1}\,M^{\,}_{1}.
\label{eq: CI 2x2 b}
\end{equation}
\end{subequations}
PHS squaring to minus unity prohibits a kinetic energy in the symmetry
class CI.

\subsection{Examples of Dirac Hamiltonian of rank 4}
\label{subsec: Examples 2}

Consider the generic Dirac Hamiltonian
of rank $r=4$ that obeys periodic boundary conditions
(ring geometry)
\begin{equation}
\begin{split}
\mathcal{H}(k):=&
\tau^{\,}_{3}\otimes\sigma^{\,}_{0}\,k
+
\tau^{\,}_{3}\otimes\sigma^{\,}_{\nu}\,A^{\,}_{1,\nu}\,
+
\tau^{\,}_{2}\otimes\sigma^{\,}_{\nu}\,M^{\,}_{2,\nu}
\\
&
+
\tau^{\,}_{1}\otimes\sigma^{\,}_{\nu}\,M^{\,}_{1,\nu} 
+
\tau^{\,}_{0}\otimes\sigma^{\,}_{\nu}\,A^{\,}_{0,\nu}.
\end{split}
\label{eq: generic 4 by 4 Dirac Hamiltonian d=1}
\end{equation}
The Fermi velocity and the Planck constants have been set to unity.
The matrices $\tau^{\,}_{0}$ and $\bm{\tau}$ were defined in 
Eq.~(\ref{eq: generic 2 by 2 Dirac Hamiltonian d=1}).
A second unit $2\times2$ matrix is denoted by $\sigma^{\,}_{0}$ and
$\bm{\sigma}=(\sigma^{\,}_{1},\sigma^{\,}_{2},\sigma^{\,}_{3})$
are another set of three Pauli matrices.
The summation convention over the repeated index $\nu=0,1,2,3$ is implied.
There are four real-valued parameters for the components
$A^{\,}_{1,\nu}$ with $\nu=0,1,2,3$ of an $U(2)$ vector potential,
eight for the components $M^{\,}_{1,\nu}$ and $M^{\,}_{2,\nu}$ with 
$\nu=0,1,2,3$ of two independent $U(2)$ masses,
and four for the components
$A^{\,}_{0,\nu}$ with $\nu=0,1,2,3$ of an $U(2)$ scalar potential.
As it should be there are 16 real-valued free parameters
(functions if we opt to break translation invariance).

\textbf{Symmetry class A:}
The Dirac Hamiltonian~(\ref{eq: generic 4 by 4 Dirac Hamiltonian d=1})
with $A^{\,}_{1,\nu}$, $M^{\,}_{1,\nu}$, $M^{\,}_{2,\nu}$, $A^{\,}_{0,\nu}$
smooth and non-vanishing functions of the position $x$ along the ring
is said to belong to the AZ symmetry class A. 
All eight mass matrices of rank $r=4$
in the 1D symmetry class A can be arranged
into the four pairs $(M^{\,}_{1,\nu},M^{\,}_{2,\nu})$ with $\nu=0,1,2,3$
of anticommuting masses.
The topological space of normalized Dirac masses obtained by adding to
the kinetic energy a mass matrix
squaring to the unit matrix is%
~\cite{morimoto-clifford13}
\begin{subequations}
\begin{equation}
\begin{split}
V^{\mathrm{A}}_{d=1,r=4}:=&\,
\left\{
\left.
\beta=
\begin{pmatrix}
0
&
U
\\
U^{\dag}
&
0
\end{pmatrix}
\right|
U\in U(2)
\right\}.
\end{split}
\label{V^A_r=4 = S1*S3}
\end{equation}
As a topological space, it is thus homeomorphic to 
$U(2)\simeq U(1) \times SU(2) \simeq S^1 \times S^3$,
an interpretation rendered plausible by the parametrization
\begin{equation}
\begin{split}
&
\begin{split}
V^{\mathrm{A}}_{d=1,r=4}&=
\left\{
\bm{M}\cdot\bm{X}\!\!+\!\!\bm{N}\cdot\bm{Y}|
\bm{M}^{2}= 
\cos^{2}\theta, 
\bm{N}=\tan\theta \bm{M}
\right\}
\\
&=:
S^{1}\times S^{3},
\end{split}
\\
&
\bm{M}:=
\left(
M^{\,}_{2,0},
M^{\,}_{1,1},
M^{\,}_{1,2},
M^{\,}_{1,3}
\right),
\\
&
\bm{N}:=
\left(
-M^{\,}_{1,0},
M^{\,}_{2,1},
M^{\,}_{2,2},
M^{\,}_{2,3}
\right),
\\
&
\bm{X}:=
\left(
\tau^{\,}_{2}\otimes\sigma^{\,}_{0},
\tau^{\,}_{1}\otimes\sigma^{\,}_{1},
\tau^{\,}_{1}\otimes\sigma^{\,}_{2},
\tau^{\,}_{1}\otimes\sigma^{\,}_{3}
\right),
\\
&
\bm{Y}:=
\left(
-\tau^{\,}_{1}\otimes\sigma^{\,}_{0},
\tau^{\,}_{2}\otimes\sigma^{\,}_{1},
\tau^{\,}_{2}\otimes\sigma^{\,}_{2},
\tau^{\,}_{2}\otimes\sigma^{\,}_{3}
\right),
\\
\end{split}
\end{equation}
\end{subequations}
We note that $V^{\mathrm{A}}_{d=1,r=4}$ is not closed
under matrix multiplication so that it does not carry the
group structure of $U(2)$.

In addition to the conservation of the fermion number, we may
impose TRS on the Dirac Hamiltonian%
~(\ref{eq: generic 4 by 4 Dirac Hamiltonian d=1})
There are two possibilities to do so.

\textbf{Symmetry class AII:} 
If charge conservation holds together with TRS through
\begin{subequations}
\begin{equation}
\mathcal{H}(k)=
+
\tau^{\,}_{1}\otimes\sigma^{\,}_{2}\,
\mathcal{H}^{*}(-k)\,
\tau^{\,}_{1}\otimes\sigma^{\,}_{2},
\label{TRS r=4}
\end{equation}
then 
\begin{equation}
\begin{split}
\mathcal{H}(k)=&\,
\tau^{\,}_{3}\otimes\sigma^{\,}_{0}\,k
+
\sum_{\nu=1,2,3}
\tau^{\,}_{3}\otimes\sigma^{\,}_{\nu}\, A^{\,}_{1,\nu}
+
\tau^{\,}_{2}\otimes\sigma^{\,}_{0}\,M^{\,}_{2,0}
\\
&\,
+
\tau^{\,}_{1}\otimes\sigma^{\,}_{0}\,M^{\,}_{1,0}
+
\tau^{\,}_{0}\otimes\sigma^{\,}_{0}\,A^{\,}_{0,0}.
\end{split}
\end{equation}
Observe that by doubling the Dirac Hamiltonian
(\ref{eq: AII 2x2 b}), we went from no mass matrix to two 
anticommuting mass matrices. 
The topological space of normalized Dirac masses obtained by adding to
the Dirac kinetic contribution a mass matrix 
squaring to unity and obeying this TRS is
obtained by imposing the TRS (\ref{TRS r=4})
on Eq.\ (\ref{V^A_r=4 = S1*S3}) to be
\begin{equation}
V^{\mathrm{AII}}_{d=1,r=4}:=
\left\{
\left.
\beta=
\begin{pmatrix}
0
&
U
\\
U^{\dag}
&
0
\end{pmatrix}
\right|
U=+\sigma^{\,}_{2}\,U^{\mathsf{T}}\sigma^{\,}_{2}\,\in U(2)
\right\}.
\end{equation}
As a topological space, 
$V^{\mathrm{AII}}_{d=1,r=4}$
can be shown to be homeomorphic to
$U(2)/Sp(1)\simeq U(1)\times SU(2)/SU(2)\simeq U(1)$,
an interpretation rendered plausible by the parametrization 
\begin{equation}
\begin{split}
&
V^{\mathrm{AII}}_{d=1,r=4}= 
\left\{
\bm{M}\cdot\bm{X}|
\bm{M}^{2}=1
\right\}=:
S^{1},
\\
&
\bm{M}:=
\left(
M^{\,}_{2,0},
M^{\,}_{1,0}
\right),
\\
&
\bm{X}:=
\left(
\tau^{\,}_{2}\otimes\sigma^{\,}_{0},
\tau^{\,}_{1}\otimes\sigma^{\,}_{0}
\right).
\end{split}
\end{equation}
\end{subequations} 

\textbf{Symmetry class AI:}
If charge conservation holds together with TRS through
\begin{subequations}
\begin{equation}
\mathcal{H}(k)=
+
\tau^{\,}_{1}\otimes\sigma^{\,}_{0}\,
\mathcal{H}^{*}(-k)\,
\tau^{\,}_{1}\otimes\sigma^{\,}_{0},
\end{equation}
then
\begin{equation}
\begin{split}
\mathcal{H}(k)=&\,
\tau^{\,}_{3}\otimes\sigma^{\,}_{0}\,k
+
\tau^{\,}_{3}\otimes\sigma^{\,}_{2}\,A^{\,}_{1,2}
+
\sum_{\nu=0,1,3}
\big(
\tau^{\,}_{2}\otimes\sigma^{\,}_{\nu}\,M^{\,}_{2,\nu} 
\\
&\,
+
\tau^{\,}_{1}\otimes\sigma^{\,}_{\nu}\,M^{\,}_{1,\nu}
+
\tau^{\,}_{0}\otimes\sigma^{\,}_{\nu}\,A^{\,}_{0,\nu}
\big).
\end{split}
\end{equation}
There are six mass matrices of rank $r=4$
in the 1D symmetry class AI that can be arranged
into the three pairs $(M^{\,}_{1,\nu},M^{\,}_{2,\nu})$ with $\nu=0,1,3$
of anticommuting masses.
The topological space of normalized Dirac masses obtained by adding to
the Dirac kinetic contribution a mass matrix 
squaring to unity and obeying this TRS is
\begin{equation}
\begin{split}
V^{\mathrm{AI}}_{d=1,r=4}:=&\,
\left\{
\left.
\beta=
\begin{pmatrix}
0
&
U
\\
U^{\dag}
&
0
\end{pmatrix}
\right|
U=+U^{\mathsf{T}}\in U(2)
\right\}.
\end{split}
\end{equation}
As a topological space, 
$V^{\mathrm{AI}}_{d=1,r=4}$
can be shown to be homeomorphic to
$U(2)/O(2)\simeq U(1)/\{\pm 1\}\times SU(2)/U(1)\simeq S^1 \times S^{2}$,
an interpretation rendered plausible by the parametrization
\begin{equation}
\begin{split}
&
\begin{split}
V^{\mathrm{AI}}_{d=1,r=4}&=
\left\{
\bm{M}\cdot\bm{X}\!+\!\bm{N}\cdot\bm{Y}|
\bm{M}^{2}= 
\cos^{2}\theta, 
\bm{N}=\tan\theta\bm{M}
\right\}
\\
&=:
S^{1}\times S^{2},
\end{split}
\\
&
\bm{M}:=
\left(
M^{\,}_{2,0},
M^{\,}_{1,1},
M^{\,}_{1,3}
\right),
\\
&
\bm{N}:=
\left(
-M^{\,}_{1,0},
M^{\,}_{2,1},
M^{\,}_{2,3}
\right),
\\
&
\bm{X}:=
\left(
\tau^{\,}_{2}\otimes\sigma^{\,}_{0},
\tau^{\,}_{1}\otimes\sigma^{\,}_{1},
\tau^{\,}_{1}\otimes\sigma^{\,}_{3}
\right),
\\
&
\bm{Y}:=
\left(
-\tau^{\,}_{1}\otimes\sigma^{\,}_{0},
\tau^{\,}_{2}\otimes\sigma^{\,}_{1},
\tau^{\,}_{2}\otimes\sigma^{\,}_{3}
\right).
\end{split}
\end{equation}
\end{subequations}

The standard symmetry classes A, AII, and AI can be further constrained
by imposing the CHS. This gives the following three possibilities.

\textbf{Symmetry class AIII:}
If charge conservation holds together with the CHS
\begin{subequations}
\begin{equation}
\mathcal{H}(k)=
-
\tau^{\,}_{1}\otimes\sigma^{\,}_{0}\,
\mathcal{H}(k)\,
\tau^{\,}_{1}\otimes\sigma^{\,}_{0},
\end{equation}
then
\begin{equation}
\mathcal{H}(k)=
\tau^{\,}_{3}\otimes\sigma^{\,}_{0}\,k
+
\sum_{\nu=0,1,2,3}\left(
\tau^{\,}_{3}\otimes\sigma^{\,}_{\nu}\,A^{\,}_{1,\nu}
+
\tau^{\,}_{2}\otimes\sigma^{\,}_{\nu}\,M^{\,}_{2,\nu}
\right).
\end{equation}
The Dirac mass matrix $\tau^{\,}_{2}\otimes\sigma^{\,}_{0}\,M^{\,}_{2,0}$
that descends from Eq.~(\ref{eq: AIII 2x2 b})
commutes with the triplet of anticommuting mass matrices
$\tau^{\,}_{2}\otimes\sigma^{\,}_{1}\,M^{\,}_{2,1}$,
$\tau^{\,}_{2}\otimes\sigma^{\,}_{2}\,M^{\,}_{2,2}$,
and
$\tau^{\,}_{2}\otimes\sigma^{\,}_{3}\,M^{\,}_{2,3}$.
The topological space of normalized Dirac masses obtained by adding to
the Dirac kinetic contribution a mass matrix 
squaring to the unit matrix and obeying the CHS is%
~\cite{morimoto-clifford13}
\begin{equation}
\begin{split}
V^{\mathrm{AIII}}_{d=1,r=4}:=\,
\Bigg\{&
\beta=
\tau^{\,}_{2}\otimes A
\Bigg|
A:=
U\,I^{\,}_{m,n}\,U^{\dag},
\\
&
m,n=0,1,2,
\quad
m+n=2,
\quad
U\in U(2),
\\
&
I^{\,}_{m,n}:=
\mathrm{diag}
(
\overbrace{-1,\ldots,-1}^{\hbox{$m$ times}},
\overbrace{+1,\ldots,+1}^{\hbox{$n$ times}}
)
\Bigg\}.
\end{split}
\end{equation}
As a topological space,  it is thus homeomorphic to
$U(2)/[U(2)\times U(0)]\cup 
U(2)/[U(1)\times U(1)]\cup 
U(2)/[U(0)\times U(2)]$,
as is also apparent from the parametrization
\begin{equation}
\begin{split}
&
V^{\mathrm{AIII}}_{d=1,r=4}=
\left\{
\pm\tau^{\,}_{2}\otimes\sigma^{\,}_{0}
\right\}
\cup
\left\{
\bm{M}\cdot\bm{X}|
\bm{M}^{2}=1
\right\},
\\
&
\bm{M}:=
\left(
M^{\,}_{2,1},
M^{\,}_{2,2},
M^{\,}_{2,3}
\right),
\\
&
\bm{X}:=
\left(
\tau^{\,}_{2}\otimes\sigma^{\,}_{1},
\tau^{\,}_{2}\otimes\sigma^{\,}_{2},
\tau^{\,}_{2}\otimes\sigma^{\,}_{3}
\right),
\end{split}
\end{equation}
\end{subequations}
(recall that $S^{2}\simeq SU(2)/U(1)$ so that
$U(2)/[U(1)\times U(1)]\simeq S^{2}$).

\textbf{Symmetry class CII:}
If charge conservation holds together with CHS and TRS
\begin{subequations}
\begin{align}
\mathcal{H}(k)&=
-
\tau^{\,}_{1}\otimes\sigma^{\,}_{0}\,
\mathcal{H}(k)\,
\tau^{\,}_{1}\otimes\sigma^{\,}_{0},
\\
\mathcal{H}(k)&=
+
\tau^{\,}_{1}\otimes\sigma^{\,}_{2}\,
\mathcal{H}^{*}(-k)\,
\tau^{\,}_{1}\otimes\sigma^{\,}_{2},
\end{align}
respectively,
then
\begin{equation}
\mathcal{H}(k)=
\tau^{\,}_{3}\otimes\sigma^{\,}_{0}\,k
+
\sum_{\nu=1,2,3}
\tau^{\,}_{3}\otimes\sigma^{\,}_{\nu}\, A^{\,}_{1,\nu}
+
\tau^{\,}_{2}\otimes\sigma^{\,}_{0}\,M^{\,}_{2,0}.
\end{equation}
There is a unique mass matrix, as was the case in Eqs.%
~(\ref{eq: AIII 2x2 b}) and (\ref{eq: BDI 2x2 b}).
The topological space of normalized Dirac masses obtained by adding to
the Dirac kinetic contribution a mass matrix 
squaring to the unit matrix and obeying CHS and
TRS squaring to minus unity is
\begin{equation}
V^{\mathrm{CII}}_{d=1,r=4}:=
\Big\{
\beta\in V^{\mathrm{AIII}}_{d=1,r=4}
\Big|
\beta=
(\tau^{\,}_{1}\otimes\sigma^{\,}_{2})\,
\beta^{*}\, 
(\tau^{\,}_{1}\otimes\sigma^{\,}_{2})
\Big\}.
\end{equation}
As a topological space, 
$V^{\mathrm{CII}}_{d=1,r=4}$
can be shown to be homeomorphic to
$Sp(1)/Sp(1)\times Sp(0)\cup 
Sp(1)/Sp(0)\times Sp(1)$,
an interpretation rendered plausible by the parametrization
\begin{equation}
V^{\mathrm{CII}}_{d=1,r=4}=\{\pm\tau^{\,}_{2}\otimes\sigma^{\,}_{0}\}.
\end{equation}
\end{subequations}

\textbf{Symmetry class BDI:}
If charge conservation holds together together with CHS and TRS
\begin{subequations}
\begin{align}
\mathcal{H}(k)&=
-
\tau^{\,}_{1}\otimes\sigma^{\,}_{0}\,
\mathcal{H}(k)\,
\tau^{\,}_{1}\otimes\sigma^{\,}_{0},
\\
\mathcal{H}(k)&=
+
\tau^{\,}_{1}\otimes\sigma^{\,}_{0}\,
\mathcal{H}^{*}(-k)\,
\tau^{\,}_{1}\otimes\sigma^{\,}_{0},
\end{align}
respectively, then
\begin{equation}
\mathcal{H}(k)=
\tau^{\,}_{3}\otimes\sigma^{\,}_{0}\,k
+
\tau^{\,}_{3}\otimes\sigma^{\,}_{2}\,A^{\,}_{1,2}
+
\sum_{\nu=0,1,3}
\tau^{\,}_{2}\otimes\sigma^{\,}_{\nu}\,M^{\,}_{2,\nu}.
\end{equation}
The Dirac mass matrix $\tau^{\,}_{2}\otimes\sigma^{\,}_{0}\,M^{\,}_{2,0}$
that descends from Eq.~(\ref{eq: BDI 2x2 b})
commutes with the pair of anticommuting mass matrices
$\tau^{\,}_{2}\otimes\sigma^{\,}_{1}\,M^{\,}_{2,1}$
and
$\tau^{\,}_{2}\otimes\sigma^{\,}_{3}\,M^{\,}_{2,3}$.
The topological space of normalized Dirac masses obtained by adding to
the Dirac kinetic contribution a mass matrix 
squaring to the unit matrix and obeying CHS and
TRS squaring to unity is
\begin{equation}
V^{\mathrm{BDI}}_{d=1,r=4}:=
\Big\{
\beta\in V^{\mathrm{AIII}}_{d=1,r=4}
\Big|
\beta=
(\tau^{\,}_{1}\otimes\sigma^{\,}_{0})\,
\beta^{*}\,
(\tau^{\,}_{1}\otimes\sigma^{\,}_{0})
\Big\}.
\end{equation}
As a topological space, 
$V^{\mathrm{BDI}}_{d=1,r=4}$
can be shown to be homeomorphic to
$O(2)/[O(2)\times O(0)]\cup O(2)/[O(1)\times O(1)]\cup O(2)/[O(0)\times O(2)]$,
an interpretation rendered plausible from the parametrization
\begin{equation}
\begin{split}
&
V^{\mathrm{BDI}}_{d=1,r=4}=
\left\{
\pm\tau^{\,}_{2}\otimes\sigma^{\,}_{0}
\right\}
\cup
\left\{
\bm{M}\cdot\bm{X}|
\bm{M}^{2}=1
\right\},
\\
&
\bm{M}:=
\left(
M^{\,}_{2,1},
M^{\,}_{2,3}
\right),
\\
&
\bm{X}:=
\left(
\tau^{\,}_{2}\otimes\sigma^{\,}_{1},
\tau^{\,}_{2}\otimes\sigma^{\,}_{3}
\right),
\end{split}
\end{equation}
\end{subequations}
(recall that $S^1\simeq O(2)/[O(1)\times O(1)]$).

Now, we move on to the four BdG symmetry classes.

\textbf{Symmetry class D:}
If we impose PHS through
\begin{subequations}
\begin{equation}
\mathcal{H}(k)=
-
\mathcal{H}^{*}(-k),
\end{equation}
then
\begin{equation}
\begin{split}
\mathcal{H}(k)=&\,
\tau^{\,}_{3}\otimes\sigma^{\,}_{0}\,k
+
\tau^{\,}_{3}\otimes\sigma^{\,}_{2}\,A^{\,}_{1,2}
+
\sum_{\nu=0,1,3}
\tau^{\,}_{2}\otimes\sigma^{\,}_{\nu}\,M^{\,}_{2,\nu}
\\
&\,
+
\tau^{\,}_{1}\otimes\sigma^{\,}_{2}\,M^{\,}_{1,2}
+
\tau^{\,}_{0}\otimes\sigma^{\,}_{2}\,A^{\,}_{0,2}.
\end{split}
\label{class D H(k) r=4}
\end{equation}
\end{subequations}
There are four Dirac mass matrices. None commutes with all remaining ones.
However, each of them is antisymmetric and so is their sum.
The topological space of normalized Dirac masses obtained by adding to
the Dirac kinetic contribution a mass matrix 
squaring to the unit matrix and obeying 
PHS squaring to unity is
\begin{equation}
\begin{split}
&
V^{\mathrm{D}}_{d=1,r=4}=
\left\{
\bm{M}\cdot\bm{X}|
\bm{M}^{2}=1
\right\}
\cup
\left\{
\bm{N}\cdot\bm{Y}|
\bm{N}^{2}=1
\right\}
,
\\
&
\bm{M}:=
\left(
M^{\,}_{2,1},
M^{\,}_{2,3}
\right),
\qquad
\bm{N}:=
\left(
M^{\,}_{2,0},
M^{\,}_{1,2}
\right),
\\
&
\bm{X}:=
\left(
\tau^{\,}_{2}\otimes\sigma^{\,}_{1},
\tau^{\,}_{2}\otimes\sigma^{\,}_{3}
\right),
\qquad
\bm{Y}:=
\left(
\tau^{\,}_{2}\otimes\sigma^{\,}_{0},
\tau^{\,}_{1}\otimes\sigma^{\,}_{2}
\right).
\end{split}
\end{equation}
As a topological space, $V^{\mathrm{D}}_{d=1,r=4}$
is homeomorphic to $O(2)$,
as $V^{\mathrm{D}}_{d=1,r=4}\simeq S^1 \cup S^1
\simeq U(1) \times \mathbb{Z}_{2} \simeq O(2)$.

Observe that
$V^{\mathrm{D}}_{d=1,r=4}$ is not closed
under matrix multiplication, i.e., it does not carry the
group structure of $O(2)$.

\textbf{Symmetry class DIII:}
If we impose PHS and TRS through
\begin{subequations}
\begin{equation}
\mathcal{H}(k)=
-
\mathcal{H}^{*}(-k),
\qquad
\mathcal{H}(k)=
+
\tau^{\,}_{2}\otimes\sigma^{\,}_{0}\,
\mathcal{H}^{*}(-k)\,
\tau^{\,}_{2}\otimes\sigma^{\,}_{0},
\end{equation}
then
\begin{equation}
\mathcal{H}(k)=
\tau^{\,}_{3}\otimes\sigma^{\,}_{0}\,k
+
\tau^{\,}_{3}\otimes\sigma^{\,}_{2}\,A^{\,}_{1,2}
+
\tau^{\,}_{1}\otimes\sigma^{\,}_{2}\,M^{\,}_{1,2}.
\end{equation}
Observe that there is only one Dirac mass matrix [there was none in 
Eq.~(\ref{eq: DIII 2x2 b})]. 
Moreover, this Dirac mass matrix is Hermitian and antisymmetric. 
The topological space of normalized Dirac masses obtained by adding to
the Dirac kinetic contribution a mass matrix 
squaring to the unit matrix and obeying this PHS and this TRS is 
\begin{equation}
V^{\mathrm{DIII}}_{d=1,r=4}=
\left\{\pm\tau^{\,}_{1}\otimes\sigma^{\,}_{2}\right\}.
\end{equation}
\label{eq: DIII 4x4}
\end{subequations}
As a topological space, $V^{\mathrm{DIII}}_{d=1,r=4}$
is homeomorphic to $O(2)/U(1)$.

\textbf{Symmetry class C:}
If we impose PHS through
\begin{subequations}
\label{eq: 1D Symmetry class C rmin}
\begin{equation}
\mathcal{H}(k)=
-
\tau^{\,}_{0}\otimes\sigma^{\,}_{2}\,
\mathcal{H}^{*}(-k)\,
\tau^{\,}_{0}\otimes\sigma^{\,}_{2},
\end{equation}
then
\begin{equation}
\begin{split}
\mathcal{H}(k)=&\,
\tau^{\,}_{3}\otimes\sigma^{\,}_{0}\,k
+
\sum_{\nu=1,2,3}
\tau^{\,}_{3}\otimes\sigma^{\,}_{\nu}\,A^{\,}_{1,\nu}
+
\tau^{\,}_{2}\otimes\sigma^{\,}_{0}\,M^{\,}_{2,0}
\\
&\,
+
\sum_{\nu=1,2,3}
\tau^{\,}_{1}\otimes\sigma^{\,}_{\nu}\,M^{\,}_{1,\nu}
+ 
\sum_{\nu=1,2,3}
\tau^{\,}_{0}\otimes\sigma^{\,}_{\nu}\,A^{\,}_{0,\nu}.
\end{split}
\end{equation}
There are four mass matrices that anticommute pairwise.
The topological space of normalized Dirac masses obtained by adding to
the Dirac kinetic contribution a mass matrix 
squaring to the unit matrix and obeying this PHS is 
\begin{equation}
\begin{split}
&
V^{\mathrm{C}}_{d=1,r=4}=
\left\{
\bm{M}\cdot\bm{X}|
\bm{M}^{2}=1
\right\}=:
S^{3},
\\
&
\bm{M}:=
\left(
M^{\,}_{2,0},
M^{\,}_{1,1},
M^{\,}_{1,2},
M^{\,}_{1,3}
\right),
\\
&
\bm{X}:=
\left(
\tau^{\,}_{2}\otimes\sigma^{\,}_{0},
\tau^{\,}_{1}\otimes\sigma^{\,}_{1},
\tau^{\,}_{1}\otimes\sigma^{\,}_{2},
\tau^{\,}_{1}\otimes\sigma^{\,}_{3}
\right).
\end{split}
\end{equation}
\end{subequations}
As a topological space, $V^{\mathrm{C}}_{d=1,r=4}$
is homeomorphic to $Sp(1)$
since we have $Sp(1)\simeq SU(2) \simeq S^{3}$.
Observe that
$V^{\mathrm{C}}_{d=1,r=4}$ is not closed
under matrix multiplication, i.e., it does not carry the
group structure of $Sp(1)$.

\textbf{Symmetry class CI:}
If we impose PHS and TRS through
\begin{subequations}
\begin{align}
\mathcal{H}(k)&=
-
\tau^{\,}_{0}\otimes\sigma^{\,}_{2}\,
\mathcal{H}^{*}(-k)\,
\tau^{\,}_{0}\otimes\sigma^{\,}_{2},
\\
\mathcal{H}(k)&=
+
\tau^{\,}_{1}\otimes\sigma^{\,}_{0}\,
\mathcal{H}^{*}(-k)\,
\tau^{\,}_{1}\otimes\sigma^{\,}_{0},
\end{align}
respectively,
then
\begin{equation}
\begin{split}
\mathcal{H}(k)=&\,
\tau^{\,}_{3}\otimes\sigma^{\,}_{0}\,k
+
\tau^{\,}_{3}\otimes\sigma^{\,}_{2}\,A^{\,}_{1,2}
+
\tau^{\,}_{2}\otimes\sigma^{\,}_{0}\,M^{\,}_{2,0}
\\
&\,
+
\sum_{\nu=1,3}\left(
\tau^{\,}_{1}\otimes\sigma^{\,}_\nu\,M^{\,}_{1,\nu}
+ 
\tau^{\,}_{0}\otimes\sigma^{\,}_\nu\,A^{\,}_{0,\nu}
\right).
\end{split}
\end{equation}
\end{subequations}
There are three mass matrices that anticommute pairwise.
The topological space of normalized Dirac masses obtained by adding to
the Dirac kinetic contribution a mass matrix 
squaring to the unit matrix and obeying this PHS and this TRS is 
\begin{equation}
\begin{split}
&
V^{\mathrm{CI}}_{d=1,r=4}=
\left\{
\bm{M}^{\,}_{i}\cdot\bm{X}^{\,}_{i}|
\bm{M}^{2}_{i}=1
\right\}=:
S^{2},
\\
&
\bm{M}:=
\left(
M^{\,}_{2,0},
M^{\,}_{1,1},
M^{\,}_{1,3}
\right),
\\
&
\bm{X}:=
\left(
\tau^{\,}_{2}\otimes\sigma^{\,}_{0},
\tau^{\,}_{1}\otimes\sigma^{\,}_{1},
\tau^{\,}_{1}\otimes\sigma^{\,}_{3}
\right).
\end{split}
\end{equation}
As a topological space, $V^{\mathrm{CI}}_{d=1,r=4}$
is homeomorphic to $Sp(1)/U(1)$
since we have the homeomorphism $Sp(1)/U(1)\simeq SU(2)/U(1)\simeq S^{2}$.

\subsection{Interpretation of $\pi^{\,}_{0}(V)=\mathbb{Z},\mathbb{Z}^{\,}_{2},0$}
\label{appendisubsec: Interpretation of ...}

Assume that we have chosen the dimensionality $d$ of space and
the AZ symmetry class. We may then increase the rank $r$
of the Dirac matrices until we reach the smallest rank $r^{\,}_{\mathrm{min}}$
for which we may write the massive Dirac Hamiltonian
\begin{equation}
\mathcal{H}^{\,}_{\mathrm{min}}=
\bm{\alpha}^{\,}_{\mathrm{min}}\cdot\frac{\partial}{\mathrm{i}\partial\bm{x}}
+
m\,\beta^{\,}_{\mathrm{min}}
+
\ldots,
\label{appeq: def massive Dirac Hamilonian}
\end{equation}
where the $(d+1)$ matrices 
$\alpha^{\,}_{\min\,1},\ldots,\alpha^{\,}_{\min\,d},\beta^{\,}_{\min}$
anticommute pairwise and square to the unit 
$r^{\,}_{\mathrm{min}}\times r^{\,}_{\mathrm{min}}$ matrix,
$m$ is a real-valued (mass) parameter,
and ``$\ldots$'' accounts for all scalar and vector gauge contributions
as well as for the possibility of additional mass terms.
Let $V^{\,}_{d,r^{\,}_{\mathrm{min}}}$
be the compact topological space associated with
the normalized Dirac masses that enter on the
right-hand side of Eq.~(\ref{appeq: def massive Dirac Hamilonian}).
There are then three possibilities when
$\pi^{\,}_{0}(V^{\,}_{d,r^{\,}_{\min}})\ne0$.

\subsubsection{Case $\pi^{\,}_{0}(V)\ne0$}
\label{appendix: interpretation of pi_{0}}

Assume that no other normalized
$r^{\,}_{\mathrm{min}}\times r^{\,}_{\mathrm{min}}$
mass matrix than $\beta^{\,}_{\mathrm{min}}$ 
enters the right-hand side of Eq.~(\ref{appeq: def massive Dirac Hamilonian}).
It then follows that 
\begin{equation}
V^{\,}_{d,r^{\,}_{\mathrm{min}}}=
\{\pm\beta^{\,}_{\mathrm{min}}\}.
\label{eq: V_rmin for pi_{0}=Z}
\end{equation}
Hence the ground state of 
$\mathcal{H}^{\,}_{\mathrm{min}}$ for $m>0$
cannot be smoothly deformed into the ground state of 
$\mathcal{H}^{\,}_{\mathrm{min}}$ for $m<0$
without closing the gap proportional to $|m|$. 
There are then three possible outcomes
upon increasing the rank in Eq.~(\ref{appeq: def massive Dirac Hamilonian})
from $r^{\,}_{\mathrm{min}}$ to $r^{\,}_{\mathrm{min}}\,N$ with $N=2,3,\ldots$.

\textbf{Case $\pi^{\,}_{0}(V^{\,}_{d})=\mathbb{Z}$:}
There is one normalized Dirac mass matrix which
commutes with all normalized Dirac mass matrices that have become available 
in the given AZ symmetry class upon increasing $r$ from
from $r^{\,}_{\mathrm{min}}$ to $r^{\,}_{\mathrm{min}}\,N$ with $N=2,3,\ldots$.
We may then write a generic Dirac mass term as 
\begin{equation}
\beta^{\,}_{\mathrm{min}}\otimes M
\equiv
\beta^{\,}_{\mathrm{min}}\otimes
\left(
\frac{2\nu}{N} \openone^{\,}_{N}
+
A
\right),
\label{appeq: m beta if pi0(V)=Z}
\end{equation}
where the $N\times N$ Hermitian matrix $M$ 
squares to the unit $N\times N$ matrix $\openone^{\,}_{N}$, 
the $N\times N$ Hermitian matrix $A$ is traceless, both
$M$ and $A$ are restricted by the AZ symmetry class,
and $\nu$ is the topological index defined in Eq.\ (\ref{eq: nu when z}).
The mass matrix 
$\beta^{\,}_{\mathrm{min}}\otimes\openone^{\,}_{N}$
is unique up to a sign.
The matrix $M$ must obey the polar decomposition
\begin{equation}
\begin{split}
&
M=
U^{\,}_{N-n,n}\,I^{\,}_{N-n,n}\,U^{\dag}_{N-n,n},\\
&
I^{\,}_{N-n,n}=
\mathrm{diag}(+\openone^{\,}_{N-n},-\openone^{\,}_{n}),
\end{split}
\label{appeq: rep for M if pi0(V)=Z}
\end{equation}
where the matrices $U^{\,}_{N-n,n}$ 
form 
either $U(N), O(N),$ or $Sp(N)$ 
depending on the AZ symmetry class.
(For the symplectic case, a matrix in $Sp(N)$ has quaternions
as matrix elements.)
The representation
(\ref{appeq: rep for M if pi0(V)=Z}) is unique
up to multiplication of $U^{\,}_{N-n,n}$ from the right by
the block diagonal
matrix $\mathrm{diag}(U^{\,}_{N-n},U^{\,}_{n})$,
where the $n\times n$ matrix $U^{\,}_{n}$ is
taken from
either $U(n), O(n),$ or $Sp(n)$
depending on the AZ symmetry class.
The representations (\ref{eq: case pi0 when z})
and (\ref{eq: nu when z})
follow.

\textbf{Case of first descendant
$\pi^{\,}_{0}(V^{\,}_{d}=R^{\,}_{1})=\mathbb{Z}^{\,}_{2}$:}
There is a pair of two anticommuting normalized Dirac mass matrices 
that span $V^{\,}_{d,2r^{\,}_{\mathrm{min}}}$
upon increasing $r$ from $r^{\,}_{\mathrm{min}}$ to $2r^{\,}_{\mathrm{min}}$.
We can choose a representation of the Dirac 
mass matrix as follows.
If $N=2,3,\ldots$ and if we define the Hermitian
$r^{\,}_{\mathrm{min}}/2\times r^{\,}_{\mathrm{min}}/2$ matrix
$\rho^{\,}_{\mathrm{min}}$ by
\begin{subequations}
\label{eq: solution to first descendent extension}
\begin{equation}
\beta^{\,}_{\mathrm{min}}=:\rho^{\,}_{\mathrm{min}}\otimes\tau^{\,}_{2},
\label{appeq: m beta if pi0(V)=Z2 first desc with r=r_min}
\end{equation}
we may then write
\begin{equation}
\rho^{\,}_{\mathrm{min}}\otimes M\equiv
\rho^{\,}_{\mathrm{min}}\otimes
\left(
\tau^{\,}_{2}\,
A^{\,}_{2}
+
\tau^{\,}_{1}\,
A^{\,}_{1}
\right)
\label{appeq: m beta if pi0(V)=Z2 first desc}
\end{equation}
\end{subequations}
for the generic normalized Dirac mass.
Here, the $2N\times2N$ matrix $M$ is Hermitian and antisymmetric 
because the $N\times N$ matrix $A^{\,}_{2}$ is Hermitian and symmetric
($A^{\,}_{2}$ commutes with the operation $\mathsf{K}$ for complex conjugation),
while the $N\times N$  matrix $A^{\,}_{1}$ is 
Hermitian and antisymmetric
($A^{\,}_{1}$ anticommutes with the operation $\mathsf{K}$ 
for complex conjugation).
The matrix $M$ is a solution of the extension problem
\begin{align}
Cl^{\,}_{1,2}=
\{
\mathrm{i}\tau^{\,}_{3};\mathsf{K},\mathrm{i}\mathsf{K} 
\}
\to 
Cl^{\,}_{1,3}=
\{
\mathrm{i}\tau^{\,}_{3};\mathsf{K},\mathrm{i}\mathsf{K},M 
\}.
\label{eq: first descendent extension}
\end{align}
Indeed, it is the matrix $\mathrm{i}\tau^{\,}_{3}$ 
that enters in the kinetic contribution to the Dirac Hamiltonian
in the example~(\ref{class D H(k) r=4})
for a one-dimensional Dirac Hamiltonian in the symmetry class D.
The explicit representation of
Eq.~(\ref{eq: case pi0 when z2}) for this example
thus follows from solving with 
Eq.~(\ref{eq: solution to first descendent extension})
the extension problem~(\ref{eq: first descendent extension}).

It is instructive to  compare
Eqs.\ (\ref{appeq: m beta if pi0(V)=Z2 first desc with r=r_min})
and (\ref{appeq: m beta if pi0(V)=Z2 first desc})
with the explicit representations
(\ref{eq: D 2x2 b}) and (\ref{class D H(k) r=4})
corresponding to the one-dimensional symmetry class D with
$r=r^{\,}_{\min}$ and $r=2r^{\,}_{\min}$, respectively. 
This comparison suggests that
the choice $A^{\,}_{2}=\pm\openone^{\,}_{N}$ and $A^{\,}_{1}=0$
is special for any odd $N$. Indeed, one may verify that
it is impossible to construct an antisymmetric matrix $A^{\,}_{1}$ of
odd rank $N$ that connects smoothly, without closing a gap,
between the choices
$A^{\,}_{2}=+\openone^{\,}_{N}$ and $A^{\,}_{1}=0$
on the one hand, and 
$A^{\,}_{2}=-\openone^{\,}_{N}$ and $A^{\,}_{1}=0$ on the other hand.
Hence the choices $A^{\,}_{2}=\pm\openone^{\,}_{N}$ and $A^{\,}_{1}=0$
represent two topologically distinct phases when $N$ is odd.

Finally, we observe that 
there is no unique Dirac mass matrix that commutes
with all other Dirac mass matrices for $N>1$.
For example, the Dirac mass matrix with $A^{\,}_{2}=\openone^{\,}_{3}$ 
and $A^{\,}_{1}=0$ neither commutes nor anticommutes
with the Dirac mass matrix for which
\begin{align}
A^{\,}_{2}=
\begin{pmatrix}
\cos \theta & 0 & 0 \\
0 & \cos \theta & 0 \\
0 & 0 & 1
\end{pmatrix}, 
\quad
A^{\,}_{1} = 
\sin \theta
\begin{pmatrix}
0 & -\mathrm{i} & 0 \\
+\mathrm{i} & 0 & 0 \\
0 & 0 & 0
\end{pmatrix}.
\end{align}

\textbf{Case of second descendant
$\pi^{\,}_{0}(V^{\,}_{d}=R^{\,}_{2})=\mathbb{Z}^{\,}_{2}$:}
There is a pair of three anticommuting normalized Dirac mass matrices 
that span $V^{\,}_{d,2r^{\,}_{\mathrm{min}}}$ upon increasing $r$
from $r^{\,}_{\mathrm{min}}$ to $2r^{\,}_{\mathrm{min}}$.
We can choose a representation of the Dirac 
mass matrix as follows.
If $N=2,3,\ldots$ and if
we define the Hermitian
$r^{\,}_{\mathrm{min}}/2\times r^{\,}_{\mathrm{min}}/2$ matrix
$\rho^{\,}_{\mathrm{min}}$ by
\begin{subequations}
\begin{equation}
\beta^{\,}_{\mathrm{min}}=:\rho^{\,}_{\mathrm{min}}\otimes\tau^{\,}_{2},
\label{appeq: m beta if pi0(V)=Z2 secon desc N=1}
\end{equation}
we may then write 
\begin{equation}
\rho^{\,}_{\mathrm{min}}\otimes M\equiv
\rho^{\,}_{\mathrm{min}}\otimes
\left(
\tau^{\,}_{2}\,
A^{\,}_{2}
+
\tau^{\,}_{3}\,
A^{\,}_{3}
+
\tau^{\,}_{1}\,
A^{\,}_{1}
+
\tau^{\,}_{0}\,
A^{\,}_{0}
\right)
\label{appeq: m beta if pi0(V)=Z2 secon desc}
\end{equation}
\end{subequations}
for the generic Dirac mass matrix.
Here, the $2N\times2N$ matrix $M$ is Hermitian and antisymmetric 
because the $N\times N$ matrix $A^{\,}_{2}$ is Hermitian and symmetric
($A^{\,}_{2}$ commutes with the operation $\mathsf{K}$ for complex conjugation),
while the $N\times N$ matrix $A^{\,}_{3}$, $A^{\,}_{1}$, and $A^{\,}_{0}$
are Hermitian and antisymmetric
($A^{\,}_{3}$ and $A^{\,}_{1}$ 
anticommute with the operation $\mathsf{K}$ for complex conjugation).
The matrix $M$ is a solution of the extension problem
\begin{align}
Cl^{\,}_{0,2}=\{; \mathsf{K}, \mathrm{i}\mathsf{K} \}
\to 
Cl^{\,}_{0,3}=\{; \mathsf{K}, \mathrm{i}\mathsf{K}, M \},
\end{align}
which constrains $M$ to be pure imaginary,
as can be verified for
Eq.~(\ref{appeq: m beta if pi0(V)=Z2 secon desc}).
The representation (\ref{eq: case pi0 when z2}) follows.

The normalized Dirac mass matrix
(\ref{appeq: m beta if pi0(V)=Z2 secon desc N=1}) should be
compared with that in Eq.~(\ref{eq: DIII 4x4})
for the one-dimensional symmetry class DIII with $r=4$,
and Eq.~(\ref{appeq: m beta if pi0(V)=Z2 secon desc}) 
should be compared with the
normalized Dirac masses for the one-dimensional symmetry class DIII 
with $r=8$ that we now present.
If we impose PHS and TRS through
\begin{subequations}
\label{eq: mathcal H if DII and r=8}
\begin{align}
\mathcal{H}(k)&=
-
\mathcal{H}^{*}(-k),
\\
\mathcal{H}(k)&=
+
\tau^{\,}_{2}\otimes\sigma^{\,}_{0} \otimes\rho^{\,}_{0} \,
\mathcal{H}^{*}(-k)\,
\tau^{\,}_{2}\otimes\sigma^{\,}_{0} \otimes\rho^{\,}_{0},
\end{align}
then
\begin{align}
\mathcal{H}(k)=&\,
\tau^{\,}_{3}\otimes\sigma^{\,}_{0} \otimes\rho^{\,}_{0}\,k  \n
&+
\sum_{\nu=0,1,3}
\tau^{\,}_{3}\otimes(\sigma^{\,}_{2}\otimes\rho^{\,}_{\nu}\,A^{\,}_{1,2,\nu}
+
\sigma^{\,}_{\nu}\otimes\rho^{\,}_{2}\,A^{\,}_{1,\nu,2}
) \n
&+
\sum_{\nu=0,1,3}
\tau^{\,}_{1}\otimes(\sigma^{\,}_{2}\otimes\rho^{\,}_{\nu}\,M^{\,}_{1,2,\nu}
+
\sigma^{\,}_{\nu}\otimes\rho^{\,}_{2}\,M^{\,}_{1,\nu,2}
).
\end{align}
\end{subequations}
The topological space of normalized Dirac masses obtained by adding to
the Dirac kinetic contribution a mass matrix 
squaring to the unit matrix and obeying 
PHS squaring to unity is
\begin{equation}
\begin{split}
&
V^{\mathrm{DIII}}_{d=1,r=8}=
\left\{
\bm{M}\cdot\bm{X}|
\bm{M}^{2}=1
\right\}
\cup
\left\{
\bm{N}\cdot\bm{Y}|
\bm{N}^{2}=1
\right\}
,
\\
&
\bm{M}:=
\left(
M^{\,}_{1,2,0},
M^{\,}_{1,1,2},
M^{\,}_{1,3,2}
\right),
\\
&
\bm{N}:=
\left(
M^{\,}_{1,0,2},
M^{\,}_{1,2,1},
M^{\,}_{1,2,3}
\right),
\\
&
\bm{X}:=
\left(
\tau^{\,}_{1}\otimes\sigma^{\,}_{2}\otimes\rho^{\,}_{0},
\tau^{\,}_{1}\otimes\sigma^{\,}_{1}\otimes\rho^{\,}_{2},
\tau^{\,}_{1}\otimes\sigma^{\,}_{3}\otimes\rho^{\,}_{2}
\right),
\\
&
\bm{Y}:=
\left(
\tau^{\,}_{1}\otimes\sigma^{\,}_{0}\otimes\rho^{\,}_{2},
\tau^{\,}_{1}\otimes\sigma^{\,}_{2}\otimes\rho^{\,}_{1},
\tau^{\,}_{1}\otimes\sigma^{\,}_{2}\otimes\rho^{\,}_{3}
\right).
\end{split}
\label{eq: DIII 8x8 normalized masses}
\end{equation}
As a topological space, 
$V^{\mathrm{DIII}}_{d=1,r=8}\simeq S^{2}\cup S^{2}$
is homeomorphic to $O(4)/U(2)$,
for the homeomorphisms
\begin{align}
O(4)/U(2)\simeq&\,
\mathbb{Z}^{\,}_{2}\times SO(4)/[U(1)\times SU(2)]
\nonumber\\
\simeq&\, 
\mathbb{Z}^{\,}_{2}\times[SO(3)\times SU(2)]/[SO(2)\times SU(2)]
\nonumber\\
\simeq&\,
\mathbb{Z}^{\,}_{2}\times SO(3)/SO(2)
\nonumber\\
\simeq&\,
S^{2}\cup S^{2}
\end{align}
hold between topological spaces.
In order to compare the Dirac mass matrices entering
Eq.~(\ref{eq: mathcal H if DII and r=8}) with
Eq.~(\ref{appeq: m beta if pi0(V)=Z2 secon desc}),
notice first that
$\rho^{\,}_{\min}=\tau^{\,}_{1}$
and second that
$\sigma^{\,}_{\nu}$ in
Eq.~(\ref{eq: DIII 8x8 normalized masses}) plays 
the role of $\tau^{\,}_{\nu}$
in Eq.~(\ref{appeq: m beta if pi0(V)=Z2 secon desc}).

\subsubsection{Case $\pi^{\,}_{0}(V)=0$}

There is at least one 
$r^{\,}_{\mathrm{min}}\times r^{\,}_{\mathrm{min}}$ 
Hermitian matrix $\beta^{\prime}_{\mathrm{min}}$ that
enters ``$\ldots$'' in Eq.~(\ref{appeq: def massive Dirac Hamilonian})
such that it anticommutes with any one of the
$(d+1)$ matrices 
$
\alpha^{\,}_{\mathrm{min}\,1},
\ldots,
\alpha^{\,}_{\mathrm{min}\,d},\beta^{\,}_{\mathrm{min}}$
and it squares to the unit $r^{\,}_{\mathrm{min}}\times r^{\,}_{\mathrm{min}}$
matrix. For any $0\leq\theta<2\pi$,
we may then define the normalized mass matrix
\begin{equation}
\beta^{\,}_{\mathrm{min}}(\theta):=
\cos\theta\,\beta^{\,}_{\mathrm{min}}
+
\sin\theta\,\beta^{\prime}_{\mathrm{min}}
\end{equation}
that provides a smooth path between 
$\beta^{\,}_{\mathrm{min}}$ 
and 
$-\beta^{\,}_{\mathrm{min}}$.
Since $\beta^{\,}_{\mathrm{min}}$ 
can be chosen arbitrarily in the compact topological space
$V$ associated with the massive Dirac Hamiltonian
(\ref{appeq: def massive Dirac Hamilonian}),
we can rule out the existence of the disconnected subspace
$\{\pm\beta^{\,}_{\mathrm{min}}\}$ in 
$V^{\,}_{d,r^{\,}_{\mathrm{min}}}$.
For general values of $N=1,2,3,\ldots$, 
the trivial zeroth homotopy group 
$\pi^{\,}_{0}(V^{\,}_{d,r^{\,}_{\min}})=0$ indicates that
for any given pair 
$(\beta,\beta')\in 
V^{\,}_{d,r^{\,}_{\min}N}\times V^{\,}_{d,r^{\,}_{\min}N}$
(not necessarily anticommuting),
we find the sequence of anticommuting pairs
$(\beta,\beta^{\,}_{1})\in 
V^{\,}_{d,r^{\,}_{\min}N}\times V^{\,}_{d,r^{\,}_{\min}N}$,
$(\beta^{\,}_{1},\beta^{\,}_{2})\in 
V^{\,}_{d,r^{\,}_{\min}N}\times V^{\,}_{d,r^{\,}_{\min}N}$,
$\ldots$,
$(\beta^{\,}_{n},\beta^{\prime})\in 
V^{\,}_{d,r^{\,}_{\min}N}\times V^{\,}_{d,r^{\,}_{\min}N}$
as occurs in Eq.~(\ref{eq: 1D Symmetry class C rmin}).

\subsection{Existence condition for the Dirac masses}
\label{appendixsubsec: Existence condition of ...}

Assume that we are given Eq.~(\ref{appeq: def massive Dirac Hamilonian}).
The question posed and solved in Sec.%
~\ref{sec: Classifying spaces of normalized Dirac masses}
was: 
\begin{enumerate}
\item[\textbf{Q1}]
What is the compact topological space spanned by the normalized Dirac
masses of rank $r=r^{\,}_{\mathrm{min}}\,N$ in the limit
$N\to\infty$? 
\end{enumerate}

By definition, the Dirac Hamiltonian~(\ref{appeq: def massive Dirac Hamilonian})
realizes a topological insulator if the compact topological space $V$
spanned by the normalized Dirac mass $\beta^{\,}_{\mathrm{min}}$
consists of no more than the set $\{\pm\beta^{\,}_{\mathrm{min}}\}$.
It then follows that,
on any $(d-1)$-dimensional boundary of $d$-dimensional space,
the Dirac Hamiltonian~(\ref{appeq: def massive Dirac Hamilonian})
reduces to a massless Dirac Hamiltonian of the form
\begin{equation}
\widetilde{\mathcal{H}}=
\widetilde{\bm{\alpha}}\cdot\frac{\partial}{\mathrm{i}\partial\bm{x}}
+
\ldots,
\label{appeq: def massless Dirac Hamilonian}
\end{equation}
where the $(d-1)$ matrices 
$\widetilde{\alpha},\ldots,\widetilde{\alpha}$
anticommute pairwise and square to the unit,
their rank is
$r=r^{\,}_{\mathrm{min}}/2$,
and ``$\ldots$'' accounts for scalar and vector gauge contributions,
but ``$\ldots$''  does not contain Dirac mass terms. Moreover, by the definition of
the minimal rank $r^{\,}_{\mathrm{min}}$, no Dirac mass is permissible
upon lowering the rank of the Dirac matrices entering
the Dirac Hamiltonian~(\ref{appeq: def massive Dirac Hamilonian})
holding the dimensionality of space and the AZ symmetry class fixed. 
These observations motivate the following question:
\begin{enumerate}
\item[\textbf{Q2}]
Given an AZ symmetry class and given the dimensionality $d$ of space,
is a Dirac mass of rank $\widetilde{r}^{\,}_{\mathrm{min}}$ 
permissible or not?
\end{enumerate}
Here, $\widetilde{r}^{\,}_{\mathrm{min}}$
is the minimal rank for which it is possible to write down the
kinetic contribution 
$-\mathrm{i}\widetilde{\bm{\alpha}}\cdot\partial/\partial\bm{x}$
to the Dirac Hamiltonian 
given the dimensionality $d$ of space and the AZ symmetry class.

We want to know if a massless Dirac Hamiltonian belonging to some prescribed
$d$-dimensional AZ symmetry class 
accommodates a Dirac mass. To answer this question, we recall that Eqs.%
~(\ref{eq: complex Clifford algebra})
and
(\ref{eq: real Clifford algebra})
define for any AZ symmetry class and for any dimension $d$
a Clifford algebra that supports at least one normalized Dirac mass,
provided the rank $r$ of the Dirac matrices is no less than the minimal rank
$r^{\,}_{\mathrm{min}}$. 
If we replace $r^{\,}_{\mathrm{min}}$ by $\widetilde{r}^{\,}_{\mathrm{min}}$,
the answer to question $\textbf{Q2}$ is given by the extension problems 
(the values of $p$ and $q$ are taken from Table \ref{table: AZ classes})
\begin{subequations}
\label{appeq: existence condition}
\begin{equation}
\begin{split}
Cl^{\,}_{q-1}&=
\{e^{\,}_{1},\ldots,e^{\,}_{q-1}\}
\\
&
\to 
\{e^{\,}_{1},\ldots,e^{\,}_{q-1},e^{\,}_{q}\}=
Cl^{\,}_{q}
\end{split}
\label{appeq: existence condition A}
\end{equation}
for the complex symmetry classes A and AIII;
\begin{equation}
\begin{split}
Cl^{\,}_{p,q-1}&=
\{e^{\,}_{1},\ldots,e^{\,}_{p};e^{\,}_{p+1},\ldots,e^{\,}_{p+q-1}\}
\\
&\to 
\{e^{\,}_{1},\ldots,e^{\,}_{p};e^{\,}_{p+1},\ldots,e^{\,}_{p+q-1},
                e^{\,}_{p+q}\}=
Cl^{\,}_{p,q}
\end{split}
\label{appeq: existence condition B}
\end{equation}
for the real symmetry classes BDI, D, DIII, CII, C and CI; and
\begin{equation}
\begin{split}
Cl^{\,}_{p-1,q}&=
\{e^{\,}_{1},\ldots,e^{\,}_{p-1};e^{\,}_{p},\ldots,e^{\,}_{p+q-1}\}
\\
&\to 
\{e^{\,}_{1},\ldots,e^{\,}_{p-1},e^{\,}_{p};e^{\,}_{p+1},\ldots,e^{\,}_{p+q}\}=
Cl^{\,}_{p,q}
\end{split}
\label{appeq: existence condition C}
\end{equation}
\end{subequations}
for the real symmetry classes AI and AII. 
In other words, for any given AZ symmetry class, we seek 
to enumerate all the distinct ways there are to construct the 
corresponding Clifford algebra from 
Eqs.~(\ref{eq: complex Clifford algebra bis})
and
(\ref{eq: real Clifford algebra bis})
out of one in the same symmetry class but with one fewer generator. 
Recall here that the Clifford algebras 
(\ref{eq: complex Clifford algebra bis})
and
(\ref{eq: real Clifford algebra bis})
are obtained after removing the generator associated with
the normalized Dirac mass in each of the Clifford algebras
~(\ref{eq: complex Clifford algebra})
and
(\ref{eq: real Clifford algebra}).  
The solutions to Eqs.~(\ref{appeq: existence condition})
are the classifying spaces
(the values of $p$ and $q$ are taken from Table \ref{table: AZ classes})
\begin{equation}
\widetilde{V}=C^{\,}_{q-1},
\qquad
\widetilde{V}=R^{\,}_{q-p-1},
\qquad
\widetilde{V}=R^{\,}_{p-q+1},
\label{appeq: solution to existence}
\end{equation}
respectively.
On the one hand,
if $\pi^{\,}_{0}(\widetilde{V})=\mathbb{Z},\mathbb{Z}^{\,}_{2}$,
then the generator $e^{\,}_{p+q}$ from the corresponding
Clifford algebra is unique in that
no additional generator exists that anticommutes with 
$e^{\,}_{p+q}$ (see Sec.~\ref{appendisubsec: Interpretation of ...}), i.e.,
no additional generator $e^{\,}_{p+q+1}$ that plays the role of
a normalized Dirac mass represented by a Hermitian matrix of rank 
$\widetilde{r}^{\,}_{\mathrm{min}}$
is allowed. On the other hand, 
if $\pi^{\,}_{0}(\widetilde{V})=0$,
then the generator $e^{\,}_{p+q}$ from the corresponding
Clifford algebra is not unique in that
there exist additional independent generators that anticommute with 
$e^{\,}_{p+q}$ (see Sec.~\ref{appendisubsec: Interpretation of ...}),
i.e., a generator $e^{\,}_{p+q+1}$ that plays the role of
a normalized Dirac mass represented by a Hermitian matrix of rank 
$\widetilde{r}^{\,}_{\mathrm{min}}$
is allowed.

For comparison, the answer to 
question $\textbf{Q1}$ was given in Table \ref{table: AZ classes}
by the extension problems
(the values of $p$ and $q$ are taken from Table \ref{table: AZ classes})
\begin{subequations}
\label{appeq: classification condition}
\begin{equation}
\begin{split}
Cl^{\,}_{q}&=
\{e^{\,}_{1},\ldots,e^{\,}_{q}\}
\\
&\to 
\{e^{\,}_{1},\ldots,e^{\,}_{q},e^{\,}_{q+1}\}=
Cl^{\,}_{q+1}
\end{split}
\label{appeq: classification condition A}
\end{equation}
for the complex symmetry classes A and AIII;
\begin{equation}
\begin{split}
Cl^{\,}_{p,q}&=
\{e^{\,}_{1},\ldots,e^{\,}_{p};e^{\,}_{p+1},\ldots,e^{\,}_{p+q}\}
\\
&\to 
\{e^{\,}_{1},\ldots,e^{\,}_{p};e^{\,}_{p+1},\ldots,e^{\,}_{p+q},
                e^{\,}_{p+q+1}\}=Cl^{\,}_{p,q+1}
\end{split}
\label{appeq: classification condition B}
\end{equation}
for the real symmetry classes BDI, D, DIII, CII, C and CI; and
\begin{equation}
\begin{split}
Cl^{\,}_{p,q}&=
\{e^{\,}_{1},\ldots,e^{\,}_{p};e^{\,}_{p+1},\ldots,e^{\,}_{p+q}\}
\\
&\to 
\{e^{\,}_{1},\ldots,e^{\,}_{p},e^{\,}_{p+1};e^{\,}_{p+2},\ldots,
                e^{\,}_{p+q+1}\}=
Cl^{\,}_{p+1,q}
\end{split}
\label{appeq: classification condition C}
\end{equation}
\end{subequations}
for the real symmetry classes AI and AII. 
The solutions to Eqs.~(\ref{appeq: classification condition})
are the classifying spaces
(the values of $p$ and $q$ are taken from Table \ref{table: AZ classes})
\begin{equation}
V=C^{\,}_{q},
\qquad
V=R^{\,}_{q-p},
\qquad
V=R^{\,}_{p-q+2},
\label{appeq: solutions to classification}
\end{equation}
respectively.
Thus, the classifying space associated with $\textbf{Q2}$ 
in $d$ dimensions coincides with 
the classifying space associated with $\textbf{Q1}$ in $d+1$ dimensions.
The existence of nontrivial topological insulators in $d+1$ dimensions
and the gapless Dirac Hamiltonian with no allowed Dirac mass in $d$ dimensions 
are equivalent.

The final answer to the question \textbf{Q2} is:
\begin{itemize}
\item 
No Dirac mass matrix of rank
$\widetilde{r}^{\,}_{\mathrm{min}}$
is permissible when
$\pi^{\,}_{0}(\widetilde V)=\mathbb{Z}$.
If $N$ channels are added by generalizing the Dirac matrices
$\widetilde{\alpha}$ 
to
$\widetilde{\bm{\alpha}}\otimes\openone^{\,}_{N}$,
then no Dirac mass of rank
$\widetilde{r}^{\,}_{\mathrm{min}}\,N$ is permissible
when
$\pi^{\,}_{0}(\widetilde V)=\mathbb{Z}$.
\item 
No Dirac mass matrix of rank 
$\widetilde{r}^{\,}_{\mathrm{min}}$ 
is permissible when
$\pi^{\,}_{0}(\widetilde V)=\mathbb{Z}^{\,}_{2}$.
If $N$ channels are added by generalizing the Dirac matrices
$\widetilde{\alpha}$ 
to
$\widetilde{\bm{\alpha}}\otimes\openone^{\,}_{N}$,
then no Dirac mass of rank
$\widetilde{r}^{\,}_{\mathrm{min}}\,N$
is permissible for odd $N$,
while a Dirac mass of rank
$\widetilde{r}^{\,}_{\mathrm{min}}\,N$
is permissible for even $N$, when
$\pi^{\,}_{0}(\widetilde V)=\mathbb{Z}^{\,}_{2}$.
\item 
A Dirac mass matrix of rank 
$\widetilde{r}^{\,}_{\mathrm{min}}$ 
is permissible when
$\pi^{\,}_{0}(\widetilde V)=0$. 
If $N$ channels are added by generalizing the Dirac matrices
$\widetilde{\alpha}$ 
to
$\widetilde{\bm{\alpha}}\otimes\openone^{\,}_{N}$,
then a Dirac mass matrix of rank 
$\widetilde{r}^{\,}_{\mathrm{min}}\,N$ 
is permissible when
$\pi^{\,}_{0}(\widetilde V)=0$.
\end{itemize}
Table \ref{table: surface localization} from
Sec.~\ref{sec: Boundaries of topological insulators} follows.

\section{One-loop renormalization group analysis in 2D space}
\label{app: OPE}

We summarize the one-loop RG flows associated with
the relevant 2D random Dirac Hamiltonians
from Refs.~\onlinecite{ludwig94,Bernard95}
that we have used to deduce the global phase diagrams
presented in Sec.~\ref{sec: Application to 2D space}.

For simplicity, we present RG flows for vanishing
mean values of all random potentials. A mean value for any
random mass is a strongly relevant perturbation. This
is the control parameter for the 2D phase diagrams
along the horizontal axis. Although we do not present the
contributions to the RG flows from non-vanishing mean values 
of the random masses, their contributions are essential
to cross phase boundaries. They are thus implicitly assumed
to be present.

\subsection{Replica limit}
\label{app: Replica limit}
 
Define the path integral
\begin{subequations}
\begin{align}
&
Z:=
\int
\mathcal{D}[\psibar] 
\mathcal{D}[\psi]\, 
\exp(-S), 
\\
&
S:=
\mathrm{i} 
\int\mathrm{d}^{2}\bm{r}\, 
\left(
\psibar\,\mathcal{H}^{\,}_{0}\,\psi 
+ 
\sum_{\iota} 
V^{\,}_{\iota}\,\phi^{\,}_{\iota}(\psibar,\psi)
\right), 
\\
&
\mathcal{H}^{\,}_{0}:=
\sigma^{\,}_{x}(-\mathrm{i}\partial^{\,}_{x}) 
+ 
\sigma^{\,}_{y}(-\mathrm{i}\partial^{\,}_{y}).
\end{align}
Here, $\phi^{\,}_{\iota}(\psibar,\psi)$
is a local quadratic form for the pair 
$\psibar$ and $\psi$
of independent Grassmann-valued spinors.
For any pair $\iota,\iota'$, the real-valued functions 
$V^{\,}_{\iota}$  and $V^{\,}_{\iota'}$ 
are random with vanishing means and with  
\begin{align}
\overline{V^{\,}_{\iota}(\bm{r})\,V^{\,}_{\iota'}(\bm{r}')}= 
\delta^{\,}_{\iota,\iota'}\,
g^{\,}_{\iota}\,
\delta(\bm{r}-\bm{r}')
\end{align}
\end{subequations}
as the only non-vanishing moments.

In order to compute the disorder-average over the correlations functions
for the Grassmann-valued spinors, it is convenient to take the replica limit
\begin{equation}
\ln Z=
\lim_{n\to0}\frac{Z^{n}-1}{n}.
\label{eq: replica limit}
\end{equation}
If we reserve the index $a,b=1,\ldots,n$, we may write 
\begin{subequations}
\label{eq: replica L}
\begin{align}
&
\overline{Z^{n}}=
\int 
\mathcal{D}[\psibar] 
\mathcal{D}[\psi] 
\exp 
\left( 
-
\int\mathrm{d}^{2}\bm{r}\,
\left(
L^{\,}_{0} 
+  
L^{\,}_{\mathrm{int}}
\right) 
\right),
\\
&
L^{\,}_{0}:= 
\mathrm{i} 
\sum_{a=1}^{n} 
\psibar^{a}
\left[
\sigma^{\,}_{x}\, 
(-\mathrm{i}\partial^{\,}_{x}) 
+ 
\sigma^{\,}_{y}\, 
(-\mathrm{i}\partial^{\,}_{y})
\right]
\psi^{a}, 
\\
&
L^{\,}_{\mathrm{int}}:=
\sum_{\iota} 
\frac{g^{\,}_{\iota}}{2}
\sum_{a,b=1}^{n} 
\phi^{a}_{\iota}\,
\phi^{b}_{\iota},
\label{quartic 1}
\end{align}
for the disorder average over the $n$th power of the partition function.
The effective action
\begin{equation}
S^{\,}_{\mathrm{eff}}:=
\int\mathrm{d}^{2}\bm{r}\,
\left(
L^{\,}_{0} 
+  
L^{\,}_{\mathrm{int}}
\right) 
\end{equation}
\end{subequations}
describes an interacting and nonunitary quantum field theory.%
~\cite{Bernard95}
Disorder averaged correlation functions 
for the Grassmann fields are computed
by first using the effective partition function
(\ref{eq: replica L})
and then by taking the replica limit (\ref{eq: replica limit}).

\subsection{One-loop renormalization group $\beta$ functions
from the operator product expansion}

Assume that space is the $d$-dimensional Euclidean space.
Define the path integral
\begin{subequations}
\begin{align}
&
Z:=
\int
\mathcal{D}[\psibar] 
\mathcal{D}[\psi]\, 
\exp(-S^{\,}_{0}-S^{\,}_{\mathrm{int}}), 
\\
&
S^{\,}_{0}:=
\mathrm{i}
\int\mathrm{d}^{d}\bm{r}\, 
\psibar\,
\bm{\alpha}
\cdot
\frac{\partial}{\partial\bm{x}}
\psi,
\\
&
S^{\,}_{\mathrm{int}}:=
\sum_{a}
\lambda^{\,}_{a}
\int \frac{\mathrm{d}^{d}\bm{r}}{2\pi\mathfrak{a}^{d-2}}\, 
\Phi^{\,}_{a}(\psibar,\psi),
\label{quartic 2}
\end{align}
where $\bm{\alpha}$ are Dirac matrices that anticommute pairwise
and square to the identity matrix,
$\mathfrak{a}$ is the short-distance cutoff,
$\Phi^{\,}_{a}(\psibar,\psi)$ is a local monomial 
of the Grassmann spinors,
and $\lambda^{\,}_{a}$ is the corresponding
dimensionless coupling. Assume the operator product expansion
(OPE)
\begin{align}
\Phi^{\,}_{a}(\bm{r})\,
\Phi^{\,}_{b}(\bm{r}')=
\frac{c^{\,}_{abc}}{|\bm{r}-\bm{r}'|^{x^{\,}_{a}+x^{\,}_{b}-x^{\,}_{c}}}\,
\Phi^{\,}_{c}(\bm{r}')
+
\ldots,
\end{align}
where the summation convention over repeated indices
is implied, the tensor $c^{\,}_{abc}$ is real valued,
the scaling exponent $x^{\,}_{a}$ of $\Phi^{\,}_{a}(\bm{r})$
is real valued, and ``$\ldots$'' contains all functions
of $|\bm{r}-\bm{r}'|$ that are subleading relative to 
$|\bm{r}-\bm{r}'|^{x^{\,}_{a}+x^{\,}_{b}-x^{\,}_{c}}$ in the limit
$\bm{r}'\to\bm{r}$. The one-loop RG flows for the coupling constants is then
\begin{equation}
\frac{\mathrm{d}\lambda^{\,}_{a}}{\mathrm{d}\ell}=
\left(d-x^{\,}_{a}\right)\,
\lambda^{\,}_{a}
+
\frac{1}{2}
c^{\,}_{abc}\,
\lambda^{\,}_{b}\,
\lambda^{\,}_{c},
\end{equation}
under the rescaling $\mathfrak{a}\to\mathfrak{a}(1+\mathrm{d\ell})$
of the short distance cutoff. 
When $d=2$, $\Phi^{\,}_{a}$ is quartic in the Grassmann spinors,
and, after performing the rescaling
\begin{equation}
\lambda^{\,}_{a}=:\pi\,g^{\,}_{a},
\end{equation}
we find the one-loop $\beta$-functions
\begin{align}
\beta^{\,}_{a}\equiv
\frac{\mathrm{d}g^{\,}_{a}}{\mathrm{d}\ell}=
\frac{\pi}{2}\,
c^{\,}_{abc}\,
g^{\,}_{b}\, 
g^{\,}_{c}.
\end{align}
\end{subequations}

\subsection{Renormalization group flows for the AZ symmetry classes}

One-loop RG flows for random masses and random gauge potentials
of 2D Dirac Hamiltonians~\cite{ludwig94,Bernard95}
are used in this paper to decide whether
a metallic phase or a critical boundary separates
topologically distinct insulating phases
in the phase diagrams of Fig.~\ref{Fig: 2D phase diagrams}.
For the symmetry classes in which metallic conductivities acquire
anti-weak localization corrections,
a metallic phase is stable for large bare values of $\mathsf{g}$.
Whether a metallic phase persists 
from large to infinitesimally small bare values of $\mathsf{g}$
along the boundary between topologically distinct insulating phases
is determined by the RG flows in the vicinity of $\mathsf{g}=0$.
If the characteristic disorder strength $\mathsf{g}$ is marginally relevant,
then it is a metallic phase that separates two topologically
distinct insulating phases for any $\mathsf{g}>0$.
(Evidently, we are assuming no intermediary fixed points
between the ones at $\mathsf{g}=\infty$ and $\mathsf{g}=0$.)
On the other hand, if the disorder strength $\mathsf{g}$
is marginally irrelevant, then it is a critical boundary that separates 
two topologically distinct insulating phases.
(In this case there has to be a critical point where the
phase boundary is terminated by a metallic phase.)

For these reasons, we revisit one-loop RG flows of random masses and 
random gauge potentials for those symmetry classes that are characterized
by anti-weak localization corrections,
i.e., the symmetry classes D, DIII, and AII.~\cite{Evers-Mirlin-rmp}
The RG equations are obtained from OPEs of operators associated
with all random masses and random gauge potentials allowed 
in each symmetry class. The OPEs are calculated by
taking contractions in a product of two fermion quartic terms
in Eqs.\ (\ref{quartic 1}) and (\ref{quartic 2})
using the two-point correlation function of free Dirac fields,
\begin{align}
\braket{\psi^{a}(\bm{r})\,\psibar^{b}(\bm{0})}=
\frac{\bm{r}\cdot\bm{\sigma}}{2\pi\,|r|^{2}}\,
\delta^{\,}_{ab}.
\end{align}
To begin with, we first consider symmetry classes A and C
and then move on to symmetry classes D, DIII, and AII.

\subsubsection{Symmetry class A}

Following Ref.~\onlinecite{ludwig94},
we consider the 2D random Dirac Hamiltonian 
\begin{align}
\mathcal{H}:=
\sigma^{\,}_{x} 
\left(
-\mathrm{i}\partial^{\,}_{x}+A^{\,}_{x}
\right) 
+ 
\sigma^{\,}_{y} 
\left(
-\mathrm{i}\partial^{\,}_{y}+A^{\,}_{y}
\right)
+
m\,
\sigma^{\,}_{z} 
+
V
\end{align}
of rank $r=2$ in the symmetry class A.

We seek the one-loop RG equations obeyed 
by the coupling constants
\begin{subequations}
\begin{align}
g^{\,}_{M}&=
\overline{m^{2}},
\\
g^{\,}_{V}&=
\overline{V^{2}}, 
\\
g^{\,}_{A}&=
\overline{A^{2}_{x}}+\overline{A^{2}_{y}}.
\end{align}
\end{subequations}
These couplings are the coefficients of  
the fermion quartic terms
\begin{subequations} 
\label{quartic terms in class A}
\begin{align}
\Phi^{\,}_{M}&=
-\sum_{a,b=1}^{n} 
:\!\psibar^{a}\,\sigma^{\,}_{z}\psi^{a}\,\psibar^{b}\,\sigma^{\,}_{z}\psi^{b}\!:,
\\
\Phi^{\,}_{V}&=
-\sum_{a,b=1}^{n} 
:\!\psibar^{a}\,\psi^{a}\,\psibar^{b}\,\psi^{b}\!:, 
\\
\Phi^{\,}_{A}&=
-\sum_{a,b=1}^{n} \frac{1}{2}
(:\!\psibar^{a}\sigma^{\,}_{x} \psi^{a} \psibar^{b} \sigma^{\,}_{x} \psi^{b}\!:
+
:\!\psibar^{a}\sigma^{\,}_{y}\psi^{a} \psibar^{b} \sigma^{\,}_{y} \psi^{b}\!:),
\end{align}
\end{subequations}
where the interactions
$\Phi^{\,}_{V}$,
$\Phi^{\,}_{A}$,
and $\Phi^{\,}_{M}$
are generated by taking the disorder average over
the random scalar potential, the random vector potentials,
and the random mass, respectively.
Their non-vanishing OPEs are~\cite{Bernard95,ryu-global-phase12}
\begin{subequations}
\label{eq: OPEs in A}
\begin{align}
\Phi^{\,}_{M}(\bm{r})\,\Phi^{\,}_{M}(\bm{0})&= 
\frac{-2}{\pi^{2}\,r^{2}}\,
\Phi^{\,}_{M}(\bm{0}),
\label{eq: OPEs in A a} 
\\
\Phi^{\,}_{V}(\bm{r})\,\Phi^{\,}_{V}(\bm{0})&= 
\frac{2}{\pi^{2}\,r^{2}}\,
\Phi^{\,}_{V}(\bm{0}), 
\label{eq: OPEs in A b}
\\
\Phi^{\,}_{V}(\bm{r})\,\Phi^{\,}_{M}(\bm{0})&= 
\frac{1}{\pi^{2}\,r^{2}}\, 
\left[
\Phi^{\,}_{V}(\bm{0})
+ 
2\Phi^{\,}_{A}(\bm{0}) 
- 
\Phi^{\,}_{M}(\bm{0})
\right],
\label{eq: OPEs in A c}
\\
\Phi^{\,}_{M}(\bm{r})\,\Phi^{\,}_{A}(\bm{0})&= 
\frac{1}{\pi^{2}\,r^{2}}\,
[\Phi^{\,}_{M}(\bm{0}) + \Phi^{\,}_{V}(\bm{0})], 
\label{eq: OPEs in A d}
\\
\Phi^{\,}_{V}(\bm{r})\,\Phi^{\,}_{A}(\bm{0})&= 
\frac{1}{\pi^{2}\,r^{2}}\, 
\left[
\Phi^{\,}_{V}(\bm{0})
+
\Phi^{\,}_{M}(\bm{0})
\right],
\label{eq: OPEs in A e}
\end{align}
\end{subequations}
where we have only kept singular parts proportional to 
$r^{-2}\equiv\bm{r}^{-2}$
on the right hand sides.
Then, the RG equations for symmetry class A read
\begin{subequations}
\label{one-loop RG for class A}
\begin{align}
\frac{\mathrm{d}g^{\,}_{M}}{\mathrm{d}\ell}&= 
\frac{1}{\pi}
\left(
-
g^{2}_{M} 
-
g^{\,}_{V}\,g^{\,}_{M} 
+
g^{\,}_{V}\,g^{\,}_{A} 
+
g^{\,}_{A}\,g^{\,}_{M}
\right),
\\
\frac{\mathrm{d}g^{\,}_{V}}{\mathrm{d}\ell}&= 
\frac{1}{\pi}
\left(
g^{2}_{V} 
+ 
g^{\,}_{V}\,g^{\,}_{M} 
+ 
g^{\,}_{V}\,g^{\,}_{A} 
+ 
g^{\,}_{A}\,g^{\,}_{M}
\right), 
\\
\frac{\mathrm{d}g^{\,}_{A}}{\mathrm{d}\ell}&= 
\frac{2}{\pi}\,g^{\,}_{V}\, g^{\,}_{M}.
\end{align}
\end{subequations}
As is well known,~\cite{ludwig94}
there is an unstable line of critical points 
$g^{\,}_{V}=g^{\,}_{M}=0$ and $g^{\,}_{A}>0$.
However, for the generic case
$g^{\,}_{V}>0$ and $g^{\,}_{M}>0$, 
the coupling constants always flow to the strong-coupling regime.
If so, the one-loop RG equations are no longer applicable.
The (marginally relevant) coupling constants are represented by
the characteristic coupling $\mathsf{g}$
in the phase diagram of 2D class A
in Fig.~\ref{Fig: 2D phase diagrams}(a).
Since a stable metallic phase cannot exist in 2D systems of class A,
topologically distinct insulating phases are
separated by a phase boundary line on which wave functions are critical.

\subsubsection{Symmetry class C}

Next, we consider the 2D random Dirac Hamiltonian 
\begin{subequations}
\label{eq: def r=2 2D mathcal H in C}
\begin{align}
\mathcal{H}:=&
-\mathrm{i}
\sigma^{\,}_{x}\tensor \tau^{\,}_{0}\,
\partial^{\,}_{x}
-\mathrm{i}
\sigma^{\,}_{y}\tensor \tau^{\,}_{0}\, 
\partial^{\,}_{y} 
+ 
m\,\sigma^{\,}_{z}\tensor \tau^{\,}_{0} 
\n
&+\sum_{i=x,y}\sum_{j=x,y,z}A_{i,j} \,\sigma^{\,}_{i}\tensor \tau^{\,}_{j}
+\sum_{i=x,y,z}V_i \,\sigma^{\,}_{0}\tensor \tau^{\,}_{i}
\end{align}
of rank $r=4$ in the symmetry class C
with the operation for charge conjugation represented by
\begin{equation}
\mathcal{C}:=\sigma^{\,}_{x}\tensor \tau^{\,}_{y}\,\mathsf{K},
\end{equation}
\end{subequations}
where $\mathsf{K}$ denotes complex conjugation.

There are ten random functions allowed by the symmetry constraints
that multiply the matrices
\begin{align}
&
\sigma^{\,}_{z}\otimes\tau^{\,}_{0},
&
(\t{mass}) 
\n
&
\sigma^{\,}_{x}\otimes\tau^{\,}_{x}, 
~
\sigma^{\,}_{x}\otimes\tau^{\,}_{y}, 
~
\sigma^{\,}_{x}\otimes\tau^{\,}_{z}, 
~
& 
(\t{vector potentials}) 
\n
&
\sigma^{\,}_{y}\otimes\tau^{\,}_{x},
~
\sigma^{\,}_{y}\otimes\tau^{\,}_{y},
~
\sigma^{\,}_{y}\otimes\tau^{\,}_{z},
&
(\t{vector potentials}) 
\\
&
\sigma^{\,}_{0}\otimes\tau^{\,}_{x}, 
~
\sigma^{\,}_{0}\otimes\tau^{\,}_{x}, 
~
\sigma^{\,}_{0}\otimes\tau^{\,}_{x}, 
&
(\t{scalar potentials}) \nonumber
\end{align}
respectively.
Taking disorder average in the replicated action yields quartic
interaction terms similar to those in Eqs.\ 
(\ref{quartic terms in class A}).
The one-loop RG equations for the quartic terms can be found
in Refs.~\onlinecite{Nersesyan95,Mudry-Chamon-Wen}.
Here we discuss only a minimal set of the RG equations
that allow us to deduce the phase diagram
in Fig.\ \ref{Fig: 2D phase diagrams}(a).
Taking
$\sigma^{\,}_{x}\otimes\tau^{\,}_{x}$ and
$\sigma^{\,}_{y}\otimes\tau^{\,}_{x}$ from the vector potentials and 
$\sigma^{\,}_{0}\otimes\tau^{\,}_{x}$ from the scalar potential 
along with the mass $\sigma^{\,}_{z}\otimes\tau^{\,}_{0}$
among the above ten terms,
we find that the one-loop RG equations for the coupling constants
$g^{\,}_{V}$, $g^{\,}_{A}$, and $g^{\,}_{M}$ take the same form as those in
the symmetry class A
in Eq.\ (\ref{one-loop RG for class A}).
Thus, the representative coupling constant $\mathsf{g}$ of disorder is
marginally relevant as in the symmetry class A.
This indicates that
the phase diagram for the symmetry class C is qualitatively the same
as that for the symmetry class A, as shown 
in Figs.~\ref{Fig: 2D phase diagrams}(a) and (b).

\subsubsection{Symmetry class D \label{appsubsec: OPE class D}}

We consider the 2D random Dirac Hamiltonian 
\begin{subequations}
\label{eq: def r=2 2D mathcal H in D}
\begin{align}
\mathcal{H}:=
-\mathrm{i}
\sigma^{\,}_{x}\,
\partial^{\,}_{x}
-\mathrm{i}
\sigma^{\,}_{y}\, 
\partial^{\,}_{y} 
+ 
m\,\sigma^{\,}_{z}
\end{align}
of rank $r=2$ in the symmetry class D
with the operation for charge conjugation represented by
\begin{equation}
\mathcal{C}:=\sigma^{\,}_{x}\,\mathsf{K},
\end{equation}
\end{subequations}
where $\mathsf{K}$ denotes complex conjugation.
Disorder only enters through the mass matrix
$m\,\sigma^{\,}_{z}$ owing to the symmetry constraint
imposed by the PHS for the symmetry class D.

The OPE of the fermion quartic term, 
\begin{align}
\Phi^{\,}_{M}:=
-
\sum_{a,b=1}^{n} 
:\!
\psibar^{a}\, 
\sigma^{\,}_{z}\,
\psi^{a}\, 
\psibar^{b}\,
\sigma^{\,}_{z}\,
\psi^{b}
\!:,
\end{align}
with itself delivers the RG equation
\cite{Bocquet00,Mildenberger07}
\begin{align}
\frac{\mathrm{d}g^{\,}_{M}}{\mathrm{d}\ell}= 
-
\frac{1}{\pi}\, 
g^{2}_{M}.
\end{align}
Hence the variance $g^{\,}_{m}$ of the random mass term is
a marginally irrelevant coupling and the clean critical point 
$g^{\,}_{M}=\mathsf{m}=0$
is stable as long as the mean $\mathsf{m}=0$.

The random Dirac Hamiltonian of rank $r=4$ in the symmetry class D
with the PHS symmetry generated by conjugation with 
$C=\sigma^{\,}_{x}\otimes\tau^{\,}_{0}\,\mathsf{K}$
depends on six random functions that multiply the matrices
\begin{align}
&
\sigma^{\,}_{z}\otimes\tau^{\,}_{0},
~
\sigma^{\,}_{z}\otimes\tau^{\,}_{x},
~
\sigma^{\,}_{z}\otimes\tau^{\,}_{z},
&
(\t{masses}) 
\n
&
\sigma^{\,}_{x}\otimes\tau^{\,}_{y}, 
~
\sigma^{\,}_{y}\otimes\tau^{\,}_{y},
&
(\t{vector potentials}) 
\\
&
\sigma^{\,}_{0}\otimes\tau^{\,}_{y}, 
&
(\t{scalar potential}) 
\nonumber
\end{align}
respectively. 

Our aim is to explain why, unlike for the case of $r=2$
in Eq.~(\ref{eq: def r=2 2D mathcal H in D}),
the massless clean critical point described by 
\begin{equation}
\mathcal{H}^{\,}_{\star}:=
\left(
-\mathrm{i}
\sigma^{\,}_{x}\,
\partial^{\,}_{x}
-\mathrm{i}
\sigma^{\,}_{y}\, 
\partial^{\,}_{y} 
\right)
\otimes
\tau^{\,}_{0}
\end{equation}
is (i) unstable to disorder, (ii) this instability depends sensitively
on the choice of the probability distribution for the disorder,
and (iii) we are led to the choice
(\ref{eq: definition of m for irrelevant g}) 
for the probability distribution and to Fig.~\ref{Fig: 2D phase diagrams}(d)
to describe qualitatively the generic phase diagram
in the 2D symmetry class D with $N=2$.

We first choose the probability distribution
for the six random functions such that they are 
independent and Gaussian distributed with the means and variances
$\mathsf{m}\in\mathbb{R}$
and $\mathsf{g}^{\,}_{0}\geq0$
for the mass matrix
$\sigma^{\,}_{z}\otimes\tau^{\,}_{0}\equiv\beta^{\,}_{0}$;
$\mathsf{m}^{\,}_{x}\in\mathbb{R}$ and $\mathsf{g}^{\,}_{x}\geq0$ 
for the mass matrix
$\sigma^{\,}_{z}\otimes\tau^{\,}_{x}$;
$\mathsf{m}^{\,}_{z}\in\mathbb{R}$ and $\mathsf{g}^{\,}_{z}\geq0$ 
for the mass matrix
$\sigma^{\,}_{z}\otimes\tau^{\,}_{z}$;
$0$ and $g^{\,}_{A}\geq0$ for the vector potentials;
and
$0$ and $g^{\,}_{V}\geq0$ for the scalar potential; respectively.
With this convention, the parameter space for the phase diagram is
no less than eight dimensional with the point
\begin{equation}
0=
\mathsf{m}=\mathsf{g}^{\,}_{0}=
\mathsf{m}^{\,}_{x}=\mathsf{g}^{\,}_{x}=
\mathsf{m}^{\,}_{z}=\mathsf{g}^{\,}_{z}=
g^{\,}_{A}=g^{\,}_{V}
\label{appeq: def multicritical point star}
\end{equation}
corresponding to the massless clean Dirac Hamiltonian 
$\mathcal{H}^{\,}_{\star}$
of rank $r=4$ in the 2D symmetry class D.

\begin{figure*}[tb]
\begin{center}
\includegraphics[width=0.8\linewidth]{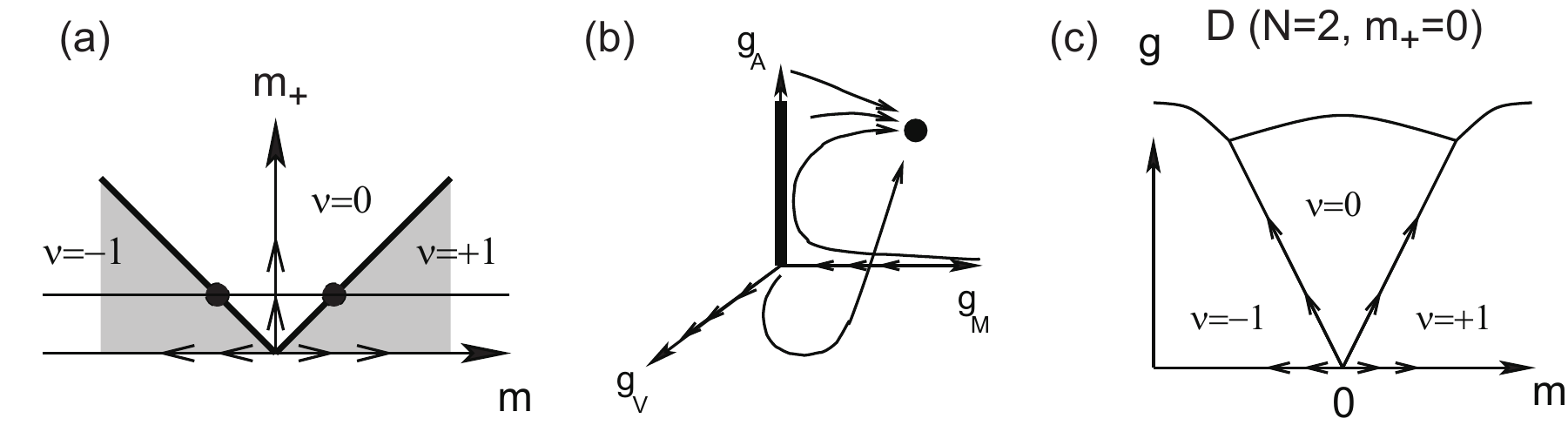}
\end{center}
\caption{\label{Fig: 2D N=2 class D}
(a) Phase diagram in the parameter space 
(\ref{appeq: def parameter space Omega}).
There are three gapped phases all meeting at the origin,
the clean massless Dirac Hamiltonian of rank $r=4$ in two-dimensional
space. The origin of parameter space is a critical point that
is unstable to both $\mathsf{m}$ or $\mathsf{m}^{\,}_{+}$
as is indicated by the arrows.
The gray region with the mass $\mathsf{m}>\mathsf{m}^{\,}_{\pm}>0$ supports
the dimensionless quantized thermal Hall conductivity $\nu=+1$.
The gray region with the mass $\mathsf{m}<-\mathsf{m}^{\,}_{+}<0$ supports
the dimensionless quantized thermal Hall conductivity $\nu=-1$.
The white region with $\mathsf{m}^{\,}_{+}>|\mathsf{m}|>0$ support
the dimensionless quantized thermal Hall conductivity $\nu=0$.
The critical lines $\mathsf{m}=\pm\mathsf{m}^{\,}_{+}$ separates
two insulating phases whose Chern numbers differ by the number 1 in magnitude.
The line parallel to the horizontal axis at fixed value of
$\mathsf{m}^{\,}_{+}$ intersects the phase boundaries at two
critical points depicted by two filled circles.
(b) Phase diagram in the parameter space 
$g^{\,}_{M},g^{\,}_{V},g^{\,}_{A}\geq0$
spanned by the variances of a random mass, a random scalar potential,
and a random vector potential (all of vanishing means)
in the 2D symmetry class A for a random
Dirac Hamiltonian of rank $r=2$.~\cite{ludwig94}
The bullet represents the stable flows 
to the critical point that describes the plateau transition
in the integer quantum Hall effect. The unstable flow
away from this critical point corresponds
to a non-vanishing mean value of the random mass 
and is thus not present in this three-dimensional parameter space.
(c) Projected phase diagram 
with the vanishing mean masses $\mathsf{m}^{\,}_{+}=0$
onto a two-dimensional cut with the same horizontal axis as in
(a) but with a vertical axis encoding multiples sources of disorder
including the variance $\mathsf{g}$ associated with the random mass of
mean $\mathsf{m}$. The direction $\mathsf{g}$
can be thought of as a cross section of 
all the directions associated with
the variances of the disorder coming out of the plane in 
(a).
}
\end{figure*}

It is natural to start with the clean limit
\begin{subequations}
\begin{equation}
0=
\mathsf{g}^{\,}_{0}=
\mathsf{g}^{\,}_{x}=
\mathsf{g}^{\,}_{z}=
g^{\,}_{A}=
g^{\,}_{V}.
\end{equation}
In this limit, we may write 
[compare with Eq.~(\ref{eq: definition of m for irrelevant g})]
\begin{equation}
\mathcal{H}=
\left(
-\mathrm{i}
\sigma^{\,}_{x}\,
\partial^{\,}_{x}
-\mathrm{i}
\sigma^{\,}_{y}\, 
\partial^{\,}_{y} 
\right)
\otimes
\tau^{\,}_{0}
+
\mathcal{V},
\end{equation}
where the Dirac mass $V$ is decomposed into
\begin{equation}
\mathcal{V}=
\mathsf{m}\,\beta^{\,}_{0}
+
\mathcal{V}^{\,}_{0}
\end{equation}
with
\begin{align}
\beta^{\,}_{0}&:=
\sigma^{\,}_{z}\otimes \tau^{\,}_{0},
&
\mathcal{V}^{\,}_{0}&:=
\sigma^{\,}_{z}\otimes M^{\,}_{0},
&
M^{\,}_{0}&:=
\mathsf{m}^{\,}_{x}\,
\tau^{\,}_{x}
+
\mathsf{m}^{\,}_{z}\,
\tau^{\,}_{z}.
\end{align}
\end{subequations}
The Hermitian $2\times2$ matrix $M^{\,}_{0}$ has the $N=2$
non-degenerate eigenvalues
\begin{equation}
\mathsf{m}^{\,}_{\pm}:=\pm\sqrt{\mathsf{m}^{2}_{x}+\mathsf{m}^{2}_{z}}
\end{equation}
provided $\mathsf{m}^{2}_{x}+\mathsf{m}^{2}_{z}>0$.
The four eigenvalues of $\mathcal{V}$ are
$\mathsf{m}^{\,}_{\pm}+\mathsf{m}$
and 
$\mathsf{m}^{\,}_{\pm}-\mathsf{m}$.
The relevant parameter space needed to identify the 
distinct topological phases is two-dimensional.
It is spanned by the parameters $\mathsf{m}$ and
$\mathsf{m}^{\,}_{\pm}$. 
There are three topologically distinct insulating phases that are defined by
\begin{subequations}
\label{appeq: r=4 2D D nu}
\begin{equation}
\nu=
\begin{cases}
+1,
&
\mathsf{m}>\mathsf{m}^{\,}_{+},
\\
\hphantom{+}0,
&
|\mathsf{m}|<|\mathsf{m}^{\,}_{\pm}|,
\\
-1,
&
\mathsf{m}<\mathsf{m}^{\,}_{-},
\end{cases}
\label{appeq: r=4 2D D nu a}
\end{equation}
where the index $\nu$ is given by [recall Eq.\ (\ref{eq: nu when z})] 
\begin{equation}
\nu=
\frac{1}{2}\,
\mathrm{sgn}\left(\mathsf{m}+|\mathsf{m}^{\,}_{+}|\right)
+
\frac{1}{2}\,
\mathrm{sgn}\left(\mathsf{m}-|\mathsf{m}^{\,}_{+}|\right).
\label{appeq: r=4 2D D nu b}
\end{equation}
The first term on the right-hand side is the 
dimensionless (thermal) Hall conductivity of a $2\times2$
massive Dirac Hamiltonian with the mass
$\mathsf{m}+|\mathsf{m}^{\,}_{+}|$.
The second term on the right-hand side is the 
dimensionless (thermal) Hall conductivity of a $2\times2$
massive Dirac Hamiltonian with the mass
$\mathsf{m}-|\mathsf{m}^{\,}_{+}|$.
A phase transition between the $|\nu|=1$ and $\nu=0$
insulating phases occurs whenever 
\begin{equation}
|\mathsf{m}|=|\mathsf{m}^{\,}_{+}|.
\label{appeq: r=4 2D D nu c}
\end{equation}
\end{subequations}
The phase boundaries
(\ref{appeq: r=4 2D D nu c})
in the two-dimensional parameter space,
\begin{equation}
\{(\mathsf{m},\mathsf{m}^{\,}_{+}) \in \mathbb{R}^2 ~|
~
\mathsf{m}^{\,}_{+} \geq 0 \},
\label{appeq: def parameter space Omega}
\end{equation}
meet at the massless Dirac point $\mathsf{m}=\mathsf{m}^{\,}_{+}=0$.
The massless Dirac point $\mathsf{m}=\mathsf{m}^{\,}_{+}=0$ 
is unstable to any non-vanishing  
$\mathsf{m}$ or $\mathsf{m}^{\,}_{+}$.
The line parallel to the horizontal axis
in Fig.~\ref{Fig: 2D N=2 class D}(a)
intersects the phase boundaries (\ref{appeq: r=4 2D D nu c})
at the two critical points $\mathsf{m}=\pm\mathsf{m}^{\,}_{+}$
at which the gap of one and only one of the two massive Dirac cones 
in the spectrum closes.

The origin of parameter space~(\ref{appeq: def parameter space Omega})
is a multicritical point, contrary to the two critical points at fixed value
of $\mathsf{m}^{\,}_{+}$ depicted by bullets in
Fig.~\ref{Fig: 2D N=2 class D}(a).
This difference manifests itself in the stability of these three points
to disorder. Our next task is to assess the effects of disorder on the
multicritical point at the origin of parameter space%
~(\ref{appeq: def parameter space Omega}).

If we choose the probability distribution with 
\begin{equation}
\begin{split}
0=
\mathsf{m}=
\mathsf{m}^{\,}_{x}=\mathsf{g}^{\,}_{x}=
\mathsf{m}^{\,}_{z}=\mathsf{g}^{\,}_{z}=0,
\quad
\mathsf{g},g^{\,}_{\mathrm{V}},g^{\,}_{\mathrm{A}}\geq0,
\end{split}
\label{appeq: vicinity star when g gA gV geq 0}
\end{equation}
then the random Dirac Hamiltonian is reducible, as
a rotation of $\tau^{\,}_{y}$ about $\tau^{\,}_{x}$ into $\tau^{\,}_{z}$ 
brings it to the representation 
\begin{equation}
\begin{split}
\mathcal{H}^{\,}_{\star\star}=&\,
\mathcal{H}^{\,}_{\star}
+
\left(
A^{\,}_{x}\,\sigma^{\,}_{x}
+
A^{\,}_{y}\,\sigma^{\,}_{y}
\right)
\otimes
\tau^{\,}_{z}
\\
&\,
+
M\,
\sigma^{\,}_{z}
\otimes
\tau^{\,}_{0}
+
V\,
\sigma^{\,}_{0}
\otimes
\tau^{\,}_{z}.
\end{split}
\label{appeq: 2D r=4 class D that is reducible}
\end{equation}
The random Dirac Hamiltonian $\mathcal{H}^{\,}_{\star\star}$
of rank $r=4$ is evidently block diagonal
with irreducible blocks of rank $r=2$ that belong to the symmetry
class A. Hence the variances 
$\mathsf{g}^{\,}_{0}$,
$g^{\,}_{V}$,
and
$g^{\,}_{A}$
obey the same one-loop RG flows as in Eq.%
~(\ref{one-loop RG for class A})
that we depict in Fig.~\ref{Fig: 2D N=2 class D}(b).
In particular, the variance $\mathsf{g}^{\,}_{0}$ 
of the random mass that commutes with
all other masses flows to the strong coupling 
as soon as either $g^{\,}_{V}$ 
or $g^{\,}_{A}$ is non-vanishing.

The reducibility of the rank $r=4$ Dirac Hamiltonian
in the 2D symmetry class D is non-generic. It is broken as soon
as any one of the two variances $\mathsf{g}^{\,}_{x}$ or 
$\mathsf{g}^{\,}_{z}$
for the anticommuting pair of Dirac mass matrices
is non-vanishing. In the parameter space
\begin{equation}
0=\mathsf{m}^{\,}_{x}=\mathsf{m}^{\,}_{z},
\qquad
|\mathsf{m}|, \mathsf{g}^{\,}_{0}, \mathsf{g}^{\,}_{x}, \mathsf{g}^{\,}_{z},
g^{\,}_{A}, g^{\,}_{V}>0,
\end{equation}
the variance $\mathsf{g}$ is then expected to inherit
the marginal relevance of the multicritical point
(\ref{appeq: def multicritical point star})
in the neighborhood (\ref{appeq: vicinity star when g gA gV geq 0}).
If the multidimensional parameter space is projected onto 
a two-dimensional plane parametrized by the probability distribution
(\ref{alternative definition 0 a}),
we conjecture by analogy to 
Figs.~\ref{Fig: 1D phase diagrams}(a)
and \ref{Fig: 2D phase diagrams}(b)
the phase diagram shown in 
Fig.~\ref{Fig: 2D N=2 class D}(c).
This phase diagram is not the only one dictated by logic.
It has the merit of simplicity, however.
For example, its RG flows are more economical than conjecturing
that the metallic phase precludes any direct phase boundary between
two topologically distinct localized phase for any non-vanishing
variance of the randomness.

The stability analysis of the two critical points intersecting
the phase boundary at a non-vanishing fixed value of $\mathsf{m}^{\,}_{+}$
in Fig.~\ref{Fig: 2D N=2 class D}(a)
in the presence of a generic disorder from the symmetry class D
is quite different from that of the multicritical point.
We assume that the fluctuations are much smaller than the means
at the two critical points defined by
$|\mathsf{m}|=\mathsf{m}^{\,}_{+}$ with $\mathsf{m}^{\,}_{+}>0$ given
in Fig.~\ref{Fig: 2D N=2 class D}(a).
This assumption justifies projecting
the Dirac Hamiltonian of rank $r=4$ onto the eigenspace of
$\overline{m^{\,}_{x}}\,\tau^{\,}_{x}+\overline{m^{\,}_{z}}\,\tau^{\,}_{z}$
with the eigenvalue
$\mathsf{m}^{\,}_{+}$.%
~\cite{ludwig94} 
The resulting projected Dirac Hamiltonian is of rank $r=2$ and
has the same form as the rank $r=2$ Hamiltonian defined in Eq.%
~(\ref{eq: def r=2 2D mathcal H in D}).
We can then apply the RG analysis for the Hamiltonian of rank $r=2$ 
in Eq.\ (\ref{eq: def r=2 2D mathcal H in D}) to find that the 
effective coupling playing the role of
$g^{\,}_{M}$ in Eq.\ (\ref{eq: def r=2 2D mathcal H in D})
is irrelevant along the phase boundary
defined by the gap closing condition 
$\mathsf{m}=\mathsf{m}^{\,}_{+}$.
This reasoning explains why a non-vanishing mean value
$\mathsf{m}^{\,}_{+}>0$ is required
in Eq.~(\ref{eq: definition of m for irrelevant g})
in order for one of the $N$ clean critical point at $\mathsf{g}=0$ 
to be the end point of a phase boundary that
separates a pair among $N+1$ topologically distinct localized phases.
This end point is the critical point that governs criticality at the phase
boundary between two topologically distinct localized phases.
No metallic phase prevents a direct transition between this pair
of topologically distinct localized phases for not too strong disorder.
The phase diagrams when $N=1$ and $N=2$ are depicted in
Figs.~\ref{Fig: 2D phase diagrams}(c)
and \ref{Fig: 2D phase diagrams}(d),
respectively. A phase diagram similar to that of 
Fig.\ \ref{Fig: 2D phase diagrams}(c) has been
obtained from numerical simulations of network models in Refs.%
~\onlinecite{chalker-classD01,Mildenberger07}.

We close the discussion of the 2D symmetry class D with a
generic $r=4$ random Dirac Hamiltonian by observing that the
Dirac mass
\begin{subequations}
\label{appeq: def vortex in mass}
\begin{equation}
\begin{split}
\mathcal{V}^{\,}_{0}(\bm{r}):=&\,
m^{\,}_{x}(\bm{r})\,
\sigma^{\,}_{z}\otimes\tau^{\,}_{x}
+
m^{\,}_{z}(\bm{r})\,
\sigma^{\,}_{z}\otimes\tau^{\,}_{z}
\\
=&\,
m^{\,}_{+}(\bm{r})\,
\left[
\cos\theta(\bm{r})\,
\sigma^{\,}_{z}\otimes\tau^{\,}_{x}
+
\sin\theta(\bm{r})\,
\sigma^{\,}_{z}\otimes\tau^{\,}_{z}
\right],
\end{split}
\label{appeq: def vortex in mass a}
\end{equation}
where
\begin{equation}
m^{\,}_{+}(\bm{r}):=
\sqrt{
m^{2}_{x}(\bm{r})
+
m^{2}_{z}(\bm{r})
     },
\qquad
\theta(\bm{r}):=
\frac{m^{\,}_{z}(\bm{r})}{m^{\,}_{x}(\bm{r})},
\label{appeq: def vortex in mass b}
\end{equation}
is generated by two anticommuting mass matrices, i.e.,
\begin{equation}
\mathcal{V}^{2}_{0}(\bm{r})=
m^{2}_{+}(\bm{r})\,
\sigma^{\,}_{0}\otimes\tau^{\,}_{0}.
\label{appeq: def vortex in mass c}
\end{equation}
\end{subequations}
The Dirac mass 
(\ref{appeq: def vortex in mass})
supports point defects defined by the two conditions that
(i) $\lim_{\bm{r}\to\infty} m^{\,}_{+}(\bm{r})=\mathsf{m}^{\,}_{+}$
and (ii) the phase $\theta(\bm{r})$ winds by an integer multiple of
$2\pi$ along any closed path $\Gamma^{\,}_{\bm{r}^{\,}_{\mathrm{vtx}}}$
belonging to any open disk centered about
the point $\bm{r}^{\,}_{\mathrm{vtx}}$ that is not too large. The point 
$\bm{r}^{\,}_{\mathrm{vtx}}$ thus represents a point-like singularity
of the vortex type. Jackiw and Rossi have shown in Ref.%
~\onlinecite{jackiw-rossi81}
that Dirac Hamiltonians of the form
\begin{equation}
\mathcal{H}^{\,}_{\mathrm{vtx}}:=
\mathcal{H}^{\,}_{\star}
+
\mathcal{V}^{\,}_{\mathrm{vtx}}(\bm{r})
\end{equation}
support $n$ normalized zero modes if 
$\mathcal{V}^{\,}_{\mathrm{vtx}}(\bm{r})$
is given by Eq.~(\ref{appeq: def vortex in mass})
with the phase $\theta(\bm{r})$ winding by $2\pi\,n$ 
along all circles of sufficiently large radii. 
These zero modes are not robust to all remaining random channels 
compatible with the 2D symmetry class D. However, if $n$ is an odd integer,
at least one zero mode survives any perturbation arising from the 
scalar gauge, vector gauge, and the last massive channel, provided
this massive channel associated to the Dirac mass matrix that commutes
with $\mathcal{V}^{\,}_{\mathrm{vtx}}(\bm{r})$
is everywhere smaller in magnitude than 
the asymptotic gap $\mathsf{m}^{\,}_{+}$. 
This is an illustration of the fact that
the fundamental homotopy group for the normalized Dirac mass matrices
in the 2D symmetry class D can be $\mathbb{Z}^{\,}_{2}$. Correspondingly,
the localized phase $\nu=0$ in the phase diagram 
\ref{Fig: 2D phase diagrams}(d)
is conjectured to support $\mathbb{Z}^{\,}_{2}$ 
quasi-zero modes that will contribute to the density of states.

\subsubsection{Symmetry class DIII}

We consider the 2D random Dirac Hamiltonian 
\begin{subequations}
\begin{align}
\mathcal{H}:=
-\mathrm{i}
\sigma^{\,}_{x}\!\otimes\!\tau^{\,}_{0}\,
\partial^{\,}_{x}
-\mathrm{i}
\sigma^{\,}_{y}\!\otimes\!\tau^{\,}_{0}\, 
\partial^{\,}_{y} 
+ 
m\, 
\sigma^{\,}_{z}\!\otimes\!\tau^{\,}_{y} 
+ 
V\, 
\sigma^{\,}_{0}\!\otimes\!\tau^{\,}_{x}
\label{eq: DIII random potentials}
\end{align}
of rank $r=4$ in the symmetry class DIII
with the operations for
reversal of time and charge conjugation represented by
\begin{align}
\mathcal{T}:=
\mathrm{i}\sigma^{\,}_{y}\otimes\tau^{\,}_{0}\,\mathsf{K}, 
\qquad 
\mathcal{C}:=
\sigma^{\,}_{x}\otimes\tau^{\,}_{z}\,\mathsf{K},
\end{align}
\end{subequations}
respectively.

There are two random functions allowed by the symmetry constraints
that multiply the matrices
\begin{equation}
\begin{split}
&
\sigma^{\,}_{z}\otimes\tau^{\,}_{y},
\qquad \hbox{(random mass)}
\\
&
\sigma^{\,}_{0}\otimes\tau^{\,}_{x}.
\qquad \hbox{(random scalar potential)}
\end{split}
\end{equation}
These couplings are the coefficients of  
the fermion quartic terms
\begin{subequations}
\begin{align}
\Phi^{\,}_{V}&=
-\sum_{a,b=1}^{n} 
:\!\psibar^{a}\, \sigma^{\,}_{0}\otimes\tau^{\,}_{x}\, \psi^{a}\,
   \psibar^{b}\, \sigma^{\,}_{0}\otimes\tau^{\,}_{x}\, \psi^{b}\!:, 
\\
\Phi^{\,}_{M}&=
-\sum_{a,b=1}^{n} 
:\!\psibar^{a}\, \sigma^{\,}_{z}\otimes\tau^{\,}_{y}\, \psi^{a}\,
   \psibar^{b}\, \sigma^{\,}_{z}\otimes\tau^{\,}_{y}\, \psi^{b}\!:, 
\end{align}
\end{subequations}
respectively. Their OPEs are
\begin{subequations}
\label{eq: OPEs in DIII}
\begin{align}
\Phi^{\,}_{M}(\bm{r})\,\Phi^{\,}_{M}(\bm{0}) &= 
\frac{-2}{\pi^{2}\,r^{2}}\,
\Phi^{\,}_{M}(\bm{0}), 
\label{eq: OPEs in DIII a}
\\
\Phi^{\,}_{V}(\bm{r})\,\Phi^{\,}_{V}(\bm{0})&= 
\frac{2}{\pi^{2}\,r^{2}}\,
\Phi^{\,}_{V}(\bm{0}), 
\label{eq: OPEs in DIII b}
\\
\Phi^{\,}_{V}(\bm{r})\,\Phi^{\,}_{M}(\bm{0})&= 
\frac{-1}{\pi^{2}\,r^{2}}\, 
\left[
\Phi^{\,}_{V}(\bm{0}) 
-
\Phi^{\,}_{M}(\bm{0})
\right].
\label{eq: OPEs in DIII c}
\end{align}
\end{subequations}
We note that
(i) the OPEs
(\ref{eq: OPEs in DIII a})
and
(\ref{eq: OPEs in A a})
are identical,
(ii) the OPEs
(\ref{eq: OPEs in DIII b})
and
(\ref{eq: OPEs in A b})
are identical,
while (iii) the OPE
(\ref{eq: OPEs in DIII c})
differs by an overall sign from the OPE
(\ref{eq: OPEs in A c})
if one ignores $\Phi^{\,}_{A}$ in the latter OPE.
The following one-loop RG equations follow from this observation, 
\begin{subequations}
\begin{align}
\beta^{\,}_{M}&= 
\frac{1}{\pi}\,
g^{\,}_{M}\,
\left(g^{\,}_{V}-g^{\,}_{M}\right), 
\\
\beta^{\,}_{V}&= 
\frac{1}{\pi}\,
g^{\,}_{V}\,
\left(g^{\,}_{V}-g^{\,}_{M}\right).
\end{align}
\end{subequations}
Addition and subtraction gives
\begin{subequations}
\label{relevant combination}
\begin{align}
&
\frac{\mathrm{d}(g^{\,}_{V}+g^{\,}_{M})}{\mathrm{d}\ell}=
\frac{1}{\pi}\,
\left(g^{2}_{V}-g^{2}_{M}\right),
\label{relevant combination a}
\\
&
\frac{\mathrm{d}(g^{\,}_{V}-g^{\,}_{M})}{\mathrm{d}\ell}=
\frac{1}{\pi}\,
\left(g^{\,}_{V}-g^{\,}_{M}\right)^{2}.
\label{relevant combination b}
\end{align}
\end{subequations}
We have found the marginally relevant linear combination 
(\ref{relevant combination b})
of the couplings associated to the disorder. 
We deduce that a metallic phase 
must separate the only two
topologically distinct localized phases in class DIII
to explain the multicritical nature of
the clean critical point $\mathsf{g}=\mathsf{m}=0$
as depicted in Fig.~\ref{Fig: 2D phase diagrams}(e),
in agreement with numerical simulations of
a network model.~\cite{Fulga-DIII-12}

\subsubsection{Symmetry class AII}

The one-loop RG analysis of a Dirac Hamiltonian of rank $r=4$
in the symmetry class AII can be found, e.g., in Ref.~\onlinecite{Mong12}.
Here we give a brief summary of the relevant results.
The $r=4$ Dirac Hamiltonian with the operation of reversal of time
represented by 
$\mathcal{T}:=\mathrm{i}\sigma^{\,}_{y}\otimes\tau^{\,}_{0}\,\mathsf{K}$ 
has symmetry-allowed random perturbations that include
those allowed in the symmetry class DIII 
[Eq.~(\ref{eq: DIII random potentials})].
Therefore, the set of one-loop RG equations for the symmetry class AII,
which is given in Ref.~\onlinecite{Mong12},
contains the one for the symmetry class DIII as a subset.
Thus, we have the same linear combination 
of coupling constants as in the symmetry class DIII
that is marginally relevant 
[Eq.~(\ref{relevant combination})].
This indicates that a metallic phase always intervenes between
two topologically distinct localized phases in class AII,
as is shown in Fig.~\ref{Fig: 2D phase diagrams}(g).%
~\cite{kraus-stern-11,Ringel12,Mong12,fu-kane-average12,Morimoto-weak14,Obuse-weakTI13}

\bibliography{localization}

\begin{thebibliography}{136}
\bibitem [{1Fo()}]{1Footnote-on-local-disorder}%
  \BibitemOpen
  \href@noop {} {}\bibinfo {note} {A local random potential is a statistical
  distribution of potentials such that all $n$-th order cumulants of the random
  potential at $n$ distinct points in space decay exponentially fast with the
  pairwise separation of any two points.}\BibitemShut {Stop}%
\bibitem [{2Fo()}]{2Footnote-on-bulk-gap}%
  \BibitemOpen
  \href@noop {} {}\bibinfo {note} {Strictly speaking, the energy spectrum is
  gapless in the presence of disorder. By bulk gap, we refer to the energy
  window that would not support energy eigenstates in the clean limit. For weak
  disorder the density of states in the bulk gap is exponentially
  small.}\BibitemShut {Stop}%
\bibitem [{foo()}]{footnote-classD}%
  \BibitemOpen
  \href@noop {} {}\bibinfo {note} {Not all topologically nontrivial phases in
  the symmetry class D can be brought to the form of the Dirac Hamiltonian
  studied here. For example, a spinless chiral $p$-wave superconducting ground
  state supports two topological phases that may be labeled by the index
  $\pm1$, each of which support $\mathbb{Z}^{\,}_{2}$ vortices that bind
  Majorana zero modes.\cite{read-green00} These $\mathbb{Z}^{\,}_{2}$ vortices
  are induced by a magnetic flux, not by a $\mathbb{Z}^{\,}_{2}$ vortex in a
  Dirac mass.}\BibitemShut {Stop}%
\bibitem [{3Fo()}]{3Footnote-on-exampleSnTe}%
  \BibitemOpen
  \href@noop {} {}\bibinfo {note} {Observe that the classifying space is the
  set of two points $\{\pm\beta\}$ for the minimal rank
  $r=r^{\,}_{\mathrm{min}}$.}\BibitemShut {Stop}%
\bibitem [{4Fo()}]{4Footnote-on-L-points}%
  \BibitemOpen
  \href@noop {} {}\bibinfo {note} {For one mirror plane, the mirror Chern
  number is 2. It protects the Dirac cones at two L points out of the four
  inequivalent L points. Here, the four inequivalent L points can be arranged
  into two pairs related by a $\pi/2$ rotation of the mirror
  plane.}\BibitemShut {Stop}%
\end{thebibliography}

\end{document}